\documentclass[prd,nofootinbib,showpacs,preprint]{revtex4}
\usepackage{graphicx}
\usepackage{epsfig}
\usepackage{amsmath}
\usepackage{amsfonts}
\usepackage{amssymb}
\usepackage{url}
\usepackage{hyperref}
\usepackage{subfigure}
\usepackage{cancel}
\newcommand{\bd}{\begin{document}}
\newcommand{\ed}{\end{document}}
\newcommand{\bc}{\begin{center}}
\newcommand{\ec}{\end{center}}
\newcommand{\beqa}{\begin{eqnarray}} 
\newcommand{\eeqa}{\end{eqnarray}} 
\newcommand{\beq}{\begin{equation}} 
\newcommand{\eeq}{\end{equation}} 
\newcommand{\lsim}{\lesssim}
\newcommand{\gsim}{\gtrsim}
\newcommand{\nn}{\nonumber}
\newcommand{\bmt}{\begin{pmatrix}}
\newcommand{\emt}{\end{pmatrix}}
\usepackage[toc,page]{appendix}
\usepackage{comment} 
\def\TeV{\mbox{TeV}}
\def\GeV{\mbox{GeV}}
\def\MeV{\mbox{MeV}}
\def\eV{\mbox{eV}}
\def\keV{\mbox{keV}}
\def\s1{\hat s}
\def\ds{\displaystyle}
\def\lb{\Lambda_b}
\def\ll{\Lambda}
\newcommand{\be}{\begin{equation}}
\newcommand{\ee}{\end{equation}}
\newcommand{\bea}{\begin{eqnarray}}
\newcommand{\eea}{\end{eqnarray}}
\newcommand{\bref}[1]{(\ref{#1})}
\def\slashi#1{\rlap{\sl/}#1}
\newcommand{\<}[1]{\langle {#1} \rangle}

\usepackage{verbatim}   
\usepackage{multirow}   

\allowdisplaybreaks[1]

\def \ket#1{ \left| #1 \right\rangle}
\def \brac#1{\left\langle #1 \right|}
\def \abs#1{ \left| #1 \right| }
\def \order#1{ {\cal O} \left( #1 \right) }
\def \dslash#1{ #1\!\!\!/}

\def \eps{\varepsilon}

\def \BtoKll{{\bar{B} \to K \bar{l}l}}
\def \gS{g_s}              
\def \alS{\alpha_s}        
\def \alE{\alpha_e}        
\def \GF{G_F}              
\def \LamConf{{\Lambda_{\rm QCD}}}    
\def \LamConfS{{\Lambda^2_{\rm QCD}}} 

\def \cA{{\cal A}}
\def \cC{{\cal C}}
\def \cF{{\cal F}}
\def \cL{{\cal L}}
\def \cM{{\cal M}}

\def \Gaml{{\Gamma_l}}
\def \Game{{\Gamma_e}}
\def \Gammu{{\Gamma_\mu}}

\def \BRl{{{\cal B}_l}}
\def \BRe{{{\cal B}_e}}
\def \BRmu{{{\cal B}_\mu}}

\def \AFBl{{A_{\rm FB}^l}}
\def \AFBmu{{A_{\rm FB}^\mu}}
\def \AFBe{{A_{\rm FB}^e}}

\def \FHl{{F_H^l}}
\def \FHe{{F_H^e}}
\def \FHmu{{F_H^\mu}}
\begin{document}
\title{\bf Rare semileptonic $B \to K(\pi)l_i^- l_j^+$ decay in vector leptoquark model } 
\author{Murugeswaran Duraisamy}
\affiliation{ Department of Physics, Virginia Commonwealth University, Richmond, VA 23284, USA}
\author{Suchismita Sahoo}
\author{Rukmani Mohanta }
\affiliation{\,School of Physics, University of Hyderabad, 
              Hyderabad - 500046, India  }      
\begin{abstract}
We investigate the consequence of vector leptoquarks on  the rare semileptonic lepton flavour violating decays of $B$ meson which are more promising and effective channels to probe the new physics signal.
We constrain the resulting new  leptoquark parameter space by using the  branching ratios of $B_{s, d} \to l^+ l^-$, $K_L \to l^+ l^-$ and $\tau^- \to l^- \gamma$ processes. We estimate the branching ratios of rare lepton flavour violating $B \to K(\pi)l_i^- l_j^+$ processes  using the  constrained leptoquark couplings. We also compute the forward-backward asymmetries and the  lepton non-universality parameters  of the LFV decays in the vector leptoquark model.  Furthermore, we study the effect of vector leptoquark on $(g-2)_\mu$ anomaly.
\end{abstract}
\maketitle
\section{Introduction}
The discovery of Higgs boson at  LHC completes the  standard model (SM) picture of particle interactions, which is quite successful in describing all the observed experimental data so far below the electroweak scale. Still we need physics beyond it in order to solve the hierarchy and  flavour problems. In this context the study of rare $B$ decay modes involving the flavour changing neutral current (FCNC) transitions, $b \to s/d$, are more captivating. The FCNC processes are highly suppressed in the SM and occur via one-loop level only. It should be noted that the current measured data by LHCb collaboration  on angular observables in rare $B$ decays show significant deviation from the SM predictions. Especially, the discrepancy of $3\sigma$  in the famous $P_5^\prime$ angular observable \cite{p5p,lhcb1} and  the  decay rate \cite{lhcb2} of rare $B \to K^* \mu^+ \mu^-$  processes  have become a tension in recent times.  In addition the ratio $R_K = {\rm Br}(B \to K \mu^+ \mu^-)/{\rm Br}(B \to K e^+ e^-)$, cancelling the hadronic uncertainties to a very large extent, has also $2.6\sigma$ deviation from the SM prediction \cite{lhcb3,bobeth2}, thus indicates the violation of the lepton flavour universality (LFU). The decay rate of  $B_s \to \phi \mu^+ \mu^-$ process is also low ($3\sigma$ deviation) compared to its SM value \cite{lhcb4}.

Within the SM of electroweak interactions, the generation lepton number is exactly conserved, since the neutrinos are deemed as massless particles. Nonetheless, the observation of neutrino oscillation has provided unambiguous evidence for lepton number violation in the neutral sector. The observation of lepton non-universality   by the LHCb collaboration generically implies the existence of lepton flavour violating (LFV) decay processes. Since the observed data on lepton non-universality is due to $25\%$ deficit in the muon channel, thus LFV is more for muonic  processes than for electronic processes \cite{glashow}. The branching ratio of $h \to \tau \mu$ LFV decay is found to be ${\rm Br}(h \to \tau \mu) = 0.84^{+0.39}_{-0.37}$ by CMS collaboration \cite{h-to-taumu}, which has a  $2.6\sigma$ deviation from the SM value, thus boosted the interest of physicists to study more LFV decay processes in charged sector such as $l_i \to l_j \gamma$, $l_i \to l_j l_k \bar{l}_k$, $B_s \to l_i^\pm l_j^\mp$ and $B \to K^{(*)} l_i^\pm l_j^\mp$ etc. Theoretically, the LFV processes are free from the non-perturbative hadronic effects and significantly contribute some additional operators in comparison with the lepton flavour conserving (LFC) processes. In the literature, there  are many attempts to analyze the LFV decays in the $B$-sector in terms of various beyond the standard model scenarios \cite{lfv, mohanta2, mohanta4, kosnik-new}.  Even though there is no direct experimental measurement on such LFV processes, but there exist upper bounds on some of these decays \cite{pdg}.  The observation of lepton flavour violating decays in the upcoming  and/or future experiments would provide evidence of new physics beyond the SM.

To settle the  observed anomalies at LHCb using a specific theoretical framework, we extend the SM by adding a single vector leptoquark (LQ), which  is a color triplet boson and arises naturally from the unification of quarks and leptons. LQs carry both baryon and lepton numbers and can be characterised by their fermion no., spin and charge. Since $1980$'s LQs had been  enthusiastically  searched for, yet without any positive results, though LQs could be produced directly at the colliders. The existence of LQ can be found in many new physics (NP) models, such as the grand unified theories \cite{georgi, georgi2}, Pati-Salam model, quark and lepton composite model \cite{kaplan} and the technicolor model \cite{schrempp}. The  lepton and baryon number violating LQs are very heavy to avoid proton decay bounds.  Nevertheless, the LQs having the baryon and lepton number conserving couplings do not allow proton decay and could be light enough to be seen in the current experiments.  The interaction of LQ with the SM fermions could be due to a scalar LQ doublet with representation $(3, 2, 7/6)$ and $(3, 2, 1/6)$ or 
 a  vector LQ triplet $V_\mu^3 (3,3,2/3)$,  singlet $V_\mu^1 (3,1,2/3)$ or doublet $V_\mu^2 (\bar{3},2,5/6)$ under the SM $SU(3)_C \times SU(2)_L \times U(1)_Y$ gauge group. In this work, we consider the vector LQ model which can produce both scalar and pseudoscalar operators in addition to the vector currents.  We assume that the LQs  conserve $B$ and $L$ quantum numbers and do not induce proton decay. 
 We investigate the  LFV $B \to K(\pi) l_i^- l_j^+$  processes in the context of vector LQ model. Even though the LFV processes occur at loop level with the presence of  massless neutrinos in one of the loop or proceed via box  diagrams, these can occur at tree level in the LQ model and are expected to have significantly large branching ratios.  We compute the branching ratios and forward-backward asymmetries in these LFV processes. In addition, we also check the existence of lepton non-universality   in the LQ model.  The complete LQ phenomenology and the additional new physics contribution to the $B$-sector has been investigated in the literature \cite{mohanta1, mohanta2, mohanta3, mohanta4, kosnik, leptoquark, 10-LQ, RD-star-LQ, tau-mu-gamma}.

The paper is organized as follows. In section II, we present the effective Hamiltonian describing the $b \to q l_i^- l_j^+$  transitions, where $q=s, d$. The angular distribution and the decay parameters of the semileptonic lepton flavour  violating decays are described in section III.  In section IV, we  discuss the new physics contribution due to the exchange of vector LQ  and the constraints on LQ couplings from $B_{s, d} \to l^+ l^-$, $K_L \to l^+ l^-$ and $\tau^- \to l^- \gamma$ processes are computed in section V. The branching ratios, forward-backward asymmetries and the lepton non-universality  of $B \to K (\pi) l^-_i l^+_j$ LFV decays are computed in section VI.
Finally in section VII  we explain the muon $g-2$ anomaly  
and the  conclusions are summarized  in section VIII.

\section{Effective Hamiltonian for $b \to q l_i^- l_j^+$ processes}
In this section we discuss the effective Hamiltonian describing the FCNC $b \to q (=d, s)l_i^- l_j^+$  transitions. Here we will focus mainly on the $b \to s l_i^- l_j^+$ Hamiltonian as the $b \to d l_i^- l_j^+$ Hamiltonian can be obtained from it with the obvious replacements. 
The effective Hamiltonian for the quark-level transition $b \to s l_i^- l_j^+~(l=e, \mu, \tau)$ in the SM is mainly given by \cite{Hamiltonian}
\bea
\mathcal{H}_{\rm eff}^{\rm SM}&=&-\frac{4G_F}{\sqrt{2}} V_{ts}^* V_{tb} \Big[ \sum_{i=1}^6 C_i\left(\mu\right) \mathcal{O}_i\left(\mu\right)+C_7^{\rm SM} \frac{e}{16 \pi^2} \left[ \bar{s} \sigma_{\mu \nu} \left(m_s P_L + m_b P_R\right)b \right]F^{\mu \nu}  \nn \\ &&+ C_V^{\rm SM} \frac{\alpha_{em}}{4\pi} \left(\bar{s}\gamma^\mu P_L b \right) L_{ij}^\mu +C_{A}^{\rm SM} \frac{\alpha_{em}}{4\pi} \left(\bar{s}\gamma^\mu P_L b \right) L_{ij}^{5\mu} \Big],
\label{SM-Ham}
\eea  
where $L_{ij}^\mu = \bar{l}_i \gamma_\mu l_j$ and $L_{ij}^{5 \mu } = \bar{l}_i \gamma_\mu \gamma_5 l_j$.  Here $G_F$ denotes the Fermi constant, $V_{qq^\prime}$ are the  Cabibbo-Kobayashi-Maskaw (CKM) matrix elements, $\alpha_{em}$ is the fine structure constant and  $P_{L, R}=\left(1\mp \gamma_5 \right)/2$ are the chirality projection operators.  The operators $\mathcal{O}_i \left(i=1,...,6\right)$ correspond to the  tree level current-current operators ($O_{1,2}$),
QCD penguin operators ($O_{3-6}$)  and $C_i$'s are the Wilson coefficients. For $i=j$, $C_{7,V,A}^{\rm SM}$ represent the SM Wilson coefficients  $C_{7,9,10}$ and for $i\neq j$ they will vanish.

The total effective Hamiltonian for processes involving  $b \to s l_i^- l_j^+$ transition, in the presence of new physics operators with all the possible Lorentz structure,   can be expressed as 
\bea
\mathcal{H}_{\rm eff}\left( b \to s l_i^- l_j^+ \right) = \mathcal{H}_{\rm eff}^{\rm SM}+\mathcal{H}_{\rm eff}^{\rm VA}+ \mathcal{H}_{\rm eff}^{\rm SP}+\mathcal{H}_{\rm eff}^T,
\label{HM}
\eea
where $\mathcal{H}_{\rm eff}^{\rm SM}$ is the SM effective Hamiltonian as given in Eqn. (\ref{SM-Ham}), and the NP contributions are given as
\bea 
\mathcal{H}_{\rm eff}^{\rm VA} &=& -N_F \Big[ C_V \left( \bar{s} \gamma^\mu P_L b \right) L_{ij}^\mu +C_A \left( \bar{s} \gamma^\mu P_L b \right) L_{ij}^{5 \mu}  + C_V^\prime \left( \bar{s} \gamma^\mu P_R b \right) L_{ij}^\mu \nn \\ && +C_A^\prime \left( \bar{s} \gamma^\mu P_R b \right) L_{ij}^{5 \mu}\Big], \\
\mathcal{H}_{\rm eff}^{\rm SP} &=& -N_F \Big[ C_S \left( \bar{s} P_R b \right) L_{ij} +C_P \left( \bar{s}  P_R b \right) L_{ij}^{5 } + C_S^\prime \left( \bar{s}  P_L b \right) L_{ij} +C_P^\prime \left( \bar{s} P_L b \right) L_{ij}^{5}\Big], 
\\
\mathcal{H}_{\rm eff}^{T} &=& -N_F \Big[ 2C_T \left( \bar{s} \sigma_{\mu \nu} b \right) L_{ij}^{\mu \nu}+i 2 C_{T5} \left( \bar{s} \sigma_{\mu \nu }  b \right) L_{ij}^{ \mu \nu 5}\Big], 
\eea 
where  $N_F = \frac{G_F \alpha_{em}}{\sqrt{2} \pi}  V_{tb} V_{ts}^*$, $L_{ij}^{5} = \bar{l}_i \gamma_5  l_j$, and $L_{ij}^{\mu \nu 5} = 2i \bar{l_i}\sigma^{\mu \nu}\gamma_5l_j$. Here we use $\sigma^{\mu \nu} \gamma_5 = -\frac{i}{2} \epsilon^{\mu \nu \alpha \beta} \sigma_{\alpha \beta}$ to calculate $L_{ij}^{\mu \nu 5}$. In the above expressions $C_i^{(\prime)}$, where $i=V, A, S, P$, and $C_{T(5)}$ are the NP effective couplings which  are negligible in the SM and can only be generated using new physics beyond the SM.

\section{Theoretical Framework for $\bar{B} \to \bar{K} (\pi) l_i l_j$ decay processes}

The semileptonic $\bar{B} \to \bar{K} l_i l_j$ decay involves the quark level $b \to s l_i^- l_j^+$ transitions as mediated by the effective Hamiltonian of the form in Eqn.(\ref{HM}). The relevant kinematical variables describing this three-body decay are the invariant mass squared of the lepton pair $q^2 = (P_B-P_K)^2$, and the polar angle $\theta_l$. Here $P_B$ and $P_K$ are the four-momenta of the $B$ meson and $K$ meson respectively and $\theta_l$ is the angle between the $K$ and lepton $l_i$ in the $l_i - l_j$ rest frame. 
The polar angle differential decay distribution in the momentum transfer squared $q^2$ for the process $\bar{B} \to \bar{K} l_i l_j$ can be written in the form 
\bea
\frac{d^2 \Gamma}{dq^2 d\cos\theta_l} = \frac{G_F^2 \alpha^2_{em} \beta_{ij} \sqrt{\lambda}|V_{tb}V_{ts}^*|^2}{2^{12} \pi^5 M_B^3} \sum_{i=1}^{12}I_i\left(\cos\theta_l\right),
\label{decayrate}
\eea   
where $\beta_{ij} = \sqrt{\left( 1-\frac{(m_i+m_j)^2}{q^2}\right) \left( 1-\frac{(m_i-m_j)^2}{q^2}\right)}$ and the kinematical factor $\lambda = M_B^4 + M_K^4 +q^4 -2\left( M_B^2 M_K^2 + M_K^2 q^2 + M_B^2 q^2 \right)$. The twelve angular coefficients $I_i(\cos\theta_l)$ appearing in the angular distribution depend on the couplings, kinematic variables, form factors and the polar angle $\theta_l$, which are defined as 
\bea
I_1 &=& 2 \Bigg[ \left( 1-\frac{(m_i-m_j)^2}{q^2}\right) \left( q^2-\left( q^2-(m_i+m_j)^2 \right) \cos^2\theta_l \right) |H_V^0|^2 \nn \\ && + 4k \frac{(m_i^2-m_j^2)}{\sqrt{q^2}} {\rm Re} \left[H_V^0 H_V^{t^*} \right] \cos\theta_l + \frac{(m_i-m_j)^2}{q^2} \left( q^2-(m_i+m_j)^2 \right) |H_V^t|^2  \Bigg], \\
I_2 &=& 2 \Bigg[ \left( 1-\frac{(m_i+m_j)^2}{q^2}\right) \left( q^2-\left( q^2-(m_i-m_j)^2 \right) \cos^2\theta_l \right) |H_A^0|^2  \nn \\ && + 4k \frac{(m_i^2-m_j^2)}{\sqrt{q^2}} {\rm Re} \left[H_A^0 H_A^{t^*} \right] \cos\theta_l  + \frac{(m_i+m_j)^2}{q^2} \left( q^2-(m_i-m_j)^2 \right) |H_A^t|^2  \Bigg], \\
I_3 &=& 2\left(q^2-(m_i+m_j)^2 \right) |H_S|^2, \\
I_4 &=& 2\left(q^2-(m_i-m_j)^2 \right) |H_P|^2,\\
I_5 &=& 8  \left( 1-\frac{(m_i-m_j)^2}{q^2}\right) \Big( (m_i+m_j)^2+ \left( q^2- (m_i+m_j)^2 \right) \cos^2 \theta_l \Big) |H_T^{0t}|^2, \\
I_6 &=& 32 \left( 1-\frac{(m_i+m_j)^2}{q^2}\right) \Big( (m_i-m_j)^2+ \left( q^2- (m_i-m_j)^2 \right) \cos^2 \theta_l \Big) |H_{TE}^{0t}|^2, \\
I_7 &=& 4{\rm Re}\Big[2k \left( m_i+m_j\right) H_V^0 H_S^* \cos\theta_l + \frac{(m_i-m_j)}{\sqrt{q^2}} \left( q^2- (m_i+m_j)^2 \right)  H_V^t H_S^* \Big], \\
I_8 &=& 4{\rm Re}\Big[2k \left( m_i-m_j\right) H_A^0 H_P^* \cos\theta_l + \frac{(m_i+m_j)}{\sqrt{q^2}} \left( q^2- (m_i-m_j)^2 \right)  H_A^t H_P^* \Big], \\
I_9 &=& -8 {\rm Re}\Big[2k \left( m_i-m_j\right) H_V^t H_T^{0t^*} \cos\theta_l + \frac{(m_i+m_j)}{\sqrt{q^2}} \left( q^2- (m_i-m_j)^2 \right)  H_V^0 H_T^{0t^*} \Big], \\
I_{10} &=& 16 {\rm Re}\Big[2k \left( m_i+m_j\right) H_A^t H_{TE}^{0t^*} \cos\theta_l + \frac{(m_i-m_j)}{\sqrt{q^2}} \left( q^2- (m_i+m_j)^2 \right)  H_0^t H_{TE}^{0t^*} \Big], \\
I_{11} &=& -16 k \sqrt{q^2} {\rm Re}[H_S H_T^{0t^*}] \cos\theta_l, \\
I_{12} &=& 32 k \sqrt{q^2} {\rm Re}[H_P H_{TE}^{0t^*}] \cos\theta_l.
\eea
Here $k=(\beta_{ij} \sqrt{q^2})/2$ is the lepton momentum and the expressions for the helicity amplitudes are given as
\bea
&&H_V^0 = \sqrt{\frac{\lambda}{q^2}} \Big[\left(C_V^{SM}+C_V+C_V^\prime \right) f_+(q^2)+2C_7^{SM} m_b \frac{f_T}{M_B+ M_K}\Big], \\
&&H_V^t= \frac{M_B^2-M_K^2}{\sqrt{q^2}} \left(C_V^{SM}+C_V+C_V^\prime \right) f_0(q^2), \\
&&H_A^0 = \sqrt{\frac{\lambda}{q^2}} \left(C_{A}^{SM}+C_A+C_A^\prime \right) f_+(q^2), \\
&&H_A^t= \frac{M_B^2-M_K^2}{\sqrt{q^2}} \left(C_{A}^{SM}+C_A+C_A^\prime \right) f_0(q^2), \\
&&H_S=\frac{M_B^2-M_K^2}{m_b} \left(C_S+C_S^\prime \right) f_0 \left( q^2 \right), \\
&&H_P=\frac{M_B^2-M_K^2}{m_b} \left(C_P+C_P^\prime \right) f_0 \left( q^2 \right), \\
&&H_T^{0t} = -2C_T \frac{\sqrt{\lambda}}{M_B+ M_K}f_T(q^2), \\
&&H_{T5}^{0t} = -2C_{T5} \frac{\sqrt{\lambda}}{M_B+ M_K}f_T(q^2).
\eea
The above expressions are calculated by  using the parametrizations of matrix elements of the various hadronic currents between the initial $B$
meson and the final $K$ meson, in terms
of the form factors $f_0$, $f_+$ and $f_T$ as \cite{bobeth2}
\bea
 &&\langle\bar{K}\left(P_K\right)|\bar{s}\gamma^\mu b|\bar{B}\left(P_B\right)\rangle = f_+\left(q^2\right) \left(P_B+P_K\right)^\mu + \left[f_0\left(q^2\right) - f_+\left(q^2\right)\right]\frac{M^2_B - M^2_K}{q^2}q^\mu, \\
&& \langle\bar{K}\left(P_K\right)|\bar{s} \sigma^{\mu\nu} b|\bar{B}\left(P_B\right)\rangle = i\frac{f_T\left(q^2\right)}{M_B + M_K} \left[\left(P_B + P_K\right)^\mu q^\nu - q^\mu \left(P_B+P_K\right)^\nu \right].
 \eea
It should be noted that in general the angular coefficients of semileptonic decays take the form 
\bea
I_i\left(\cos\theta_l \right) =  a_i+b_i\cos\theta_l +c_i\cos^2\theta_l.
\eea
The differential decay rate for the decay $\bar{B} \to \bar{K} l_i l_j$ can be found by integrating over the polar angle in Eqn. (\ref{decayrate}) to get 
\bea
\frac{d\Gamma}{dq^2} = \frac{G_F^2 \alpha^2_{em} \beta_{ij} \sqrt{\lambda}|V_{tb}V_{ts}^*|^2}{2^{12} \pi^5 M_B^3} \sum_{i=1}^{10} J_i,
\eea 
where the coefficients $J_i = \int_{-1}^{1}I_i\left( \cos\theta_l \right) d\cos\theta_l$ are given below as
\bea
J_1 &=& 4\Bigg[ \left( 1-\frac{(m_i-m_j)^2}{q^2}\right) \frac{1}{3} \left( 2q^2+\left(m_i+m_j\right)^2 \right) |H_V^0|^2  \nn \\ && + \frac{(m_i-m_j)^2}{q^2} \left(q^2-\left(m_i+m_j\right)^2 \right) |H_V^t|^2 \Bigg],  \\
J_2 &=&4\Bigg[ \left( 1-\frac{(m_i+m_j)^2}{q^2}\right) \frac{1}{3} \left( 2q^2+\left(m_i-m_j\right)^2 \right) |H_A^0|^2 \nn \\ && + \frac{(m_i+m_j)^2}{q^2} \left( q^2-\left(m_i-m_j\right)^2 \right) |H_A^t|^2 \Bigg],   \\
J_3 &=& 4 \left( q^2 -(m_i+m_j)^2 \right) |H_S|^2, \\
J_4 &=& 4 \left( q^2 -(m_i-m_j)^2 \right) |H_P|^2, \\
J_5 &=& 16  \left( 1-\frac{(m_i-m_j)^2}{q^2}\right) \frac{1}{3} \Big( 2\left(m_i + m_j \right)^2 +q^2 \Big) |H_T^{0t}|^2, \\
J_6 &=& 64  \left( 1-\frac{(m_i+m_j)^2}{q^2}\right) \frac{1}{3} \Big( 2\left(m_i - m_j \right)^2 +q^2 \Big) |H_{TE}^{0t}|^2, \\
J_7 &=& 8 \frac{\left(m_i-m_j \right)}{\sqrt{q^2}} \left( q^2 -(m_i+m_j)^2 \right) {\rm Re}[H_V^t H_S^*], \\
J_8 &=& 8 \frac{\left(m_i+m_j \right)}{\sqrt{q^2}} \left( q^2 -(m_i-m_j)^2 \right) {\rm Re}[H_A^t H_P^*], \\
J_9 &=& -16 \frac{\left(m_i+m_j \right)}{\sqrt{q^2}} \left( q^2 -(m_i-m_j)^2 \right) {\rm Re}[H_V^0 H_T^{0t^*}], \\
J_{10} &=& 32 \frac{\left(m_i-m_j \right)}{\sqrt{q^2}} \left( q^2 -(m_i+m_j)^2 \right) {\rm Re}[H_0^t H_{TE}^{0t^*}].
\eea
Here the coefficients $J_{11}=J_{12}=0$. 
Next we define the forward-backward asymmetry ($A_{FB}$) for the leptons by integrating over $\cos\theta_l$ in Eqn. (\ref{decayrate}) as
\bea
A_{FB}(q^2) = \Big (\int_0^1 d\cos\theta_l \frac{d^2 \Gamma}{dq^2 d\cos\theta_l}-\int_{-1}^0 d\cos\theta_l \frac{d^2 \Gamma}{dq^2 d\cos\theta_l} \Big ) \Big / \frac{d\Gamma}{dq^2}.
\eea
After integration, we obtain 
\bea
A_{FB}(q^2) = \frac{X}{\sum_{i=1}^{10} J_i},
\eea
where the quantity $X$ is defined as
\bea 
X &= &8 k {\rm Re} \Big[ \frac{\left(m_i^2-m_j^2 \right)}{\sqrt{q^2}} \left( H_V^0 H_V^{t *} + H_A^0 H_A^{t *} \right) + (m_i+m_j) \left( H_V^0 H_S^{*} +4 H_A^t H_{TE}^{0t^*} \right) \nn \\ &+& (m_i-m_j) \left( H_A^0 H_P^{*} -2 H_V^t H_{T}^{0t^*} \right)-2\sqrt{q^2} \left( H_S^0 H_T^{0t^*} -2 H_P  H_{TE}^{0t^*} \right) \Big].
\eea
Another interesting observable is the lepton non-universality parameter, which has been recently observed by LHCb in $B^+ \to K^+ l^+ l^-$ process and has a $2.6 \sigma$ discrepancy from the SM prediction  in the dilepton invariant mass bin $(1 \leq q^2 \leq 6)~{\rm GeV^2}$. Analogously we  would like to see whether it is possible to observe    non-universality  in  the  LFV decays. Hence,  we define the ratios of branching ratios of  various LFV decays as
\bea
&&R_{K l}^{\mu e} = \frac{{\rm Br}\left( \bar{B} \to \bar{K } \mu^- e^+ \right)}{{\rm Br}\left( \bar{B} \to \bar{K } l^+ l^- \right)}, \\
&&R_{Kl}^{\tau e} = \frac{{\rm Br}\left( \bar{B} \to \bar{K } \tau^- e^+ \right)}{{\rm Br}\left( \bar{B} \to \bar{K } l^+ l^- \right)}, \\
&&R_{Kl}^{\tau \mu} = \frac{{\rm Br}\left( \bar{B} \to \bar{K } \tau^- \mu^+ \right)}{{\rm Br}\left( \bar{B} \to \bar{K } l^+ l^- \right)}, \\
&&R_{K}^{\mu \mu} = \frac{{\rm Br}\left( \bar{B} \to \bar{K } \mu^+ \mu^- \right)}{{\rm Br}\left( \bar{B} \to \bar{K} e^+ e^- \right)}, \\
&&R_{K l}^{\tau \tau} = \frac{{\rm Br}\left( \bar{B} \to \bar{K } \tau^+ \tau^- \right)}{{\rm Br}\left( \bar{B} \to \bar{K } l^+ l^- \right)}, 
\eea
where $l=\mu, e$. Similarly, one can obtain the branching ratios and other physical observables in $B \to \pi l_i^- l_j^+$ processes by  incorporating the appropriate CKM matrix elements, form factors and the NP effective couplings. 
Recently LHCb has  measured the ratio of branching fractions of $B^+ \to \pi^+ \mu^+ \mu^-$ over  $B^+ \to K^+ \mu^+ \mu^-$ processes  \cite{Rplus}, given  as
\bea
\frac{{\rm Br}(B^+ \to \pi^+ \mu^+ \mu^-)}{{\rm Br}(B^+ \to K^+ \mu^+ \mu^-)}=0.053 \pm 0.014 ({\rm stat}) \pm 0.001~ ({\rm syst}).
\eea
In the same context, we also define the ratio of branching fractions of $B^+ \to \pi^+ l_i^- l_j^+$ and $B^+ \to K^+ l_i^- l_j^+$ LFV processes as 
\bea
R_{+}^{l_i l_j} = \frac{{\rm Br}\left( B^+ \to \pi^+ l_i^- l_j^+ \right)}{{\rm Br}\left( B^+ \to K^+ l_i^- l_j^+ \right)}.
\eea

\section{New physics contributions due to the exchange of  vector leptoquark}
There are  $10$ different  LQ multiplets under the $SU(3)_C\times SU(2)_L \times U(1)_Y$ SM gauge group \cite{10-LQ}, of these one half have scalar nature and the rest have   vectorial nature under the Lorenz transformation. 
Vector LQs have spin 1 which exist in grand unified theories, $SO(10)$ including Pati-Salam color $SU(4)$ and larger gauge groups. The scalar and vector LQ multiplets are differ by their weak-hypercharge and fermion number. The strongest bounds on the vector LQs can be avoid by demanding  chirality and diagonality of the coupling and  diquark coupling have to be forbidden to evade proton decay. There are three relevant  vector LQ multiplets, $(3,3,2/3)$, $(3,1,2/3)$ and $(\bar{3},2,5/6)$ \cite{kosnik}, out of which only $(3,3,2/3)$ leptoquark conserves both baryon and lepton numbers.

\subsubsection{$Q=2/3$ vectors}
There are two vector LQ multiplets $V^3(3,3,2/3)$ and $V^1(3,1,2/3)$  having fermion number zero and electric charge $Q=2/3$. 
The  interaction Lagrangian of isotriplet state $V^{(3)}$ with the SM fermions is given by \cite{kosnik}
\begin{equation}
  \mathcal{L}^{(3)} = g_L \,\overline{Q}\,  \pmb{\tau} \cdot V^{(3)}_\mu
  \gamma^\mu L + h.c.,
\end{equation}
which conserves both lepton and baryon number and contributes new Wilson coefficients, $C_{V, A}^{\rm LQ}$  as
\begin{equation}
  C_{V}^{\rm LQ} = -C_{A}^{\rm LQ}  = \frac{\pi}{\sqrt{2} G_F V_{tb}V_{ts}^* \alpha_{em}}
  \frac{(g_L)_{s l} (g_L)^*_{bl}}{M^2_{V^{(3)}}}\,.
  \label{u3c10}
\end{equation}
Here $Q(L)$ is the left handed quark (lepton) doublet, $g_L$ is the LQ coupling having left handed quark current and ${\pmb \tau}$ represents the Pauli matrices.

The Lagrangian for isosinglet state, $V^{(1)}$  is given by
\begin{equation}
  \mathcal{L}^{(1)} = \left(g_L\, \overline{Q} \gamma^\mu L  +  g_R\, \overline{d_R}
    \gamma^\mu l_R \right)\, V^{(1)}_\mu + h.c.,
\end{equation}
where $d_R$ and $l_R$ are the right handed down quark and lepton singlets respectively and $g_R$ is the LQ coupling with down quarks and right handed leptons.  This LQ violates baryon number and has the coupling to both left and right handed fermions  \textit{ i.e.} it is a non-chiral LQ.  In addition to $C_{V,A}$ new Wilson coefficients, these non-chiral LQ contributes  scalar and pseudoscalar operators given by
\begin{subequations}
\bea
 && C_V^{NP} = -C_{A}^{NP} = \frac{\pi}{\sqrt{2} G_F V_{tb}V_{ts}^* \alpha_{em}}
  \frac{(g_L)_{s l} (g_L)^*_{bl}}{M_{V^{(1)}}^2}\,,\label{u1c10np}\\
&&  C_V^{\prime NP} = C_{A}^{\prime NP} = \frac{\pi}{\sqrt{2} G_F V_{tb}V_{ts}^*
    \alpha_{em}} \frac{(g_R)_{sl} (g_R)^*_{bl}}{M_{V^{(1)}}^2}\,\label{u1c10pnp} \\ 
&&  -C_P^{NP} = C_{S}^{NP} = \frac{\sqrt{2}\pi}{G_F V_{tb}V_{ts}^* \alpha_{em}}
  \frac{(g_L)_{sl} (g_R)^*_{bl}}{M_{V^{(1)}}^2}\,,\label{u1csnp} \\
&&  C_P^{\prime NP} = C_{S}^{\prime  NP}= \frac{\sqrt{2}\pi}{G_F V_{tb}V_{ts}^* \alpha_{em}} \frac{(g_R)_{sl} (g_L)^*_{bl}}{M_{V^{(1)}}^2}\, \label{u1cspnp}. 
\eea
\end{subequations}
\subsubsection{$Q=4/3$ vectors}

The vector LQ with charge $Q=4/3$ has one  isospin doublet state $V^2(\bar{ 3},2,5/6)$, whose coupling with fermion bilinear  is given by \cite{kosnik}
\begin{equation}
\mathcal{L}^{(2)}= g_R\, \overline{Q^C}\, i\tau_2 V^{(2)}_\mu \gamma^\mu l_R +
  g_L\, \overline{d_R^C}\, \gamma^\mu\, \tilde{V}^{(2)\dagger}_\mu L + h.c. 
\end{equation}
This LQ also has both left handed and right handed  lepton couplings and violates baryon number. Now performing the Fierz transformation, the additional Wilson coefficients contribution to the $b \to q l^- l^+$  processes as
\begin{subequations}
\bea
&&C_V^{NP} = C_{A}^{NP} = \frac{-\pi}{\sqrt{2} G_F V_{tb}V_{ts}^* \alpha_{em}}
  \frac{(g_R)_{bl} (g_R)^*_{sl}}{M_{V^{(2)}}^2}, \\
&&-C_V^{\prime NP} = C_{A}^{\prime NP}= \frac{\pi}{\sqrt{2} G_F V_{tb}V_{ts}^* \alpha_{em}}
  \frac{(g_L)_{bl} (g_L)^*_{sl}}{M_{V^{(2)}}^2}, \\
&&C_P^{NP} = C_{S}^{NP} = \frac{\sqrt{2}\pi}{G_F V_{tb}V_{ts}^* \alpha_{em}}   \frac{(g_R)_{bl} (g_L)^*_{sl}}{M_{V^{(2)}}^2}, \\
&&-C_P^{\prime NP} = C_{S}^{\prime  NP}= \frac{\sqrt{2}\pi}{G_F V_{tb}V_{ts}^* \alpha_{em}}   \frac{(g_L)_{bl} (g_R)^*_{sl}}{M_{V^{(2)}}^2}. 
\eea
\end{subequations}
\section{Constraint on the leptoquark couplings }
After having  an idea about all possible new physics contributions to the SM,  we now  proceed to constrain the new Wilson coefficients by comparing the theoretical and experimental branching ratios of various rare decay processes.

\subsection{$B_{s, d} \to l^+ l^-$ processes}
The rare leptonic $B_{s, d} \to \mu^+ \mu^-$ processes  are mediated by the FCNC $b \to (s, d)$ transitions and in the SM the branching ratios depend only on the Wilson coefficient $C_{A}$. In addition to $C_{V, A}^{(\prime)}$ Wilson coefficients, vector LQ  also  contributes   scalar and pseudoscalar $(C_{S, P}^{(\prime)})$ Wilson coefficients to the SM.  However, there is no additional contributions of tensor Wilson coefficients $C_{T, T_5}$ due to the exchange of vector LQ.

The branching ratio of $B_q \to \mu^+ \mu^-$ process in the LQ model is given by \cite{BR-leptonic, BR-leptonic-2}
\bea
{\rm Br}(B_q \to \mu^+ \mu^-) = \frac{G_F^2}{16 \pi^3} \tau_{B_q} \alpha_{em}^2 f_{B_q}^2 M_{B_q} m_{\mu}^2 |V_{tb} V_{tq}^*|^2
\left |C_{A}^{SM}\right |^2 \sqrt{1- \frac{4 m_\mu^2}{M_{B_q}^2}} \times \left(|P|^2 + |S|^2 \right),
\eea
where 
\bea
&&P \equiv \frac{C_A^{SM}+ C_{A}^{LQ}-C_{A}^{\prime LQ}}{C_{A}^{SM}}+\frac{M_{B_q}^2}{2m_{\mu}} \Big(\frac{m_b}{m_b+m_s} \Big) \Big(\frac{C_{P}^{LQ}-C_{P}^{\prime LQ}}{C_{A}^{SM}}\Big) \equiv |P|e^{i\phi_P} ,\nn \\ 
&&S \equiv \sqrt{1- \frac{4 m_\mu^2}{M_{B_q}^2}} \frac{M_{B_q}^2}{2m_{\mu}} \Big(\frac{m_b}{m_b+m_s} \Big) \Big(\frac{C_{S}^{LQ}-C_{S}^{\prime LQ}}{C_{A}^{SM}}\Big) \equiv |S|e^{i\phi_S}. 
\label{P-S}
\eea
Here $C_{A}^{(\prime)\rm LQ}$ and $C_{S, P}^{(\prime) \rm LQ}$ Wilson coefficients are generated due to the vector LQ exchange and are negligible  in the SM, which implies $P^{\rm SM}=1$ and $S^{\rm SM}=0$. The experimental result is related to the theoretical predictions as  \cite{BR-leptonic-2}
\bea
{\rm Br}^{\rm th}(B_q \to \mu^+ \mu^-) = \Bigg[\frac{1-y_q^2}{1+A_{\Delta \Gamma}y_q} \Bigg] {\rm Br}^{\rm exp}(B_q \to \mu^+ \mu^-),
\eea
where $y_q = \tau_{B_q} \Delta \Gamma_q/2$ and the observables $A_{\Delta \Gamma}$ is the mass eigenstate rate asymmetry equals to $+1$ in the SM. For calculational conveniene, we define the parameter $R_q$ as
\bea
R_{q}=\frac{{\rm Br}^{\rm th}(B_q \to \mu^+ \mu^-)}{{\rm Br}^{\rm SM}(B_q \to \mu^+ \mu^-)} = |P|^2+|S|^2.
\label{R-q}
\eea
If we apply  chirality on vector LQ, then the $C_{S, P}^{(\prime)\rm LQ}$ Wilson coefficients will vanish and there will be additional contribution of only $C_{V, A}^{(\prime)\rm LQ}$ Wilson coefficients  to the SM.  Hence, the $R_q$ parameter  can be given as \cite{mohanta1, mohanta2}
\bea
R_q=\left | 1+ \frac{C_{A}^{\rm LQ} - C_{A}^{\prime \rm LQ}}{C_{A}^{\rm SM}} \right |^2
\equiv \left | 1+ r e^{i \phi^{ NP}} \right |^2,
\eea
where  the parameters $r$ and $\phi^{NP}$ are related to the new  Wilson coefficients  as
\be
r e^{i \phi^{NP}}=\frac{C_{A}^{\rm LQ} - C_{A}^{\prime \rm LQ}}{C_{A}^{SM}}\;.
\ee
Now comparing the theoretical \cite{bobeth1} branching ratios  of $B_q \to \mu^+ \mu^-$ processes with the $1\sigma$ range of experimental values \cite{average}, the constraint on $r$ and $\phi^{NP}$  is computed for scalar LQ model in  our previous work \cite{mohanta1, mohanta2}. If we assume that both the scalar and  vector LQs have same order mass, $M_{\rm LQ} = 1$ TeV, one can use the same constraint on $r$ and $\phi^{NP}$ parameters to study the processes mediated via vector LQ.   For $B_s \to \mu^+ \mu^-$ process, the constraints are found to be  \cite{mohanta2}
 \bea
 0\leq r \leq 0.35\;, ~~~~{\rm with}~~~~\pi/2 \leq \phi^{NP} \leq 3 \pi/2\;,\label{r-bound1}
 \eea
and for $B_d \to \mu^+ \mu^-$ process  \cite{mohanta2}
\bea
 0.5\leq r \leq 1.3\;, ~~~{\rm for}~~~\left (0 \leq \phi^{NP} \leq  \pi/2\right )~~{\rm or}~~\left (3 \pi/2 \leq \phi^{NP} \leq  2\pi 
\right ).\label{r-bnd}
 \eea
Using Eqns. (\ref{u3c10}, \ref{u1c10np}, \ref{u1c10pnp}), this  can be translated to obtain the bounds on LQ couplings (for $M_{\rm LQ} = 1$ TeV) as
\bea
 0 \leq |(g_L)_{s\mu} (g_L)^*_{b \mu}| \leq 2.3 \times 10^{-3}~ , \hspace*{2.0 true cm}
  \eea
 \bea
 0.7 \times 10^{-3} ~ \leq|(g_L)_{d\mu} (g_L)^*_{b \mu}| \leq 1.81 \times 10^{-3}\;.  
 \eea
Similarly using the theoretical predictions \cite{bobeth1} and the experimental upper limits \cite{CDF, tau-expt} on $B_q \to e^+ e^- (\tau^+ \tau^-)$ processes, the constraint on the product of scalar LQ couplings  are presented in Table I, which are found to be rather loose as the measured branching ratios of $B_{d,s} \to \tau^+ \tau^- (e^+ e^+)$ are  not very precise.
\begin{table}[htb]
\begin{center}
\caption{Constraints on leptoquark couplings obtained from  various leptonic $B_{s,d} \to l^+ l^-$ decays.}
\vspace*{0.1 true in}
\begin{tabular}{|c|c|c|}
\hline
Decay Process ~& ~Couplings involved ~&~ Upper bound of  \\
             &  &~the couplings   \\
\hline

$B_s \to e^\pm e^\mp $ &~ $|(g_L)_{s e} (g_L)^*_{b e}|$ ~& ~$ < 11.8 $~\\

\hline

$B_s \to \tau^\pm \tau^\mp $ &~ $|(g_L)_{s\tau} (g_L)^*_{b \tau}|$ ~& ~$ < 0.4 $~\\

\hline
$B_d \to e^\pm e^\mp $ &~ $|(g_L)_{d e} (g_L)^*_{b e}|$ ~& ~$ < 8.0  $~\\

\hline

$B_d \to \tau^\pm \tau^\mp $ &~ $|(g_L)_{d \tau} (g_L)^*_{b \tau}|$ ~& ~$ < 0.593 $~\\

\hline
\end{tabular}
\end{center}
\end{table}

 For simplicity we can neglect the NP contributions to the $C_{V, A}^{(\prime)\rm LQ}$ Wilson coefficients, as the $C_{S, P}^{(\prime)\rm LQ}$ Wilson coefficients are enhanced by the factor $M_{B_q}^2/m_l$. Now using Eqns. (\ref{P-S}), (\ref{u1csnp}), (\ref{u1cspnp}) and (\ref{R-q}), the $R_q$ parameter becomes
\bea
R_q=\frac{|C_S^{\rm LQ}-C_S^{\prime \rm LQ}|^2}{r_q^2} +\Big |1-\frac{|C_S^{\rm LQ}+C_S^{\prime \rm LQ} |}{r_q}\Big |^2 
\label{R-q1}
\eea 
where 
\bea
r_q=\frac{2m_l \left(m_b+m_q \right)C_{A}^{SM}}{M_{B_q}^2 }.
\eea
\begin{figure}[h]
\centering
\includegraphics[scale=0.6]{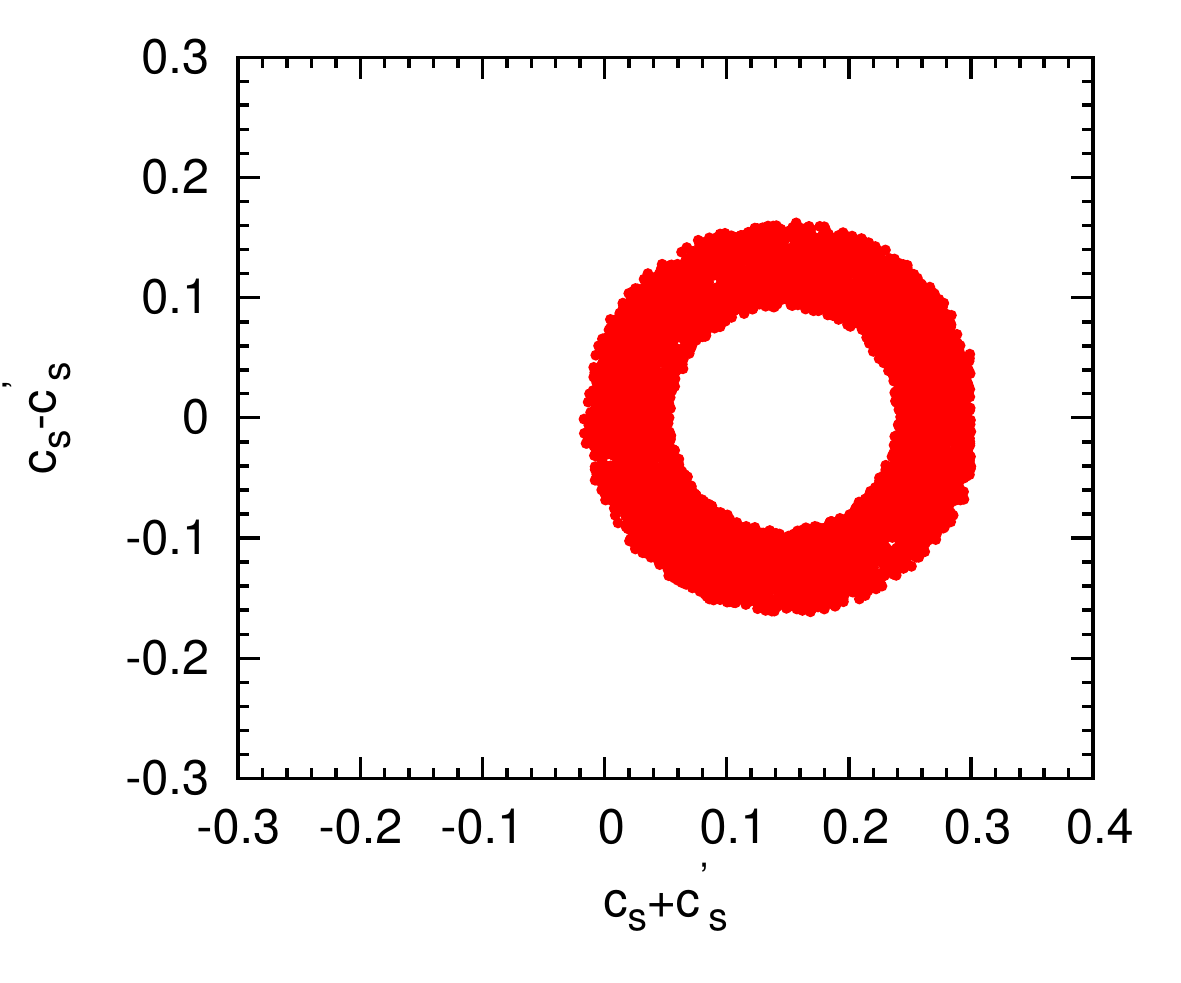}
\quad
\includegraphics[scale=0.6]{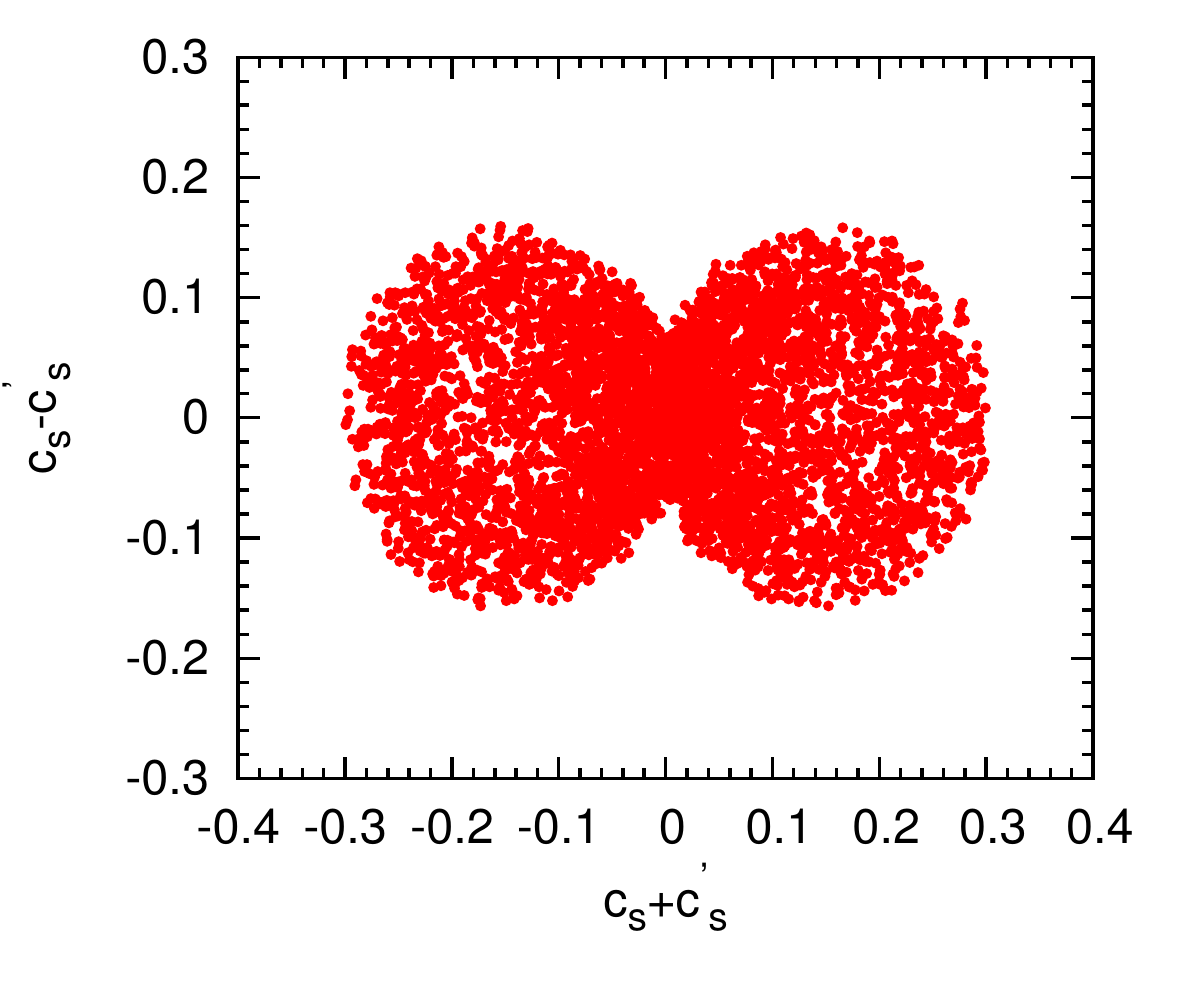}
\caption{Constraint on the combination of scalar  Wilson coefficient from  $B_s \to \mu^+ \mu^-$ process. The left panel is for real $C_S^{\rm LQ} \pm C_S^{\prime \rm LQ}$ Wilson coefficients and right panel is for complex Wilson coefficients.}
\end{figure}
\begin{figure}[h]
\centering
\includegraphics[scale=0.6]{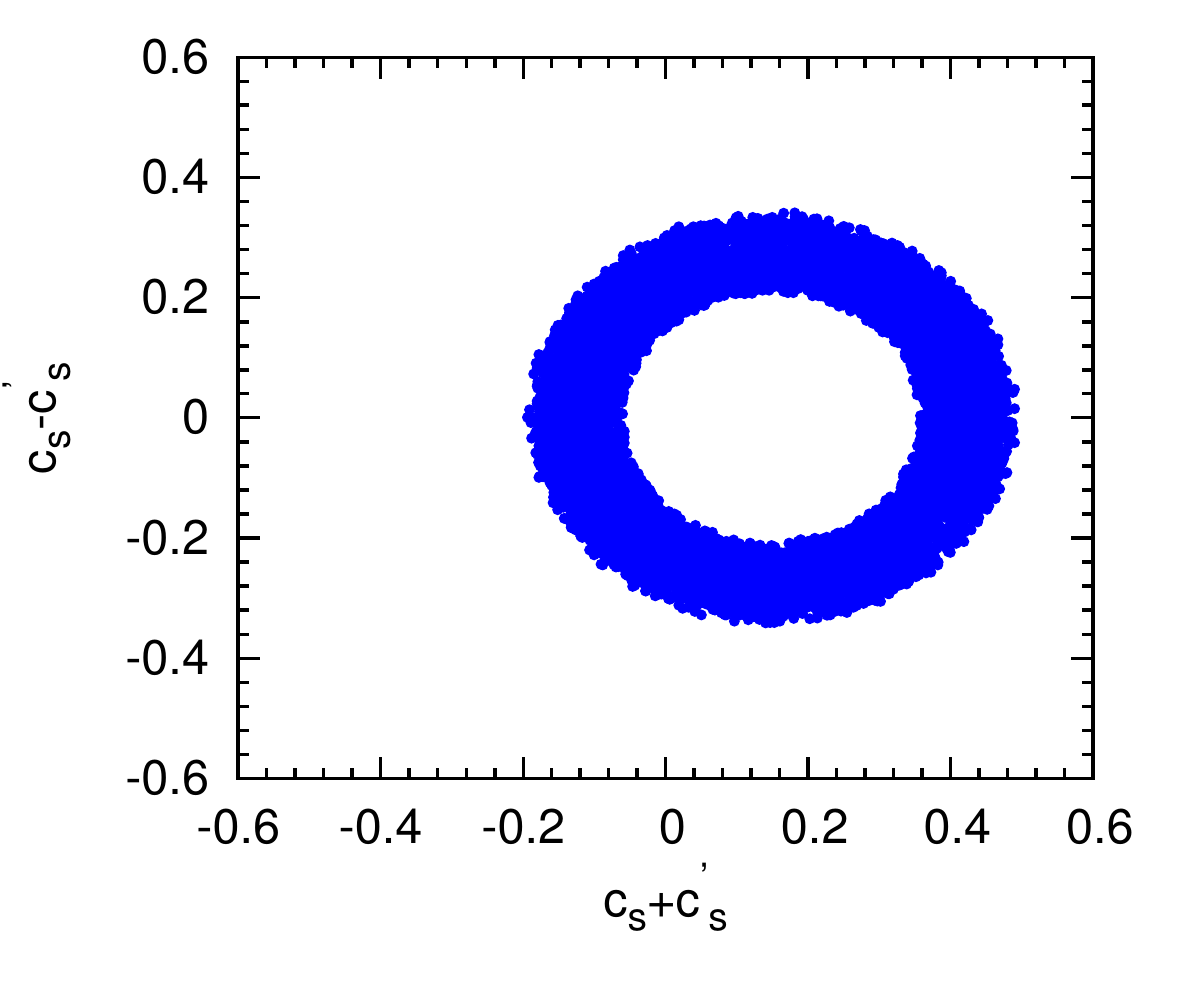}
\quad
\includegraphics[scale=0.6]{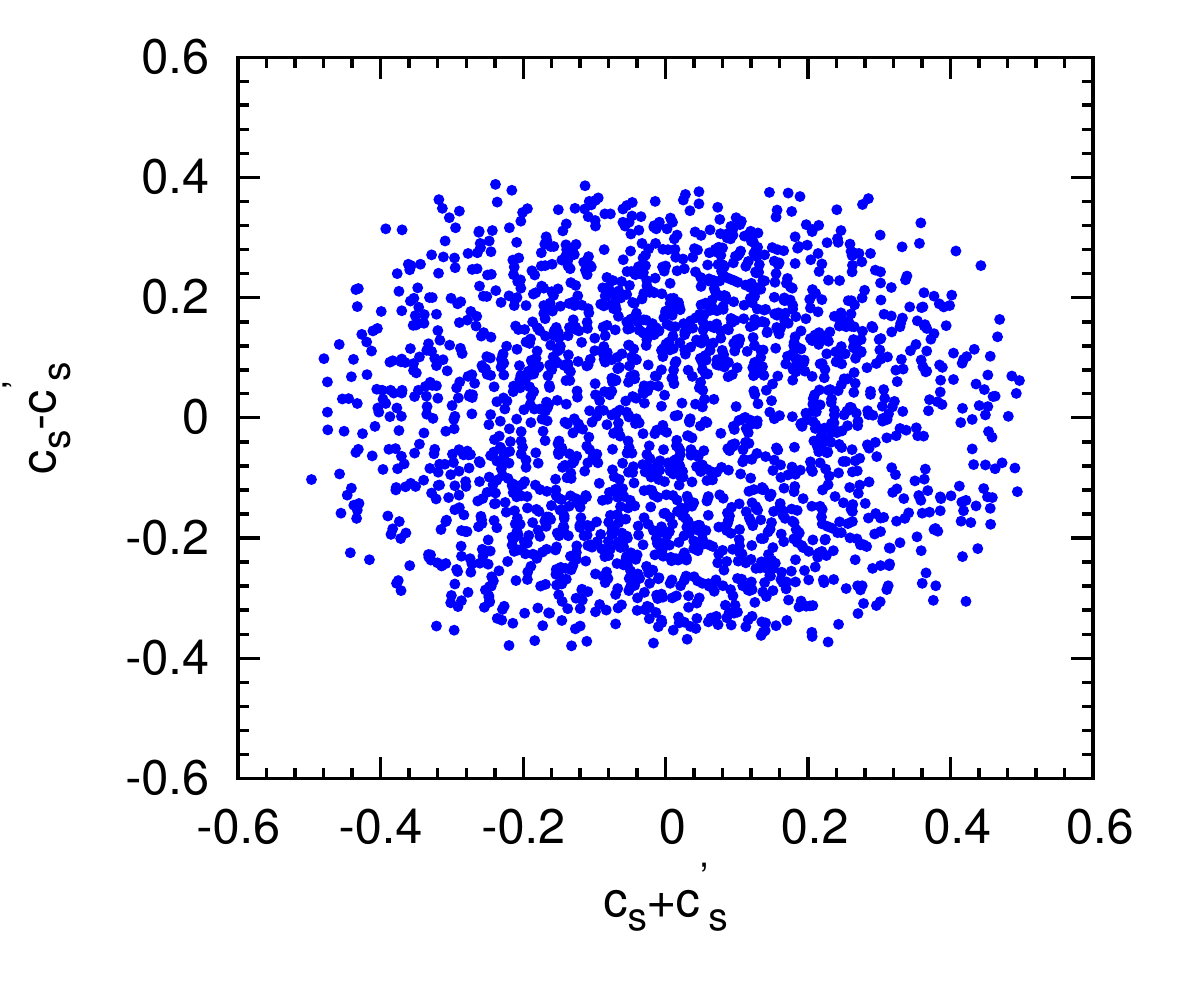}
\caption{Constraint on $C_S^{\rm LQ} \pm C_S^{\prime \rm LQ}$  Wilson coefficients from  $B_d \to \mu^+ \mu^-$ process. The left panel is for real Wilson coefficients and right panel is for complex Wilson coefficients.}
\end{figure}

Now comparing the theoretical  and experimental values of $B_q \to l^+ l^-$ decays, we calculate the allowed region of $C_S^{LQ}\pm C_S^{\prime LQ}$ Wilson coefficients.  If the  Wilson coefficients are real, Eqn. (\ref{R-q1}) will be a circle of radius $ |r_q| \sqrt{R_q^{\rm expt}}$  with center at $ \left(C_S^{LQ}+C_S^{\prime LQ}, C_S^{LQ}-C_S^{\prime LQ} \right) = (r_q, 0)$. The left panel of Fig. 1 represents the constraint  on real $C_S^{LQ}\pm C_S^{\prime LQ}$ Wilson coefficients from $B_s \to \mu^+ \mu^-$ process and the right panel is  for complex Wilson coefficients. Similarly in  Fig. 2, we show the constraint on real  (left panel)  and complex (right panel) Wilson coefficients  for $B_d \to \mu^+ \mu^-$ process.  The allowed range of real Wilson coefficients from  $B_{s} \to e^+ e^-$ (left panel) and  $B_{d} \to e^+ e^-$ (right panel) processes are shown in Fig. 3. In Fig. 4, we present the constraint obtained from $B_{s} \to \tau^+ \tau^-$ (left panel) and $B_{d} \to \tau^+ \tau^-$ (right panel) processes. 
 The allowed region of $C_S^{LQ}\pm C_S^{\prime LQ}$  real Wilson coefficients obtained  from $B_q \to l^+ l^-$ processes are presented in Table II. Now using the constrained Wilson coefficients, one can calculate the bound on the product of various LQ couplings from Eqns. (\ref{u1csnp}, \ref{u1cspnp}). 
\begin{figure}[h]
\centering
\includegraphics[scale=0.6]{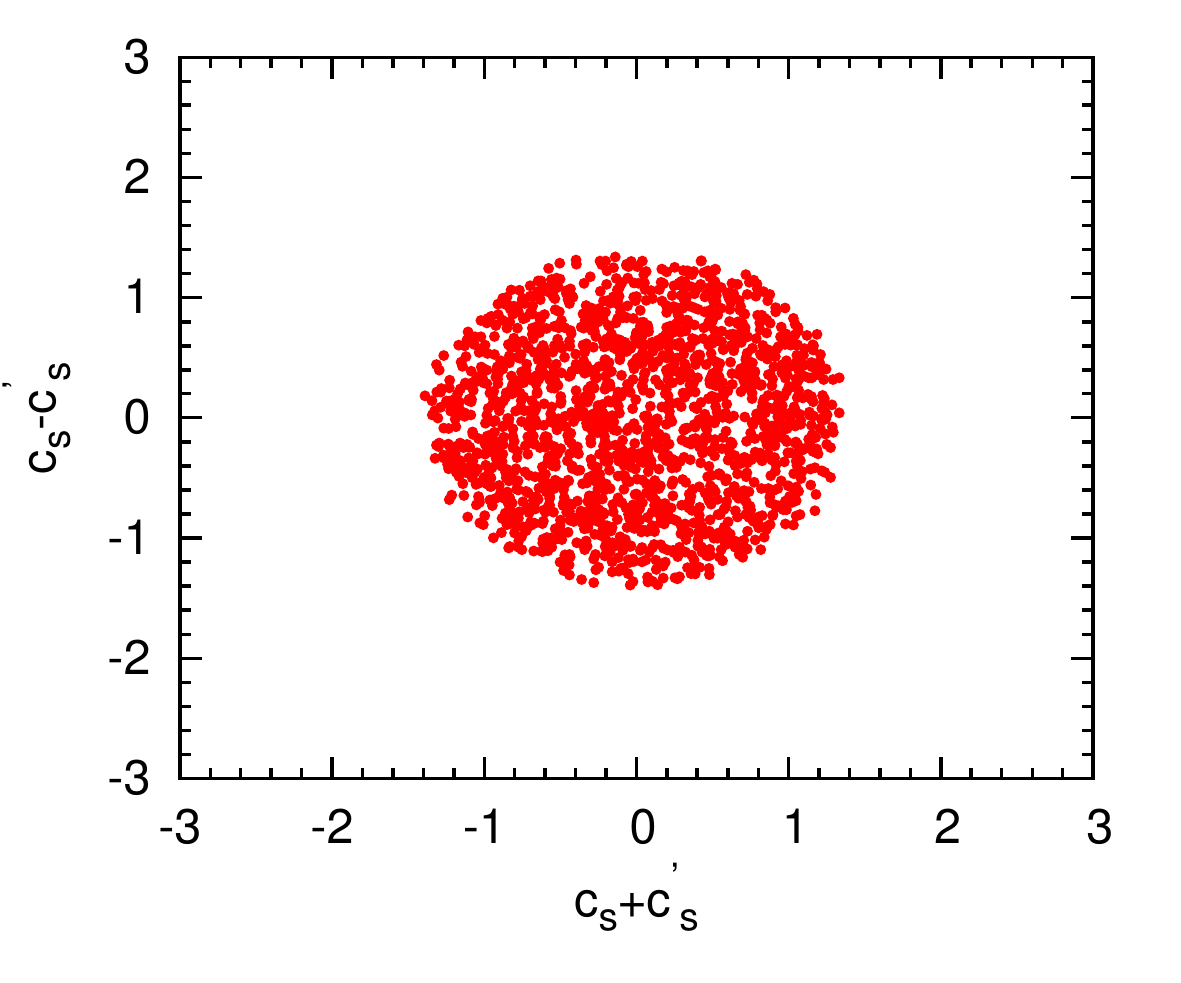}
\quad
\includegraphics[scale=0.6]{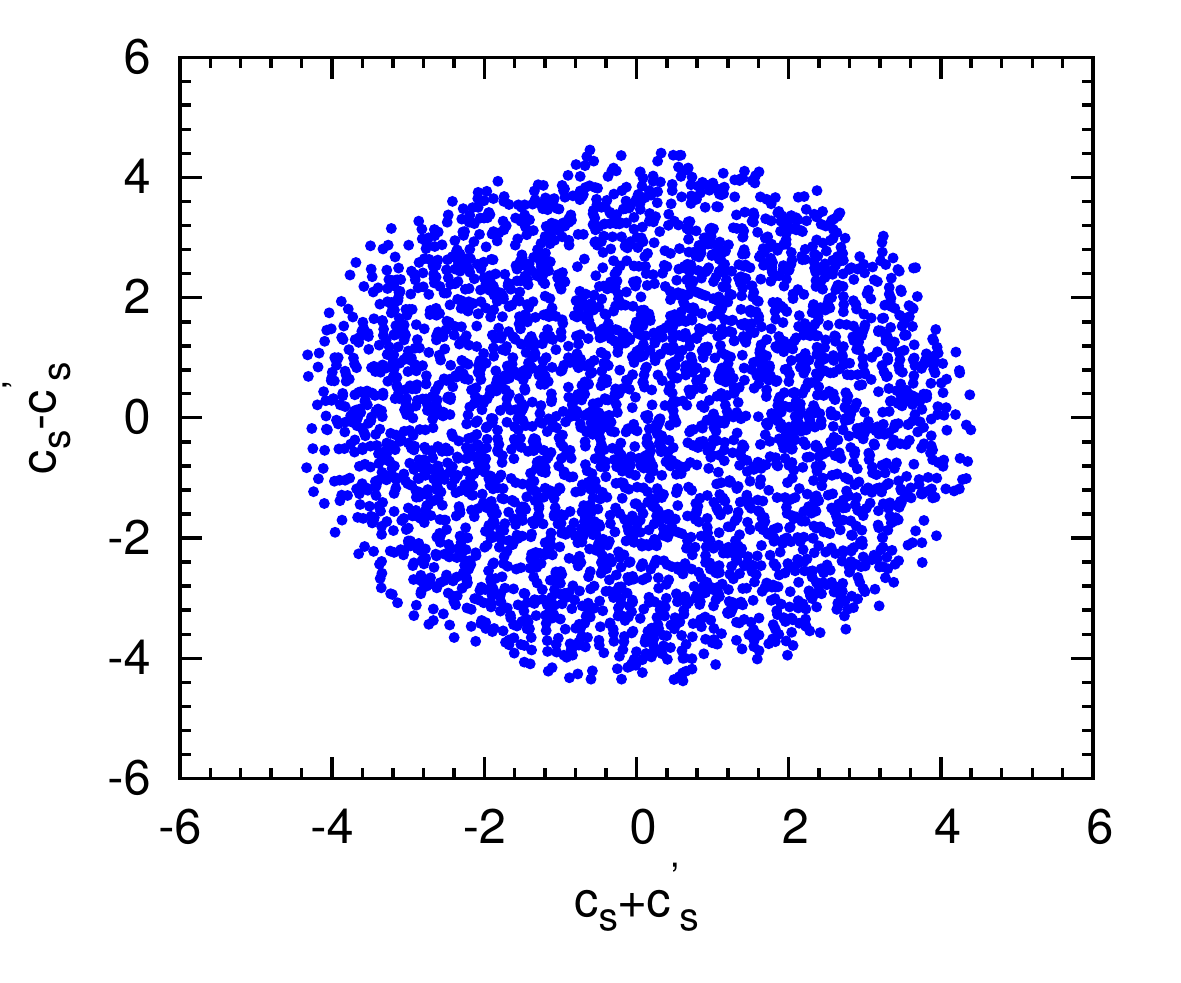}
\caption{The allowed region of $C_S^{\rm LQ}-C_S^{\prime \rm LQ}$ and $C_S^{\rm LQ}+C_S^{\prime \rm LQ}$ Wilson coefficients from $B_s \to e^+ e^-$ (left panel) and  $B_d \to e^+ e^-$ (right panel) processes.}
\end{figure}
\begin{figure}[h]
\centering
\includegraphics[scale=0.6]{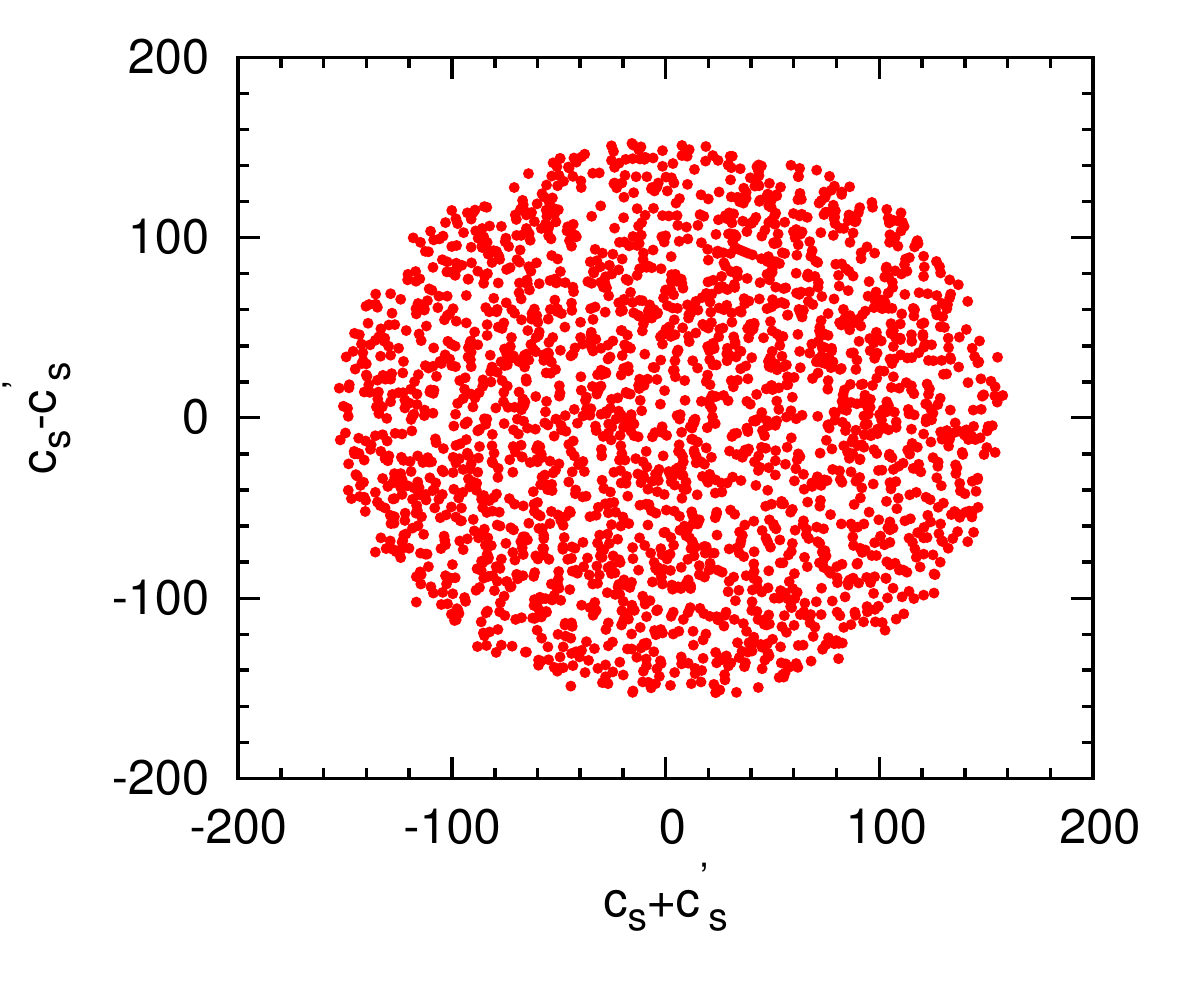}
\quad
\includegraphics[scale=0.6]{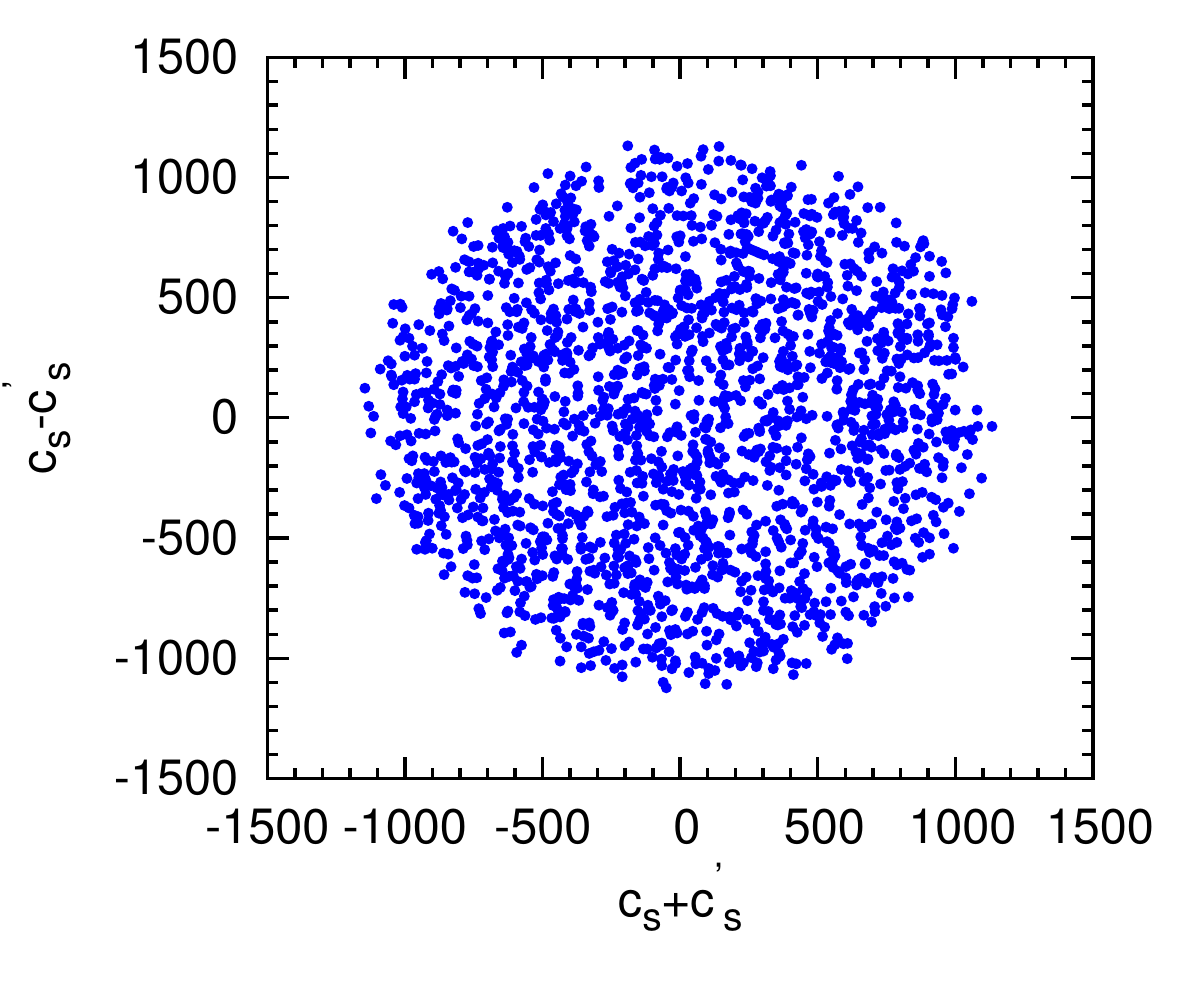}
\caption{The allowed region of $C_S^{\rm LQ}-C_S^{\prime \rm LQ}$ and $C_S^{\rm LQ}+C_S^{\prime \rm LQ}$  Wilson coefficients from $B_s \to \tau^+ \tau^-$ (left panel) and $B_d \to \tau^+ \tau^-$ (right panel) processes.}
\end{figure}
\begin{table}[htb]
\begin{center}
\caption{Constraint on  combinations of $C_S^{(\prime)LQ}$ Wilson coefficients from various leptonic $B_{s,d} \to l^+ l^-$ decays.}
\vspace*{0.1 true in}
\begin{tabular}{|c|c|c|}
\hline
Decay Process ~& ~Bound on $C_S^{LQ}+ C_S^{\prime LQ}$   ~&~Bound on $C_S^{LQ}- C_S^{\prime LQ}$  \\

\hline
$B_s \to \mu^\pm \mu^\mp $~~ &~~ $0.0 \to 0.32$ ~~& ~~$0.1 \to 0.18$~\\

\hline

$B_s \to e^\pm e^\mp $ &~ $-1.4 \to 1.4$ ~& ~$-1.4 \to 1.4$ ~\\

\hline

$B_s\to \tau^\pm \tau^\mp $ &~ $-150 \to 150$ ~& ~$-150 \to 150$~\\

\hline
$B_d \to \mu^\pm \mu^\mp $ &~ $-0.16 \to 0.44$ ~& ~$0.2 \to 0.36$~\\

\hline
$B_d \to e^\pm e^\mp $ &~ $-4 \to 4$ ~& ~$-4 \to 4$~\\

\hline

$B_d \to \tau^\pm \tau^\mp $ &~ $-1000\to 1000$ ~& ~$-1000 \to 1000$~\\

\hline
\end{tabular}
\end{center}
\end{table}

\subsection{$K_L \to \mu^+ \mu^- (e^+ e^-)$ process}
The constraint on  the product of various LQ couplings from the rare leptonic decays of $K$ meson are discussed in this subsection. The  rare $K_L \to \mu^+ \mu^-$ decay   mode has both the long and short distance contributions and  the dominant contribution comes  from the long-distance two photon intermediates state $K_L \to \gamma^* \gamma^* \to \mu^+ \mu^-$. 
Only the short distance (SD) part can be calculated reliably and the estimated branching ratio of the SD part is ${\rm Br}(K_L \to \mu^+ \mu^-)|_{\rm SD} ~\textless ~ 2.5 \times 10^{-9}$ \cite{K-Isdori}. In the SM the effective Hamiltonian for the $K_L \to \mu^+ \mu^-$ process is given by \cite{K-Buchalla}
\bea
\mathcal{H}_{eff} &=& \frac{G_F}{\sqrt{2}} \frac{\alpha}{2 \pi \sin^2\theta_W} \Big(\lambda_c Y_{NL} + \lambda_t Y(x_t) \Big)  \left( \bar{s} \gamma^\mu (1-\gamma_5)d \right) \left(\bar{\mu} \gamma_\mu (1-\gamma_5)\mu \right), \\
&=& \frac{G_F}{\sqrt{2}} \frac{\alpha}{2 \pi} \lambda_u C_{\rm SM}^K \left( \bar{s} \gamma^\mu (1-\gamma_5)d \right) \left(\bar{\mu} \gamma_\mu (1-\gamma_5)\mu \right),
\eea
where $\lambda_i = V_{id} V_{is}^*$,  $x_t = m_t^2/M_W^2$ and $\sin^2 \theta_W = 0.23$ and $C_{\rm SM}^K$ is the  SM Wilson coefficient  given as
\bea
C_{\rm SM}^K = \frac{\lambda_c Y_{NL} + \lambda_t Y(x_t) }{\sin^2\theta_W \lambda_u}.
\eea
 The functions $Y_{NL}$ and $Y(x_t)$ are the contributions from charm and top quark respectively and the $Y(x_t)$ function 
in the next-to-leading order (NLO) is  \cite{K-Misiak}
\bea
Y(x_t) = \eta_Y \frac{x_t}{8} \Bigg( \frac{4-x_t}{1-x_t} + \frac{3x_t}{\left(1-x_t \right)^2} \ln x_t \Bigg).
\eea
 The branching ratio  for the  SD part of $K_L \to \mu^+ \mu^-$ process  in the SM  is   given by
\bea
{\rm Br}(K_L \to \mu^+ \mu^-)|_{\rm SD} = \tau_{K_L}  \frac{G_F^2 }{2\pi }  |\lambda_u|^2
\sqrt{1-\frac{4m_\mu^2}{M_K^2}} f_K^2 M_K m_\mu^2 \Big |C_{\rm SM}^K \Big |^2.
\eea
Now including  the contribution of  $V^{(1)}(3,1,2/3)$ leptoquark, the total branching ratio of $K_L \to \mu^+ \mu^-$ process is given by
\bea \label{br-KL}
{\rm Br}(K_L \to \mu^+ \mu^-) = \frac{G_F^2}{8 \pi^3} \tau_{K_L} \alpha_{em}^2 f_{K}^2 M_{K} m_{\mu}^2 |\lambda_u|^2
\left |C_{\rm SM}^{K}\right |^2 \sqrt{1- \frac{4 m_\mu^2}{M_{K}^2}} \times \left(|P_K|^2 + |S_K|^2 \right),
\eea
where 
\bea
&&P_K \equiv \frac{C_{\rm SM}^K+C_{A}^{LQ}-C_{A}^{\prime LQ}}{C_{\rm SM}^{K}}+\frac{M_{K}^2}{2m_{\mu}} \Big(\frac{m_s}{m_s+m_d} \Big) \Big(\frac{C_{P}^{LQ}-C_{P}^{\prime LQ}}{C_{\rm SM}^{K}}\Big),\nn \\ 
&&S_K \equiv \sqrt{1- \frac{4 m_\mu^2}{M_{K}^2}} \frac{M_{K}^2}{2m_{\mu}} \Big(\frac{m_s}{m_s+m_d} \Big) \Big(\frac{C_{S}^{LQ}-C_{S}^{\prime LQ}}{C_{\rm SM}^{K}}\Big).
\label{P-S}
\eea
It should be noted that for  $K_L \to  \mu^+ \mu^- $ decay process, CP
violation in $K - \bar K $ mixing is irrelevant and  $K_L$ can be treated as a pure CP-odd state. Therefore, we have to take into account the contributions of both $K^0$  and $\bar K^0$ 
amplitudes, which can be done by replacing the leptoquark couplings 
$ (g_L)_{d\mu}(g_L)_{s\mu}^* \to \sqrt 2{\rm Re}[(g_L)_{d\mu}(g_L)_{s\mu}^*]$.
Thus, the  new $C_i^{LQ}$ coefficients arise due to the exchange of  vector leptoquark and are defined as 
\begin{subequations}
\bea
&&C_A^{LQ} = -\frac{\pi}{G_F \alpha_{em} \lambda_u} \frac{{\rm Re}[(g_L)_{d\mu}(g_L)_{s\mu}^*]}{M_{V^{(1)}}^2}, \\
&&C_A^{\prime LQ} = -\frac{\pi}{G_F \alpha_{em} \lambda_u} \frac{{\rm Re}[(g_R)_{d\mu}(g_R)_{s\mu}^*]}{M_{V^{(1)}}^2},\\
&&C_S^{LQ} =-C_P^{LQ} = \frac{\pi}{2 G_F \alpha_{em} \lambda_u} \frac{{\rm Re}[(g_L)_{d\mu}(g_R)_{s\mu}^*]}{M_{V^{(1)}}^2}, \\
&&C_S^{\prime LQ} =C_P^{\prime LQ} = \frac{\pi}{2 G_F \alpha_{em} \lambda_u} \frac{{\rm Re}[(g_R)_{d\mu}(g_L)_{s\mu}^*]}{M_{V^{(1)}}^2}.
\eea
\end{subequations}
In the presence of $V^{(3)}(3,3,2/3)$ leptoquark, the branching is given by
\bea \label{br-KL3}
{\rm Br}(K_L \to \mu^+ \mu^-) = \frac{G_F^2}{8 \pi^3} \tau_{K_L} \alpha_{em}^2 f_{K}^2 M_{K} m_{\mu}^2 |\lambda_u|^2
 \sqrt{1- \frac{4 m_\mu^2}{M_{K}^2}} \times \Big | C_{\rm SM}^K+\frac{ C_A^{LQ}}{2} \Big |^2.
\eea
For muonic decay the experimentally measured branching ratio  is ${\rm Br} (K_L \to \mu^+ \mu^-) = (6.84 \pm 0.11) \times 10^{-9}$ \cite{pdg} and for $K_L \to e^+ e^-$ process the branching ratio is ${\rm Br} (K_L \to e^+ e^-) = 9^{+6}_{-4} \times 10^{-12}$ \cite{pdg}.
If we apply chirality on the leptoquark, then only $C_A^{(\prime) LQ}$ Wilson coefficients will contribute. Now comparing  Eqn. (\ref{br-KL}) with the experimental branching ratio of $K_L \to \mu^+ \mu^-(e^+ e^-)$ processes, the constraint on the leptoquark couplings for $M_{LQ}= 1$ TeV are given by
\bea
&&1.3 \times 10^{-3} ~ \leq ~{\rm Re}[(g_L)_{d e} (g_L)_{s e}^*] ~\leq~ 2.35 \times 10^{-3}, \\ 
&&1.4 \times 10^{-4}~ \leq~ {\rm Re}[(g_L)_{d \mu} (g_L)_{s \mu}^* ] ~\leq~ 1.5 \times 10^{-4}.\label{KL-bound}
\eea
Now by neglecting the $C_A^{LQ}$ coefficients, the constraints on $(C_{S}^{LQ} \pm C_{ S}^{\prime LQ})$ Wilson coefficients from $K_L \to e^+ e^-$ (left panel) and $K_L \to \mu^+ \mu^-$ (right panel) are shown in Fig. 5.  From the figure the allowed regions of LQ couplings  for $K_L \to e^+ e^-$ process  are given by 
\bea
&&-2 \times 10^{-4} ~ \leq ~C_S^{LQ}+C_S^{ \prime LQ}~\leq~ 2 \times 10^{-4}, \\ 
&&1.25 \times 10^{-4} ~ \leq ~C_S^{LQ}-C_S^{ \prime LQ}~\leq~ 2 \times 10^{-4},
\eea
and for $K_L \to \mu^+ \mu^-$ process
\bea
&&-6 \times 10^{-3} ~ \leq C_S^{LQ}+C_S^{ \prime LQ}~\leq~ 3 \times 10^{-3}, \\ 
&&5\times 10^{-5} ~ \leq ~C_S^{LQ}-C_S^{ \prime LQ}~\leq~ 5.6 \times 10^{-3}.
\eea
\begin{figure}[h]
\centering
\includegraphics[scale=0.5]{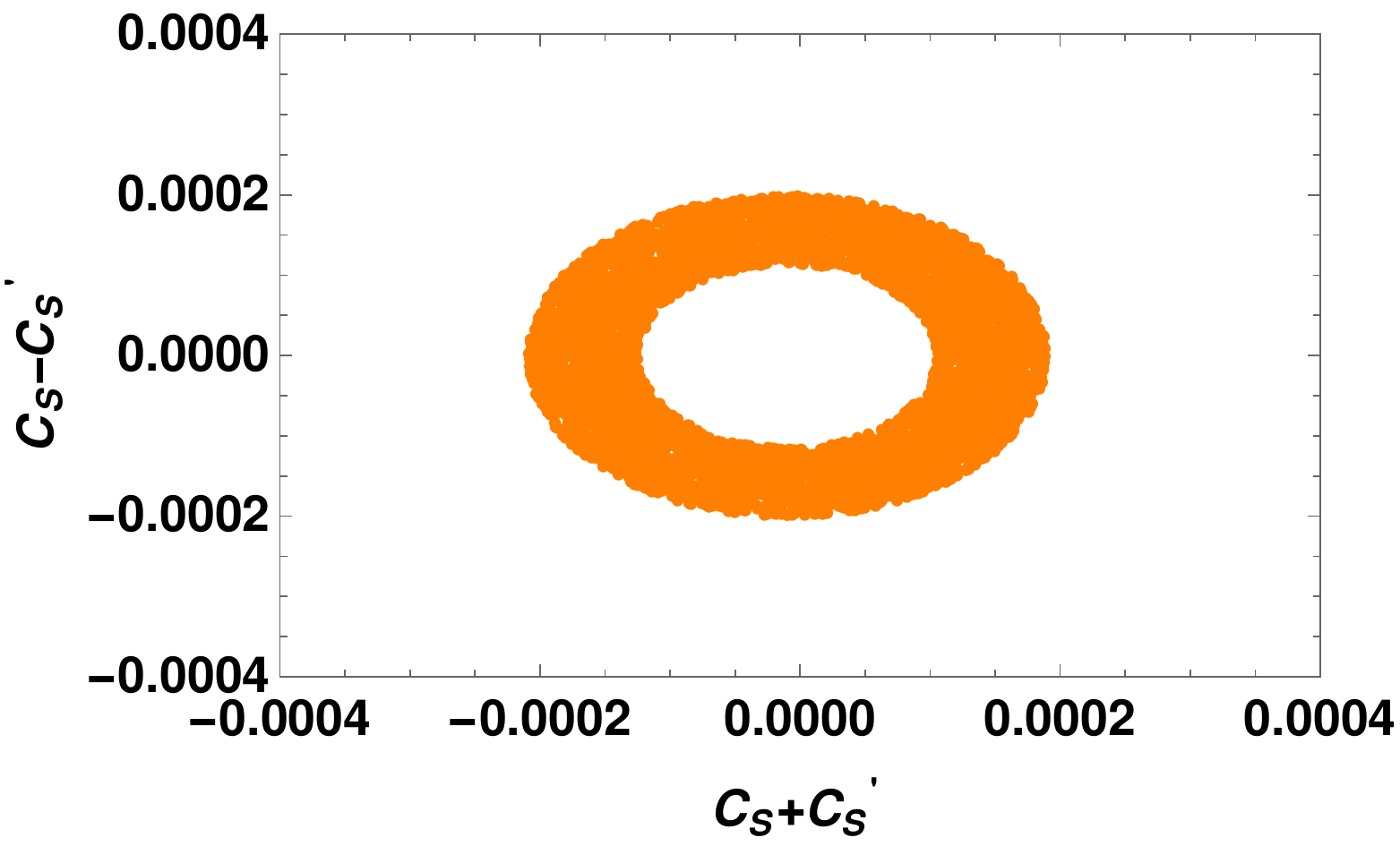}
\quad
\includegraphics[scale=0.5]{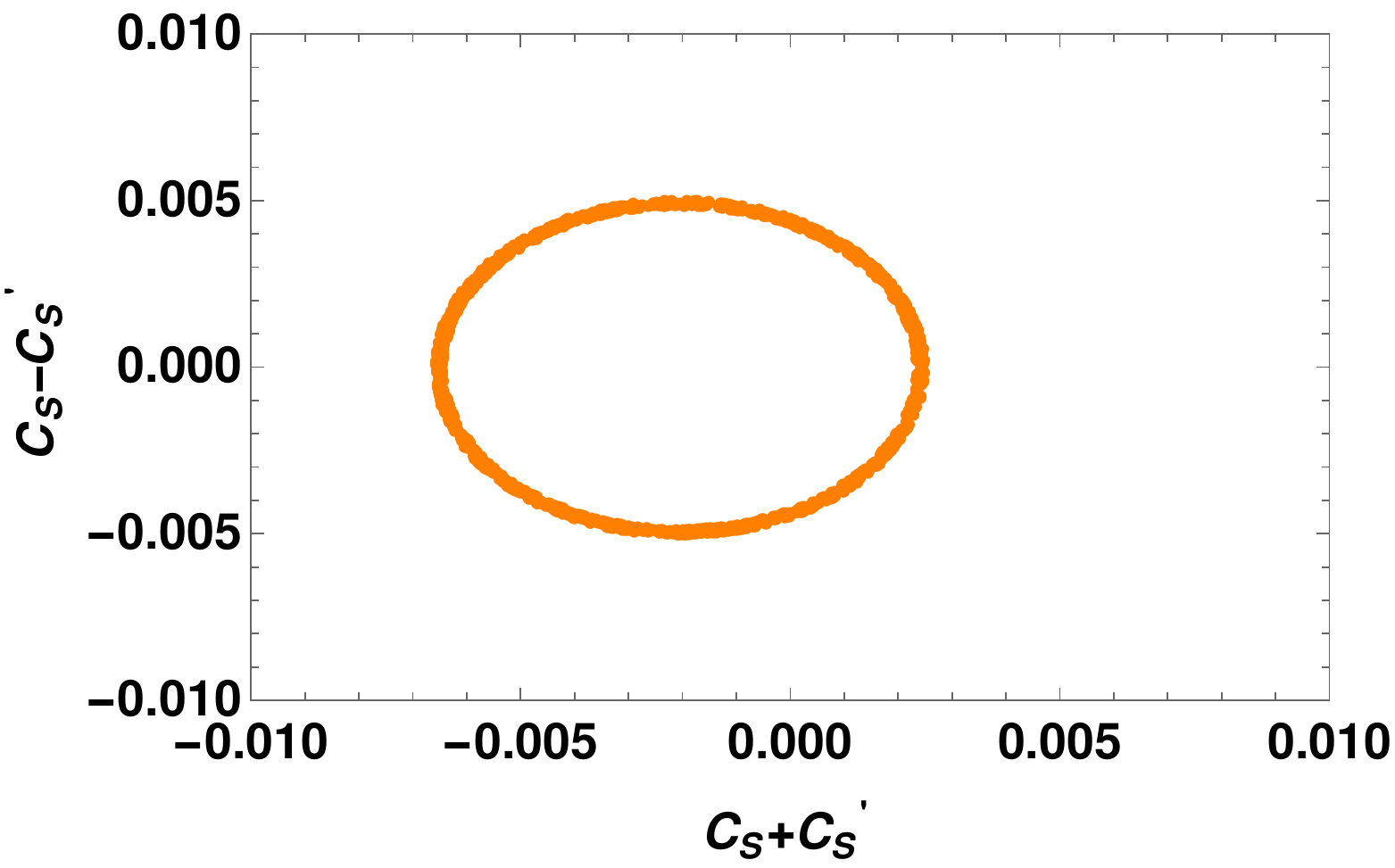}
\caption{The allowed region of  $C_S^{LQ}\pm C_S^{\prime LQ}$ Wilson coefficients   from $K_L \to e^+ e^-$ (left panel) and $K_L \to \mu^+ \mu^-$ (right panel) processes.}
\end{figure}
\subsection{$\tau^- \to \mu^- \gamma$ process}
In this subsection we compute the constraint on  $(3,1,2/3)$ vector  LQ couplings from the charged lepton flavour violating processes like $\tau^- \to l^- \gamma$, where $l=\mu, e$. These radiative decays provide an important testing ground for many new physics beyond the SM. The similar analysis in the context of scalar LQs can be found in the literature \cite{tau-mu-gamma}.  In the Ref. \cite{Lavoura}, the authors have given the general loop formulas for the radiative decay modes. The effective Hamiltonian for $\tau^- \to \mu^- \gamma$ process is given by 
\bea
\mathcal{H}_{\rm eff} = e \Bigg (C_L \bar{\mu}_R \sigma^{\mu \nu} F_{\mu \nu} \tau_L + C_R \bar{\mu}_L \sigma^{\mu \nu} F_{\mu \nu} \tau_R \Bigg),
\eea
where $\sigma^{\mu \nu}$ is the photon field strength tensor and  the Wilson coefficients $C_{L, R}$ are expressed as
\bea
C_L  &=& \frac{N_c }{16 \pi^2 M_{V^{(1)}}^2} \Bigg ( -\frac{1}{3} \Bigg [ (g_L)_{b\tau} (g_L)_{b\mu} ^ * f_2(x_b)+(g_R)_{b \tau} (g_R)_{b\mu} ^* f_1(x_b) \nn \\ &&~~~~~~~~~~~~~~~~~~~~~+(g_L)_{b\tau} (g_R)_{b\mu} ^* f_3(x_b)+(g_R)_{b\tau} (g_L)_{b\mu} ^* f_4(x_b) \Bigg ]\nn \\ &&~~~~~~~~~~~~~~+ \frac{2}{3}\Bigg [ (g_L)_{b\tau} (g_L)_{b\mu} ^ * \bar{f_2}(x_b)+(g_R)_{b\tau} (g_R)_{b\mu} ^ * \bar{f_1}(x_b)\nn \\&&~~~~~~~~~~~~~~~~~~~~~+(g_L)_{b\tau} (g_R)_{b\mu} ^* f_3(x_b)+(g_R)_{b\tau} (g_L)_{b\mu} ^* f_4(x_b) \Bigg ]\Bigg ) \nn \\ && + \frac{N_c }{16 \pi^2 M_{V^{(3)}}^2}  (g_L)_{b \tau} (g_L)_{b \mu} ^* \Bigg [ -\frac{1}{6} f_2^{(3)}(x_b) +\frac{1}{3} \bar{f_2}^{(3)}(x_b) \Bigg ],
\eea
\bea
C_R  &=& \frac{N_c }{16 \pi^2 M_{V^{(1)}}^2} \Bigg ( -\frac{1}{3} \Bigg [ (g_L)_{b\tau} (g_L)_{b\mu} ^ * f_1(x_b)+(g_R)_{b\tau} (g_R)_{b\mu} ^ * f_2(x_b) \nn \\ && ~~~~~~~~~~~~~~~~~~~~~(g_L)_{b\tau} (g_R)_{b\mu} ^* f_4(x_b)+(g_R)_{b\tau} (g_L)_{b\mu} ^* f_3(x_b) \Bigg ]\nn \\ &&~~~~~~~~~~~~~~+ \frac{2}{3}\Bigg [ (g_L)_{b\tau} (g_L)_{b\mu} ^ * \bar{f_1}(x_b)+(g_R)_{b\tau} (g_R)_{b\mu} ^ * \bar{f_2}(x_b) \nn \\ && ~~~~~~~~~~~~~~~~~~~~~(g_L)_{b\tau} (g_R)_{b\mu }^* \bar{f_4}(x_b)+(g_R)_{b\tau} (g_L)_{b\mu} ^* \bar{f_3}(x_b) \Bigg ]\Bigg ) \nn \\ && + \frac{N_c }{16 \pi^2 M_{V^{(3)}}^2}  (g_L)_{b \tau} (g_L)_{b \mu} ^*  \Bigg [ -\frac{1}{6} f_1^{(3)}(x_b) +\frac{1}{3} \bar{f_1}^{(3)}(x_b) \Bigg ].
\eea
Here $N_c=3$ is the color factor, $x_b = m_b^2/M_{\rm LQ}^2$ (where $M_{\rm LQ}= M_{V^{(1)}}~ {\rm or}~M_{V^{(3)}} $)  and the loop functions are  given as \cite{Lavoura}
\begin{subequations}
\bea \label{loop-f}
&& f_1(x_b) = m_\tau \Bigg [ \frac{-5x_b^3+9x_b^2-30x_b+8}{12(x_b-1)^3} + \frac{3x_b^2\ln x_b}{2(x_b-1)^4}\Bigg ], \\
&& f_2(x_b) = m_\mu \Bigg [ \frac{-5x_b^3+9x_b^2-30x_b+8}{12(x_b-1)^3} + \frac{3x_b^2\ln x_b}{2(x_b-1)^4}\Bigg ], \\
&& f_3(x_b) = m_b \Bigg [ \frac{x_b^2+x_b +4 }{2(x_b-1)^2} - \frac{3x_b\ln x_b}{(x_b-1)^3} \Bigg ], \\
&&f_4(x_b) = - \frac{m_\tau m_\mu m_b}{m_{V^{(1)}}^2}\Bigg [ \frac{-2x_b^2+7x_b-11}{6(x_b-1)^3} + \frac{\ln x_b}{(x_b-1)^4} \Bigg ], \\
&& \bar{f_1}(x_b) = m_\tau \Bigg [\frac{-4x_b^3+45x_b^2-33x_b+10}{12(x_b-1)^3} -\frac{3x_b^3 \ln x_b}{2(x_b-1)^4} \Bigg ], \\
&& \bar{f_2}(x_b) = m_\mu \Bigg [\frac{-4x_b^3+45x_b^2-33x_b+10}{12(x_b-1)^3} -\frac{3x_b^3 \ln x_b}{2(x_b-1)^4} \Bigg ], \\
&& \bar{f_3}(x_b) = m_b \Bigg [ \frac{x_b^2-11x_b +4 }{2(x_b-1)^2} + \frac{3x_b^2\ln x_b}{(x_b-1)^3} \Bigg ],\\
&& \bar{f_4}(x_b) = \frac{m_\tau m_\mu m_b}{m_{V^{(1)}}^2}\Bigg [\frac{x_b^2-5x_b-6-6x_b(1+x_b) \ln x_b}{6(x_b-1)^3} + \frac{x_b^3\ln x_b}{(x_b-1)^4} \Bigg ].
\eea
\end{subequations}
The branching ratio of $\tau^- \to \mu^- \gamma$ process is given by  
\bea \label{br-tau-mu}
{\rm Br} (\tau^- \to \mu^- \gamma) = \frac{\tau_\tau\left(m_\tau ^2 - m_\mu ^2 \right)^3}{16\pi m_\tau^3} \Big [ | C_L |^2 + |C_R |^2 \Big ].
\eea
 This expression can be applied to study other LFV radiative decays like, $\tau^- \to e^- \gamma$ and $\mu^- \to e^- \gamma$. The current upper bounds on the branching ratios of $\tau^- \to \mu^-(e^-) \gamma$ is given by \cite{pdg}
\bea
&&{\rm Br} (\tau^- \to \mu^- \gamma) ~ \textless ~ 4.4 \times 10^{-8}, \nn\\
&&{\rm Br} (\tau^- \to e^- \gamma)~ \textless ~3.3 \times 10^{-8}.
\label{tau-bound}
\eea
 Comparing Eqn. (\ref{br-tau-mu}) with the current experimental bounds
 (\ref{tau-bound}), the allowed regions of  $(g_{L(R)})_{bl_i} (g_{L(R)})_{bl_j }^*$ couplings  from $\tau^- \to e^- \gamma$ (left panel) and $\tau^- \to \mu^- \gamma$ (right panel) are shown in Fig. 7, and the constraints on the $(g_{L(R)})_{bl_i} (g_{R(L)})_{bl_j}^ *$ leptoquark couplings  from $\tau^- \to e^- \gamma$ (left panel) and $\tau^- \to \mu^- \gamma$ (right panel) are presented  in Fig. 8. The numerical values of  the constraints on leptoquark couplings are given in Table III. These bounds are rather weak in comparison to $B_{s,d} \to \mu^+ \mu^-$ processes and  they also involve the coupling only to $b$ quark in both the LQ coupling-parameters.  
\begin{figure}[h]
\centering
\includegraphics[scale=0.5]{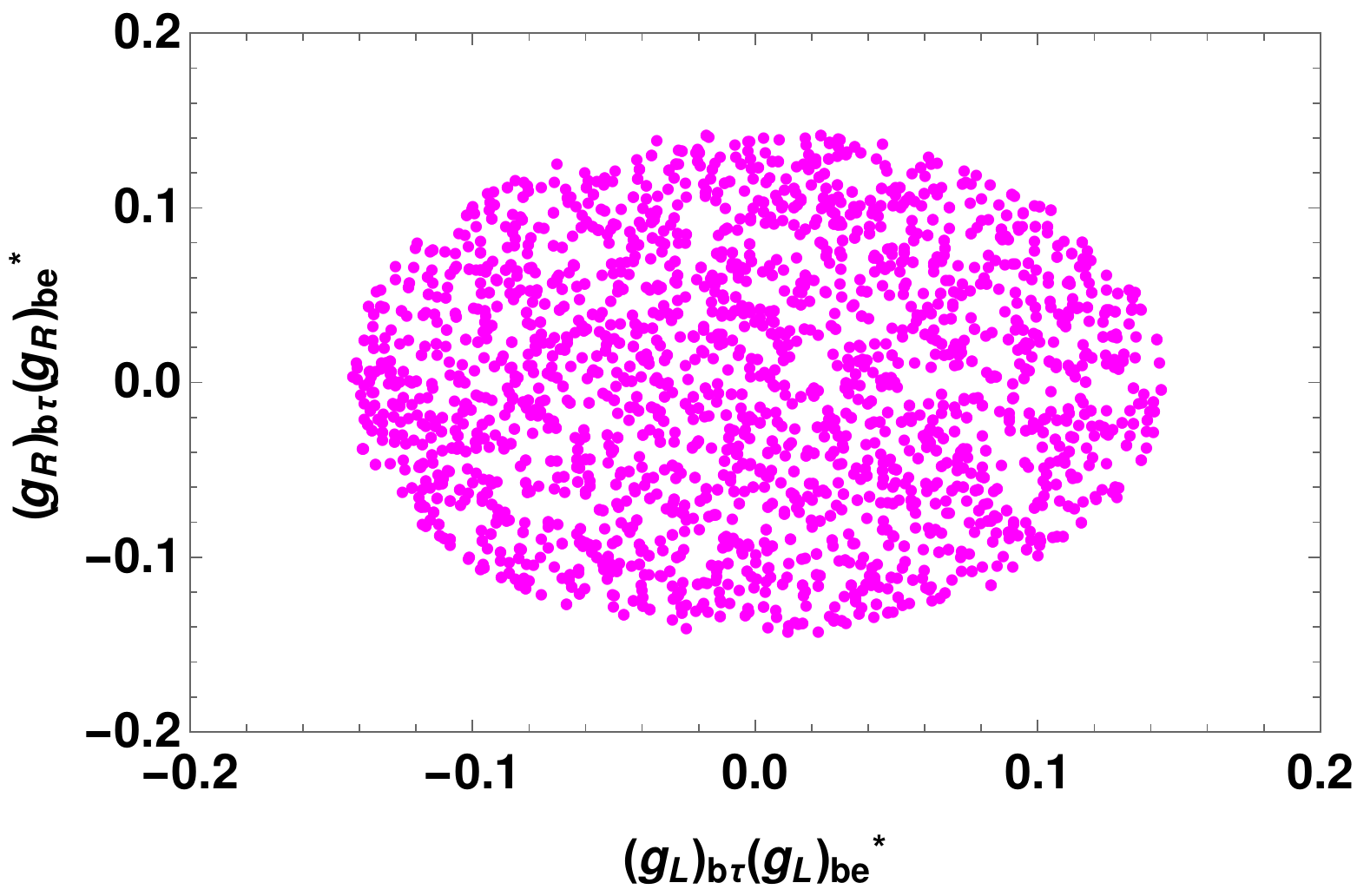}
\quad
\includegraphics[scale=0.5]{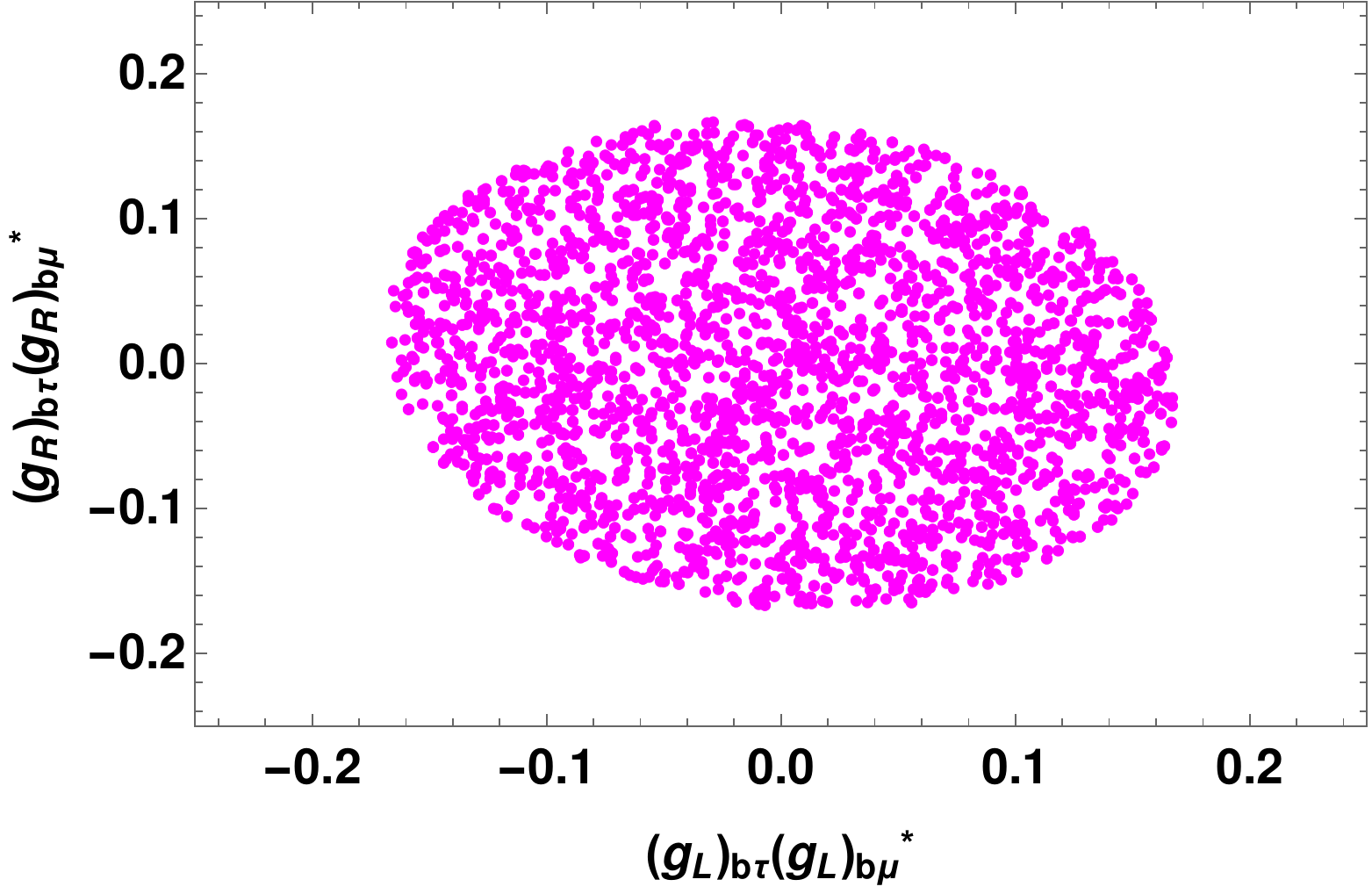}
\caption{The constraint on $(g_L)_{b l_i} (g_L)_{bl_j}^ *$ and $(g_R)_{bl_i} (g_R)_{bl_j}^ *$  leptoquark couplings  from $\tau^- \to e^- \gamma$ (left panel) and $\tau^- \to \mu^- \gamma$ (right panel) processes.}
\end{figure}
\begin{figure}[h]
\centering
\includegraphics[scale=0.5]{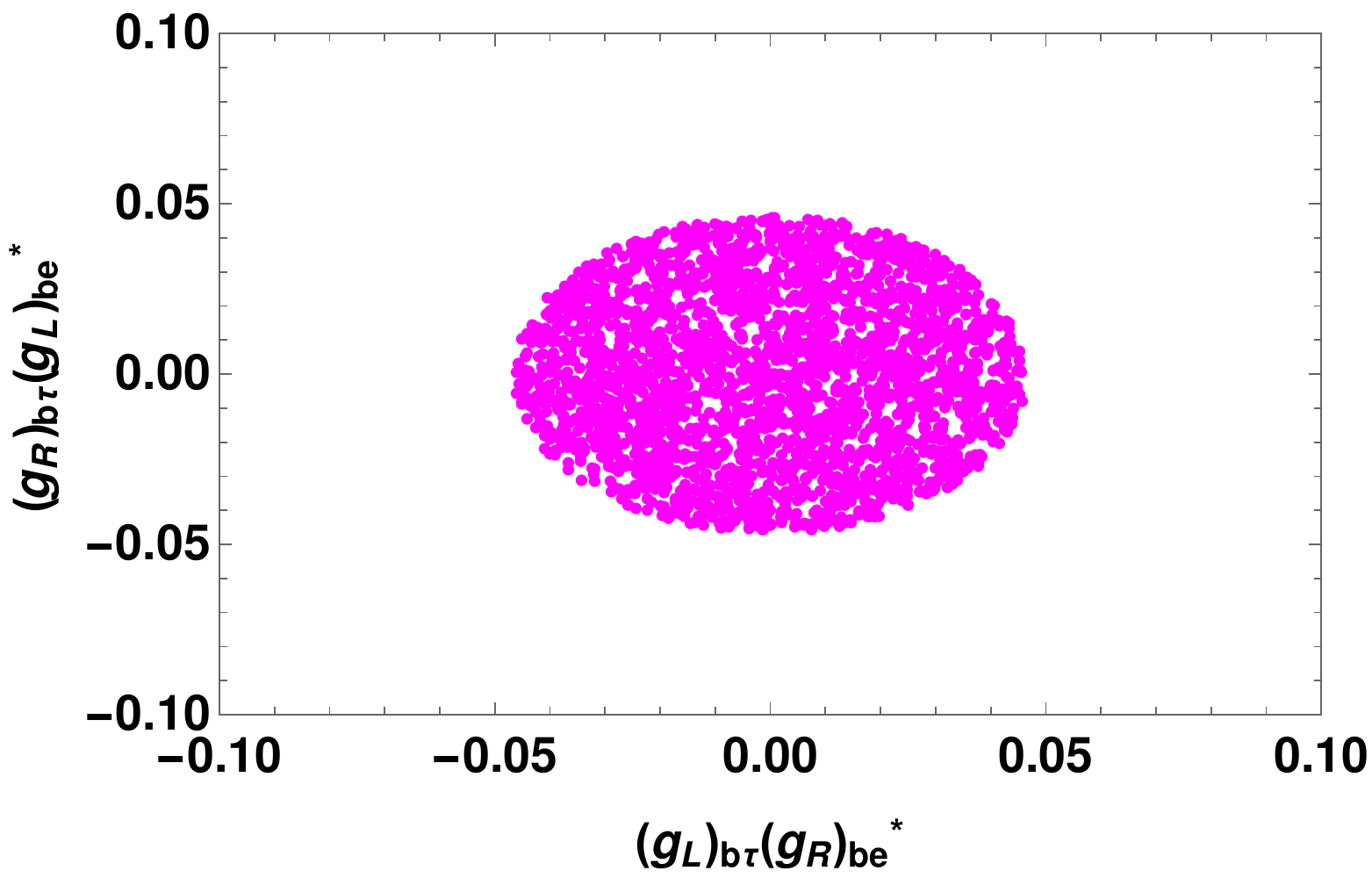}
\quad
\includegraphics[scale=0.5]{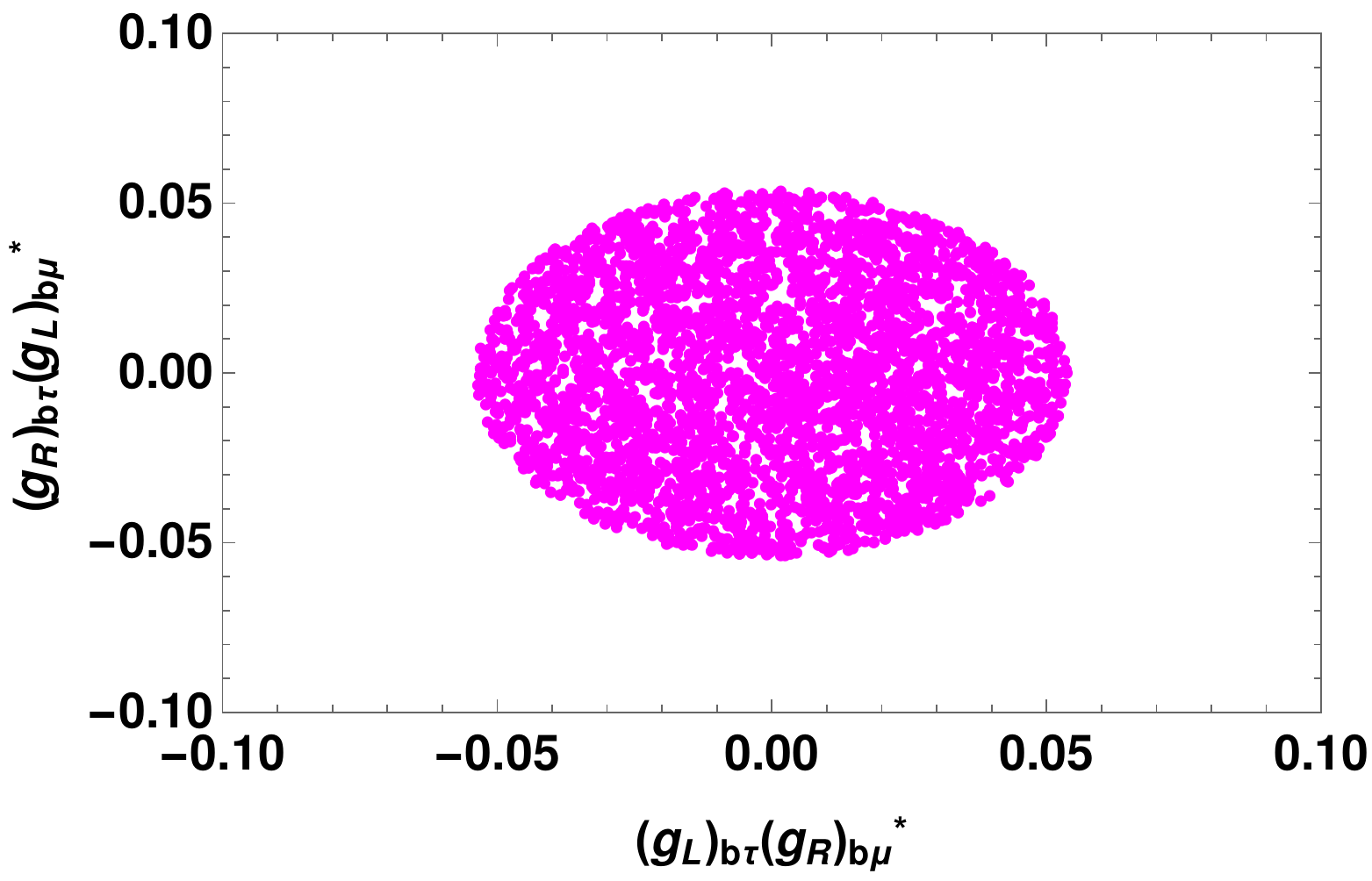}
\caption{The constraint on $(g_L)_{bl_i} (g_R)_{bl_j} ^*$ and $(g_R)_{bl_i} (g_L)_{bl_j}^ *$  leptoquark couplings  from $\tau^- \to e^- \gamma$ (left panel) and $\tau^- \to \mu^- \gamma$ (right panel) processes.}
\end{figure}
\begin{table}[htb]
\begin{center}
\caption{Constraints on leptoquark couplings obtained from   $\tau^- \to l^- \gamma$ processes.}
\vspace*{0.1 true in}
\begin{tabular}{|c|c|c|}
\hline
Couplings involved ~& ~$\tau^- \to e^- \gamma$ process ~&~ $\tau^- \to \mu^- \gamma$ process \\
\hline

$(g_L)_{b\tau} (g_L)_{b l}^*$ &~ $-0.14 \to 0.14$ ~& ~$ -0.16 \to 0.16 $~\\

\hline

$(g_R)_{b\tau} (g_R)_{b l}^*$ &~ $-0.14 \to 0.14$ ~& ~$ -0.16 \to 0.16 $~\\

\hline
$(g_L)_{b \tau} (g_R)_{bl}^* $ &~ $-0.04 \to 0.04$ ~& ~$ -0.05 \to 0.05  $~\\

\hline

$(g_R)_{b\tau} (g_L)_{bl}^* $ &~ $-0.04 \to 0.04$ ~& ~$ -0.05 \to 0.05 $~\\

\hline
\end{tabular}
\end{center}
\end{table}
\section{Numerical analysis of LFV decays}

After having detailed knowledge about the observables and the bound on new Wilson coefficients, we now proceed for numerical analysis  of LFV decays in the LQ model.  Though LFV decays are extremely rare in the SM due to loop suppression and the presence of tiny neutrino mass  in  the loop, still they can occur at tree level and are expected to have significantly large branching ratios in the LQ model. There will be no contributions from SM Wilson coefficients in the LFV decays of $B$ meson.
 
\begin{figure}[h]
\centering
\includegraphics[scale=0.4]{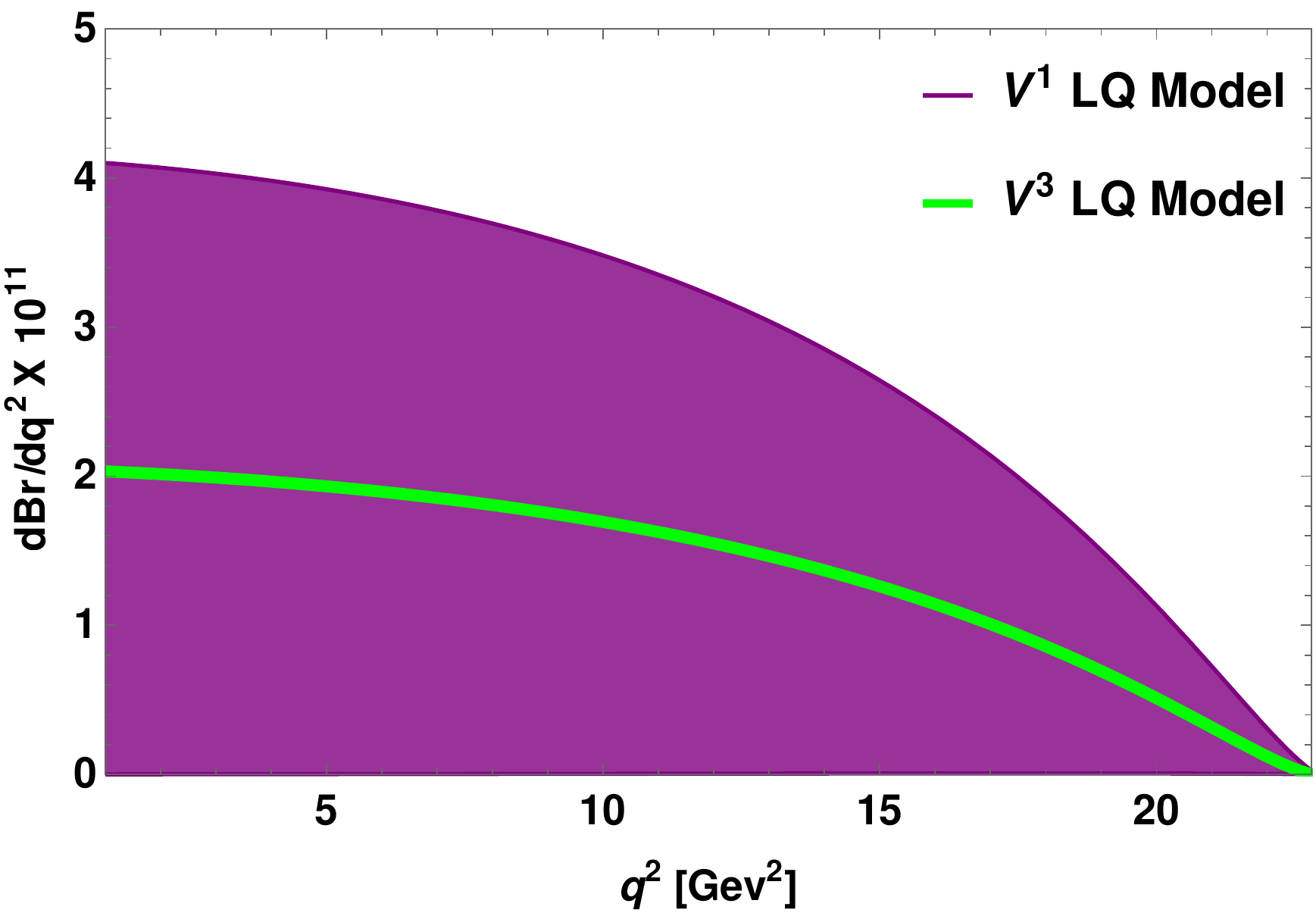}
\quad
\includegraphics[scale=0.4]{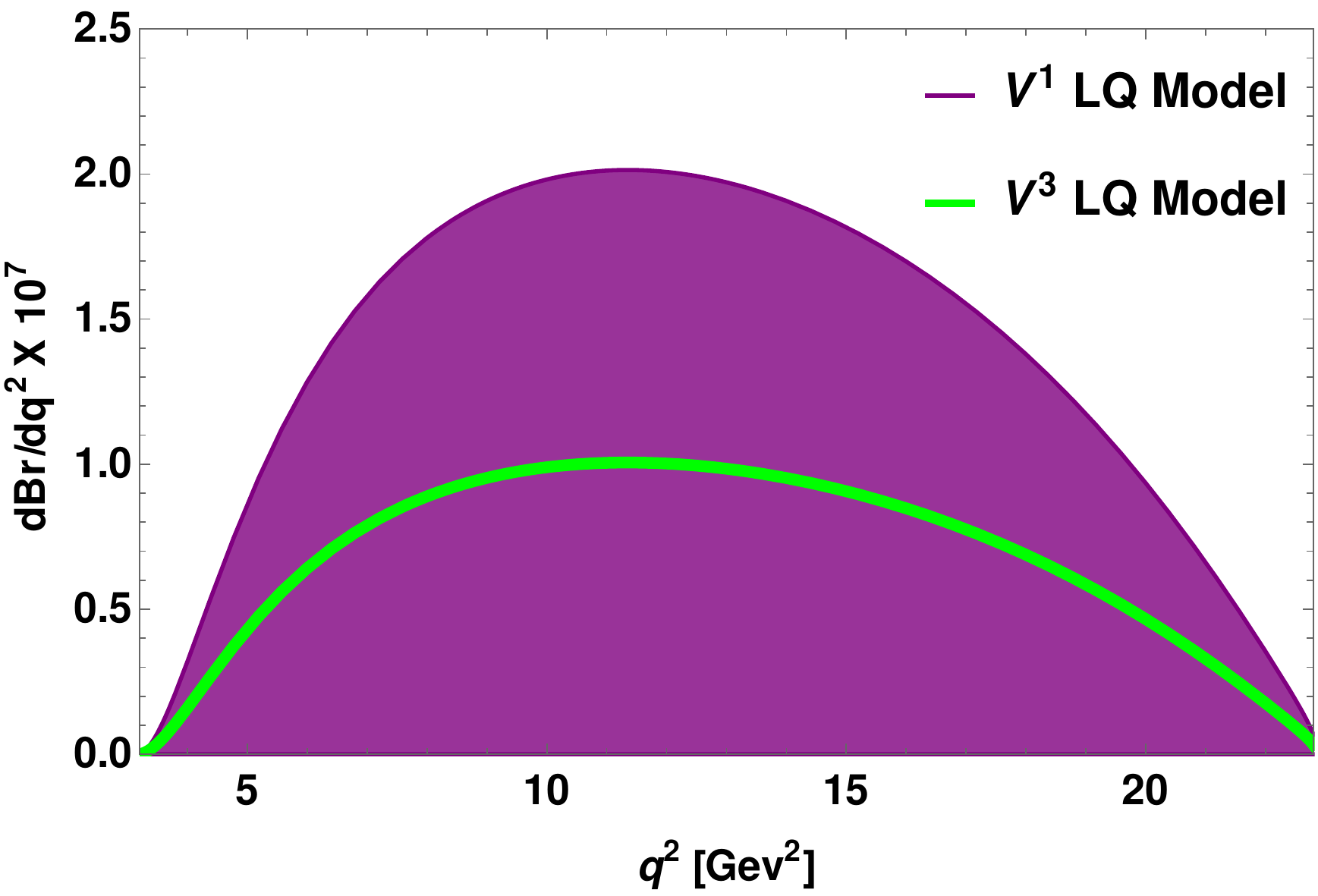}
\quad
\includegraphics[scale=0.4]{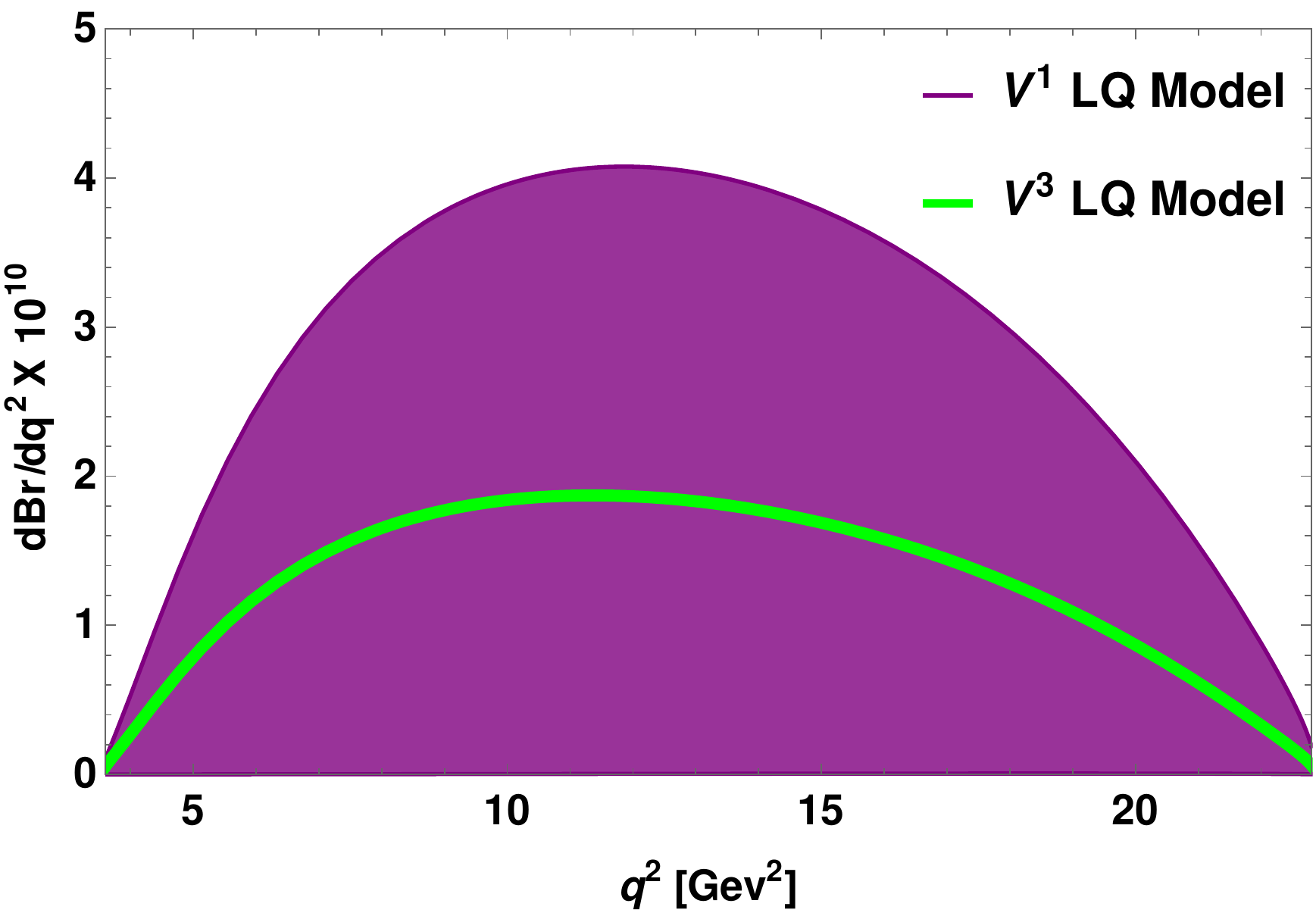}
\caption{The variation of branching ratios of $B^+ \to K^+ \mu^- e^+$ (top left panel), $B^+ \to K^+ \tau^- e^+$ (top right panel) and $B^+ \to K^+ \tau^- \mu^+$ (bottom panel) processes (in units of ${\rm GeV}^{-2}$) in the $V^{1, 3}$  vector leptoquark model. Here purple bands represent the contribution from $V^1$ leptoquark model and green solid lines are for $V^3$ leptoquark.}
\end{figure}

In the presence of LQ, the modified helicity amplitudes are given as
\bea
&&H_V^{0 \rm LQ} = \sqrt{\frac{\lambda}{q^2}} \left(C_V^{\rm LQ}+C_V^{\prime \rm LQ} \right) f_{+}(q^2),  \\
&&H_V^{t\rm LQ} = \frac{M_{B}^2-M_K^2}{\sqrt{q^2}} \left(C_{V}^{\rm LQ}+C_{V}^{\prime \rm LQ} \right) f_{0}(q^2),  \\
&&H_A^{0 \rm LQ} = \sqrt{\frac{\lambda}{q^2}} \left(C_{A}^{\rm LQ}+C_{A}^{\prime \rm LQ} \right) f_{+}(q^2), \\
&&H_A^{t \rm LQ} = \frac{M_{B}^2-M_K^2}{\sqrt{q^2}} \left(C_{A}^{\rm LQ}+C_{A}^{\prime \rm LQ} \right) f_{0}(q^2),  \\
&&H_S^{\rm LQ} = \frac{M_{B}^2-M_K^2}{m_b} \left(C_{S}^{\rm LQ}+C_{S}^{' \rm LQ} \right) f_{0}(q^2),  \\
&&H_P^{\rm LQ} = \frac{M_{B}^2-M_K^2}{m_b} \left(C_{P}^{\rm LQ}+C_{P}^{'\rm  LQ} \right) f_{0}(q^2).
\eea
\begin{figure}[h]
\centering
\includegraphics[scale=0.4]{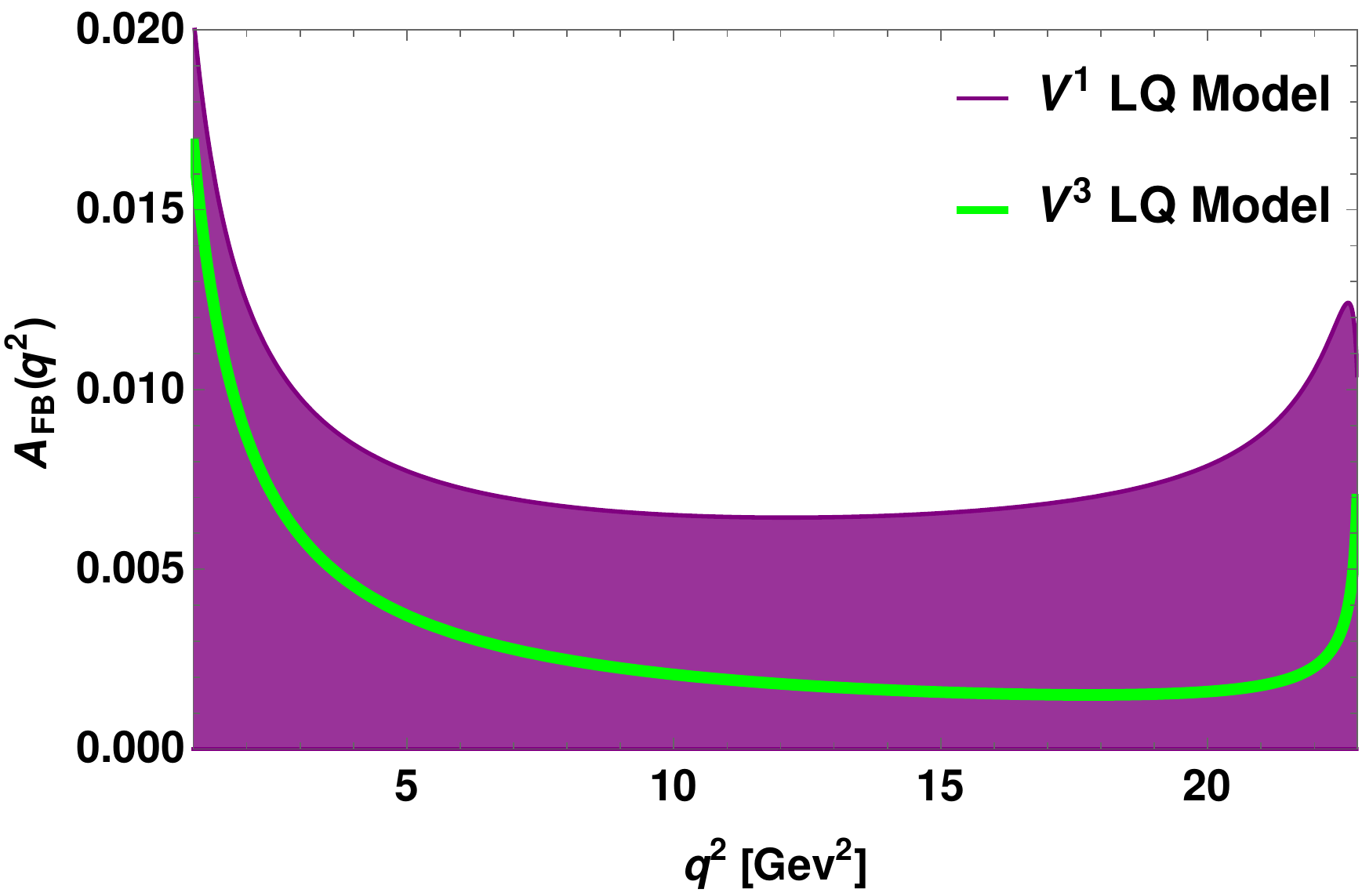}
\quad
\includegraphics[scale=0.4]{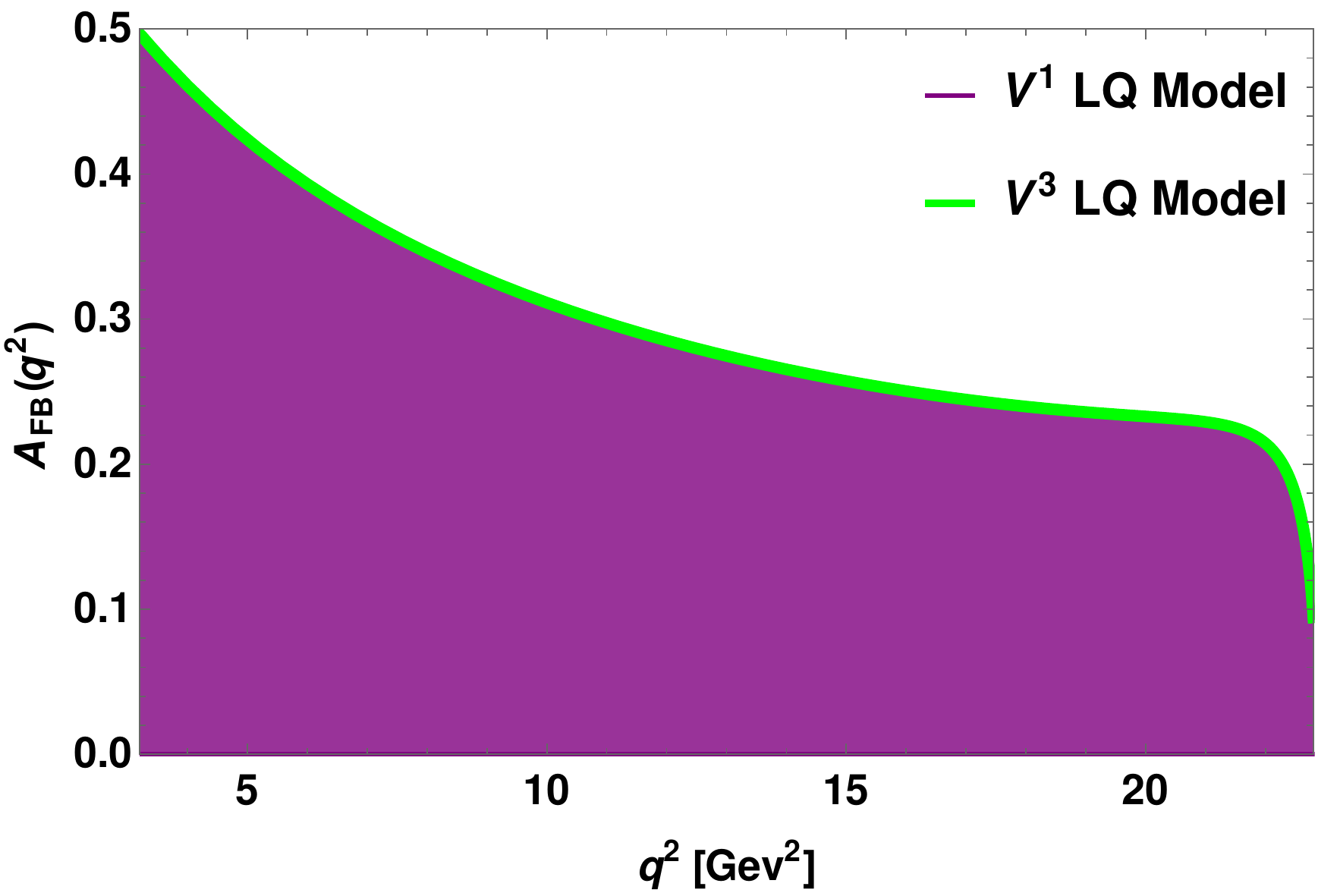}
\quad
\includegraphics[scale=0.4]{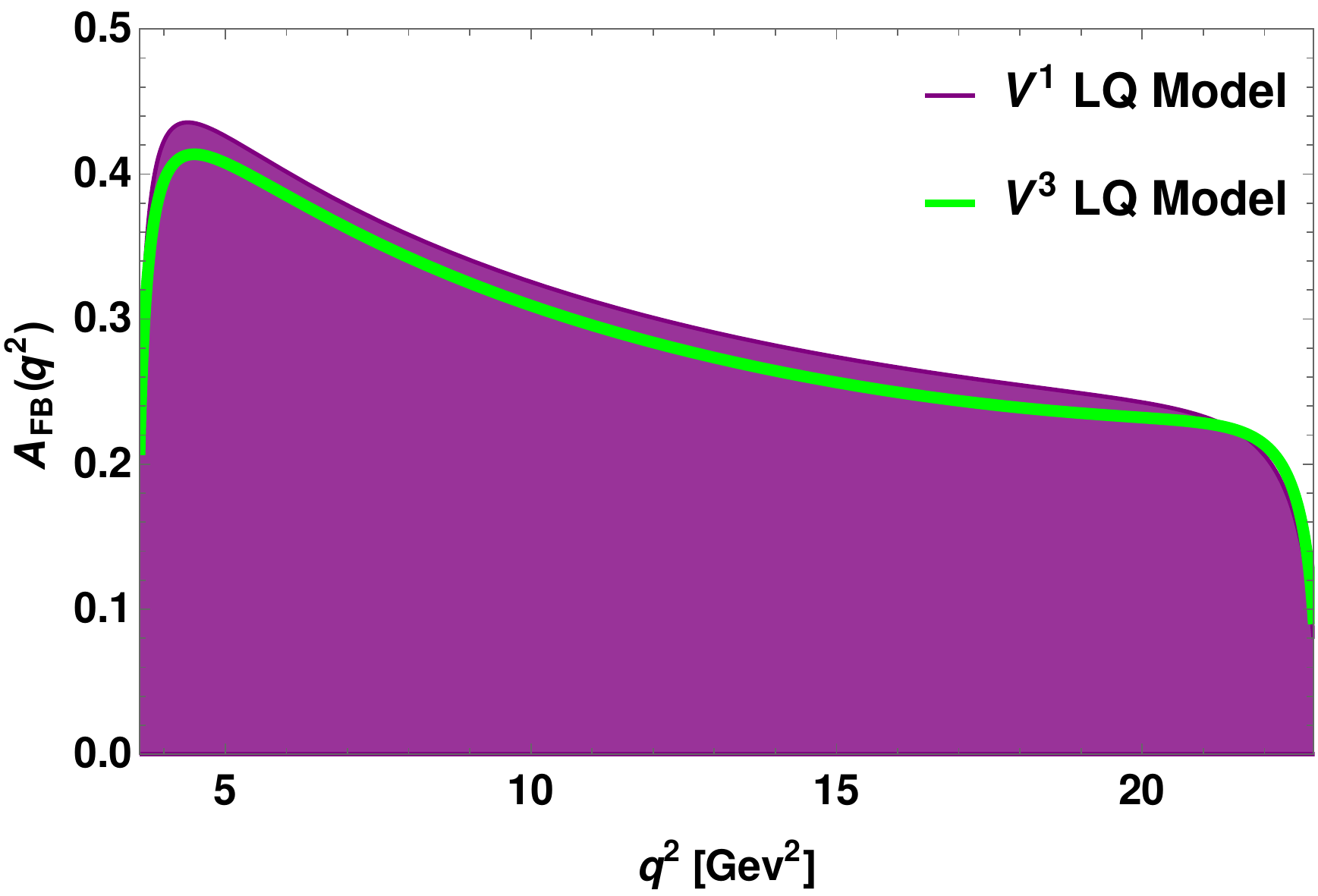}
\caption{The variation of forward-backward asymmetries of  $B^+ \to K^+ \mu^- e^+$ (top left panel), $B^+ \to K^+ \tau^- e^+$ (top right panel) and $B^+ \to K^+ \tau^- \mu^+$ (bottom panel) processes   in the leptoquark model. }
\end{figure}
\begin{figure}[h]
\centering
\includegraphics[scale=0.4]{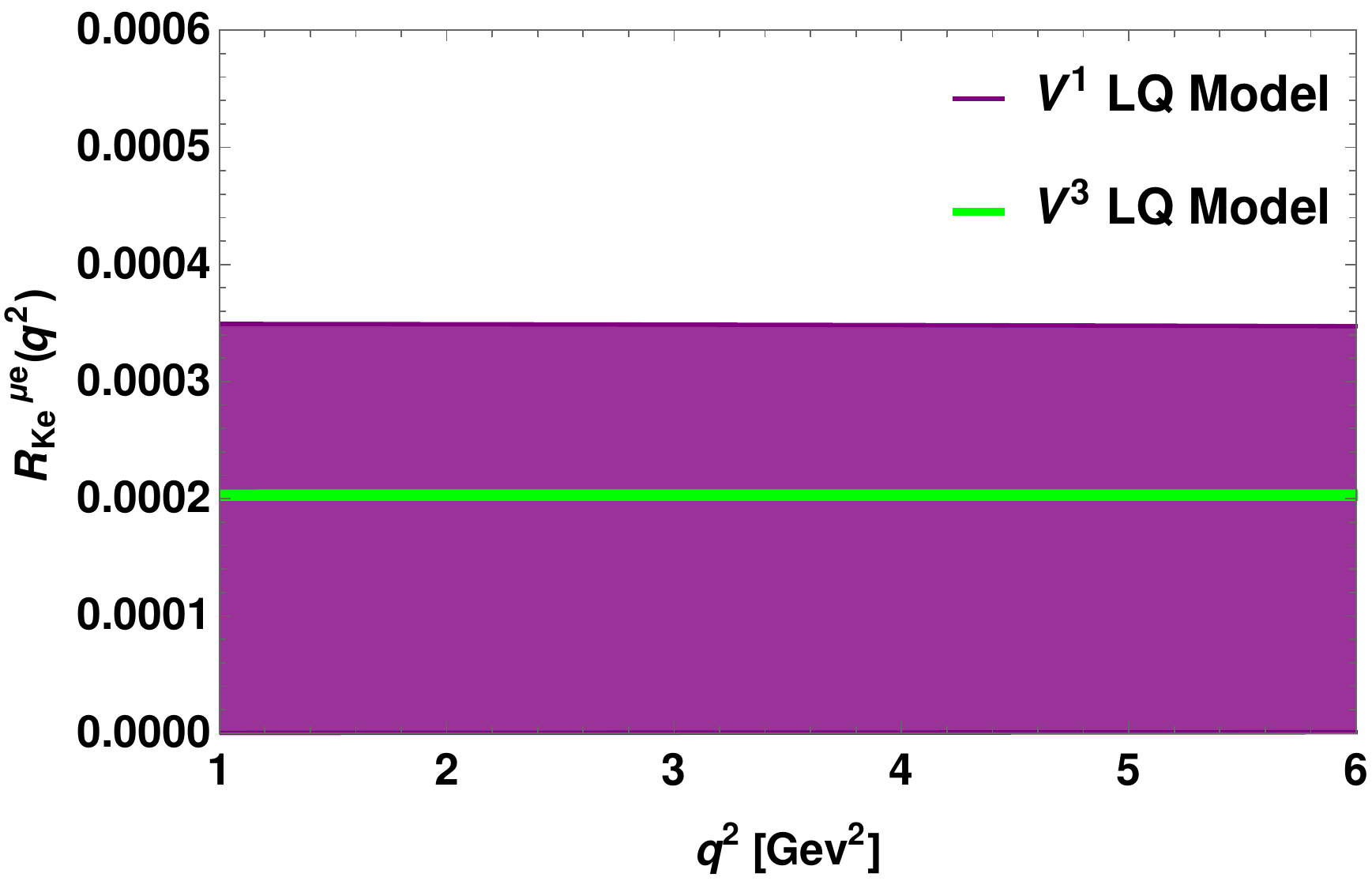}
\quad
\includegraphics[scale=0.4]{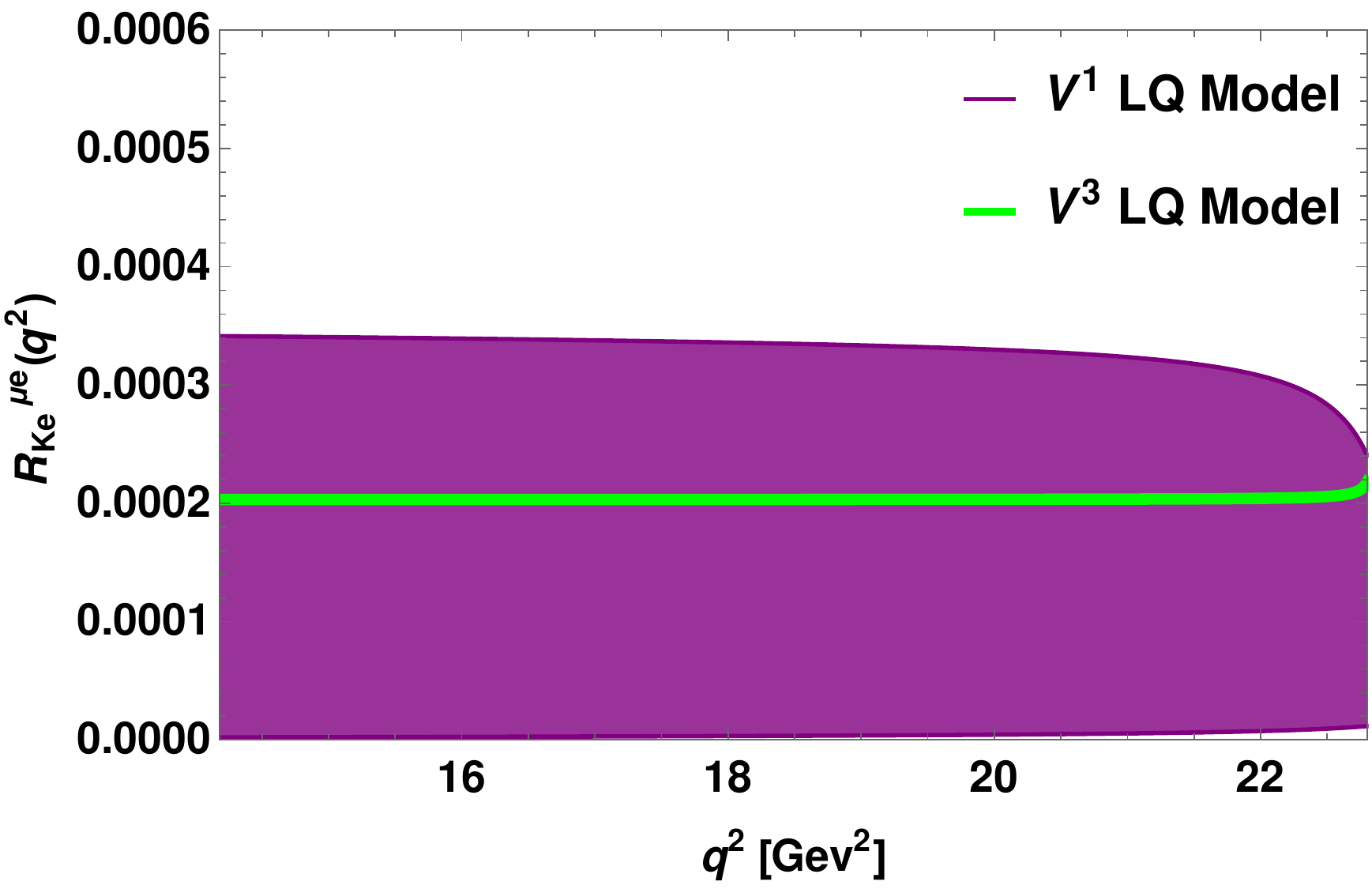}
\quad
\includegraphics[scale=0.4]{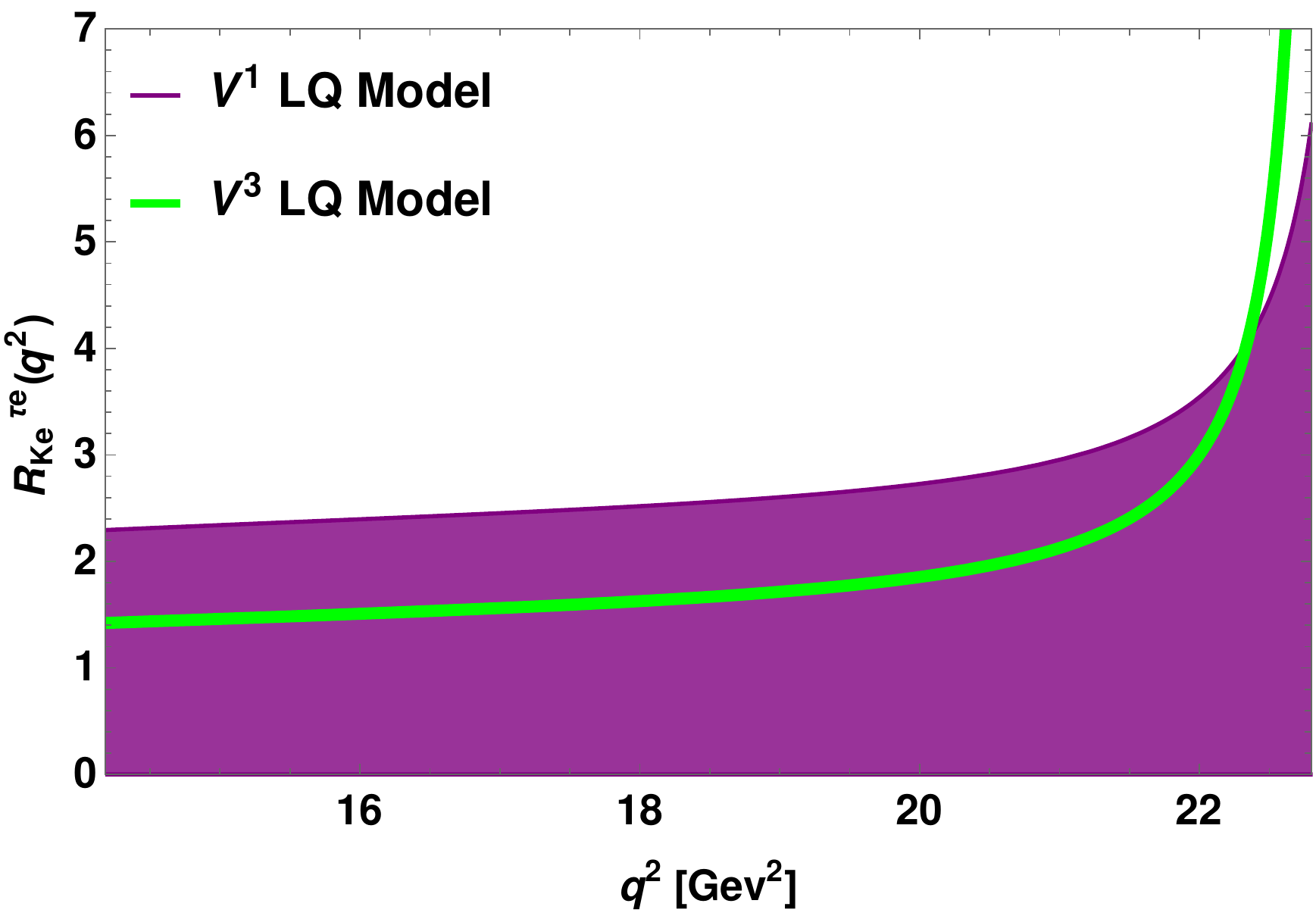}
\quad
\includegraphics[scale=0.4]{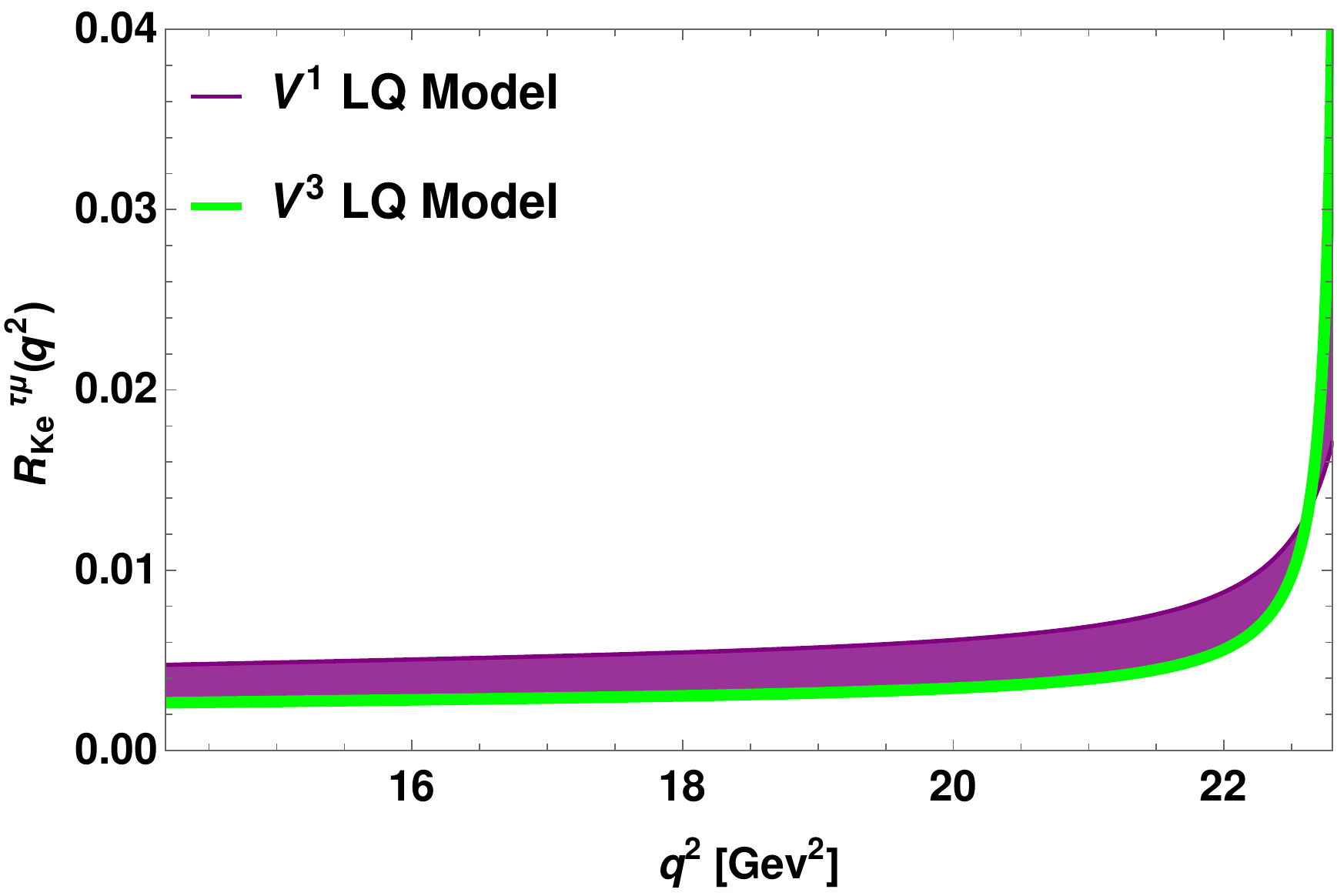}
\caption{The plots for lepton non-universality  parameters,  $R_{K e}^{\mu e}$ (top right panel), $R_{K e}^{\tau e}$ (bottom left panel) and $R_{K e}^{\tau \mu}$ (bottom  right panel) in high $q^2$ region. Here the top left panel shows the non-universality  $R_{K e}^{\mu e}$ in low $q^2 \in [1, 6]$ region.}
\end{figure}
\begin{figure}[h]
\centering
\includegraphics[scale=0.4]{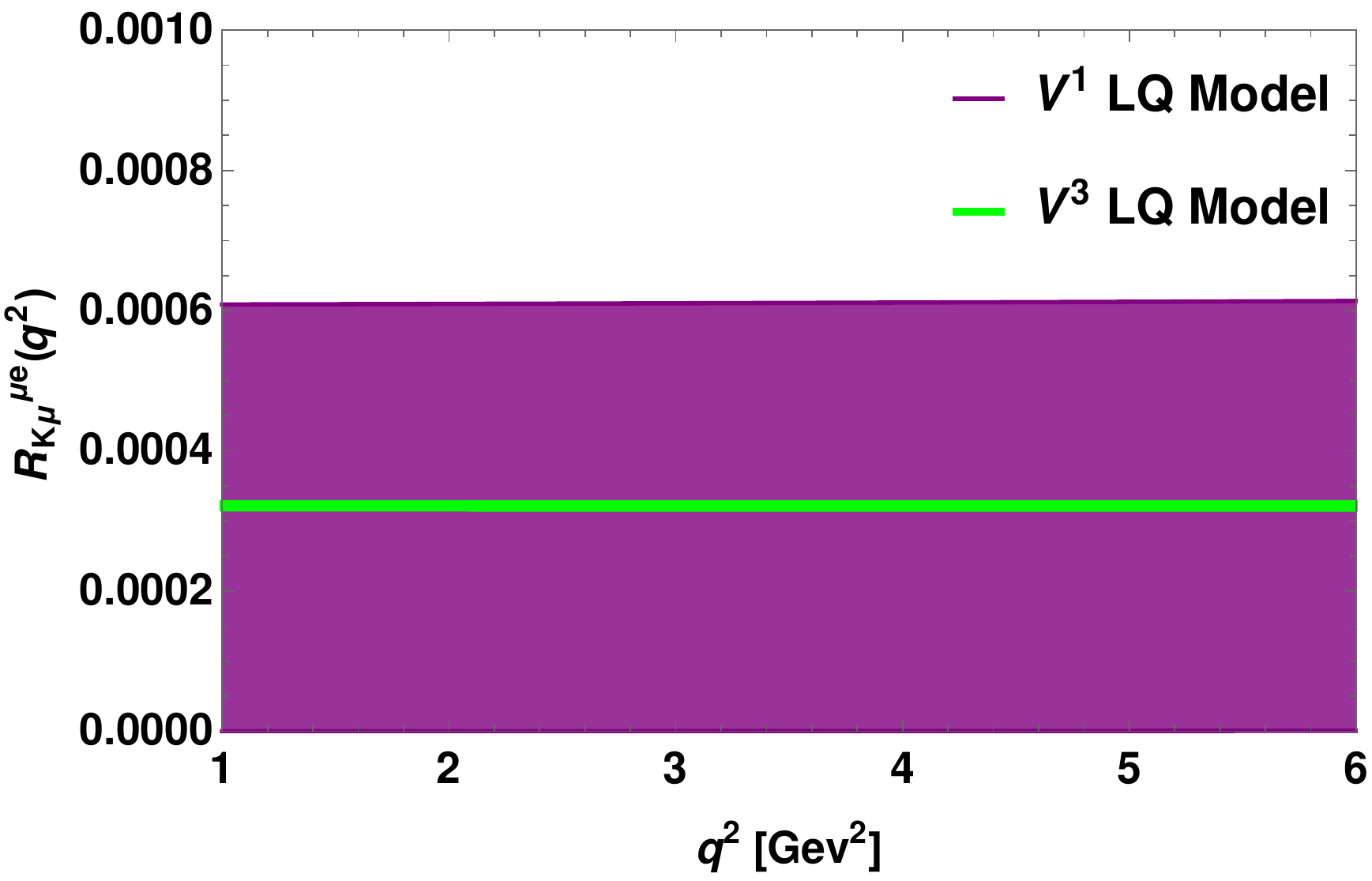}
\quad
\includegraphics[scale=0.4]{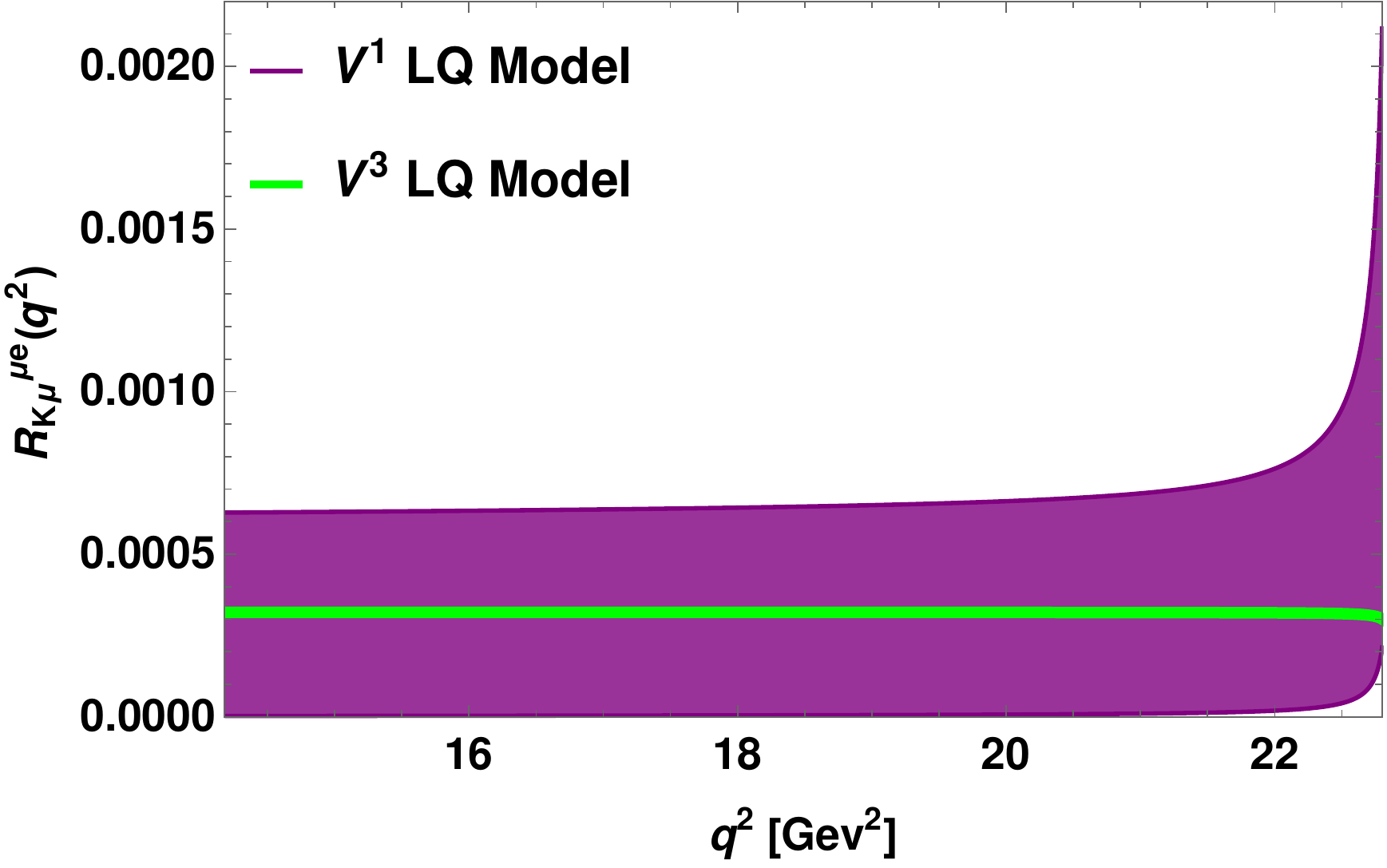}
\quad
\includegraphics[scale=0.4]{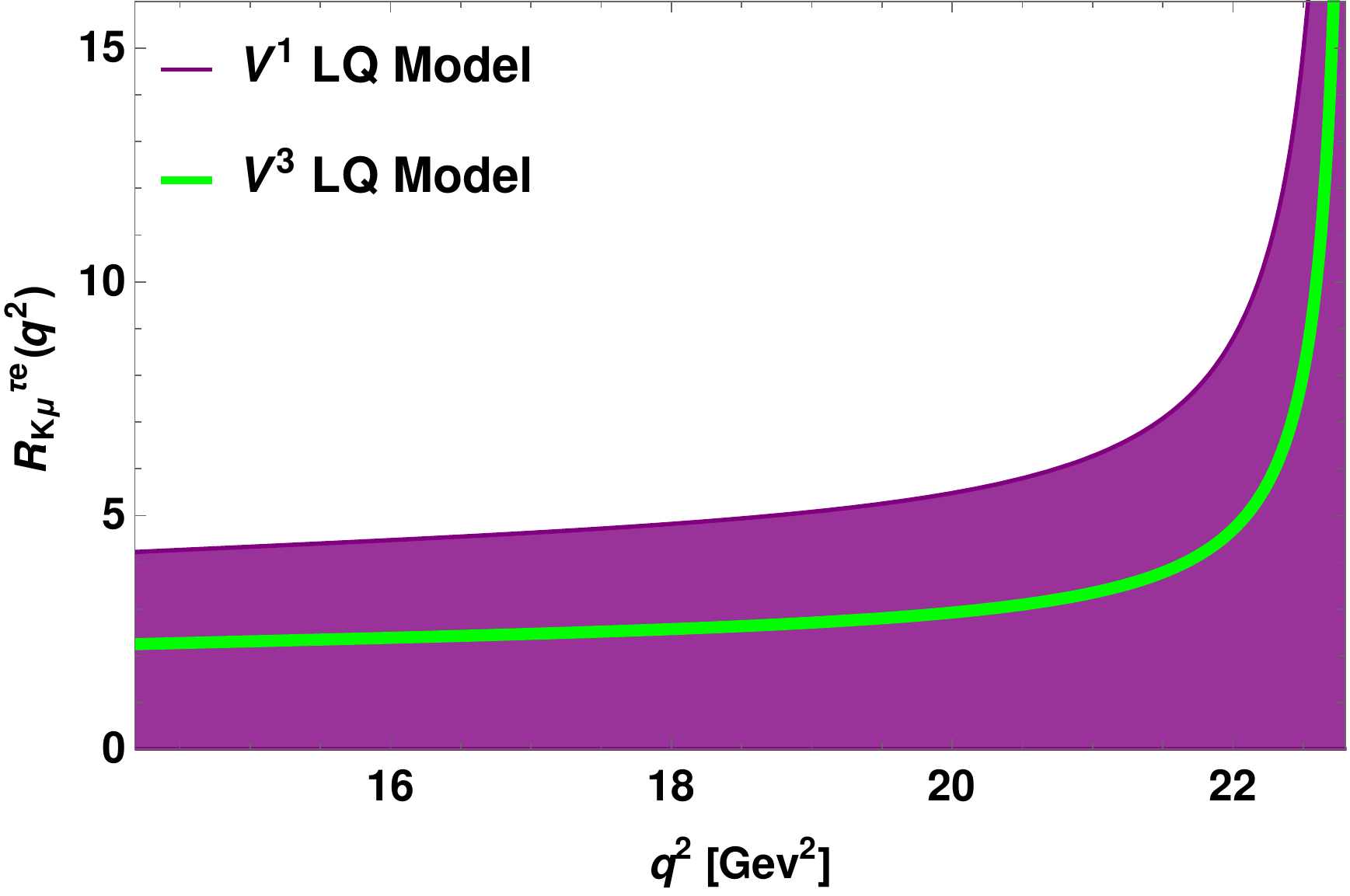}
\quad
\includegraphics[scale=0.4]{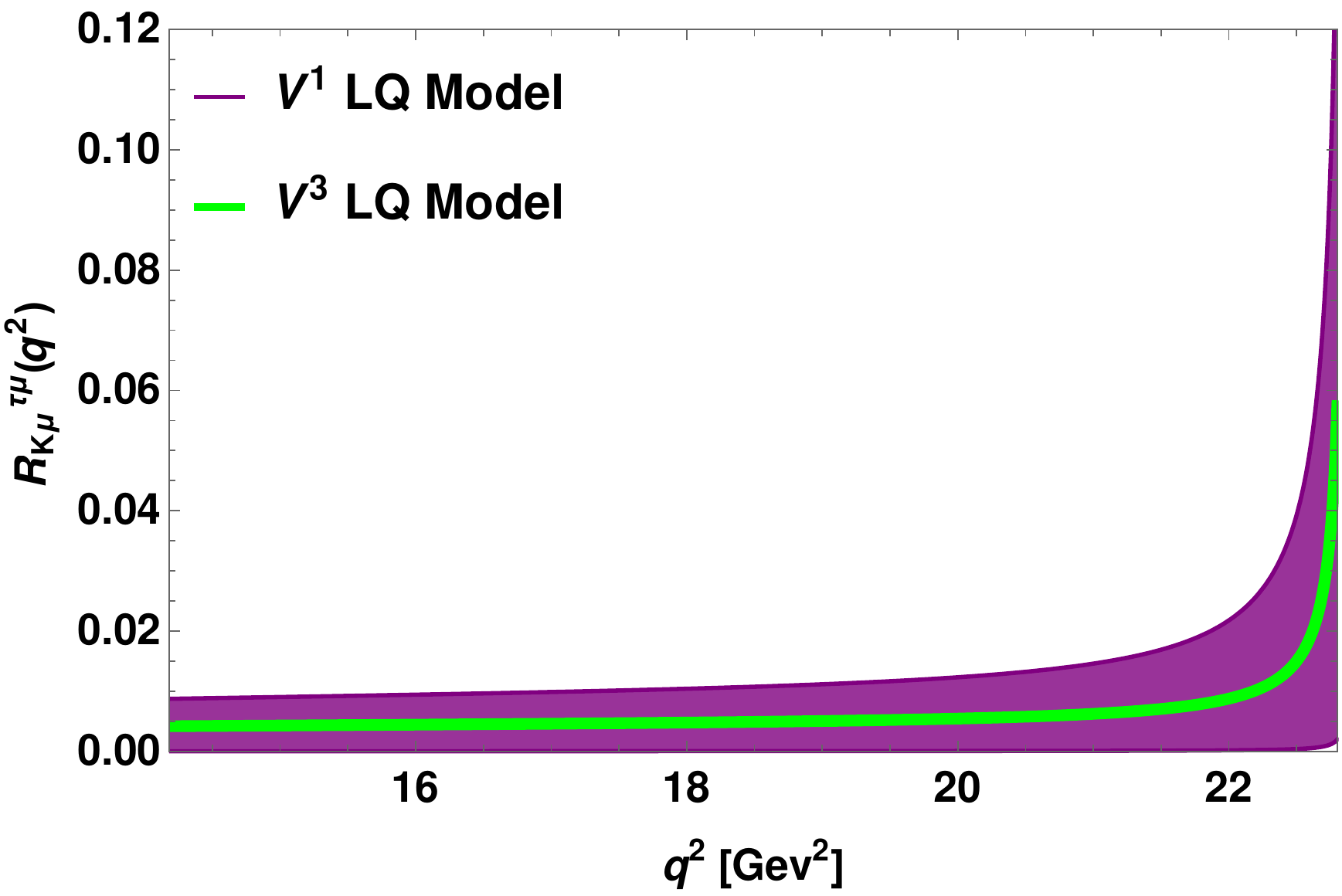}
\caption{The plots for lepton non-universality  parameters,  $R_{K \mu}^{\mu e}$ (top right panel), $R_{K \mu}^{\tau e}$ (bottom left panel) and $R_{K \mu}^{\tau \mu}$ (bottom  right panel) in high $q^2$ region. Here the top left panel shows the non-universality  $R_{K \mu}^{\mu e}$ in low $q^2 \in [1, 6]$ region.}
\end{figure}
\begin{figure}[h]
\centering
\includegraphics[scale=0.4]{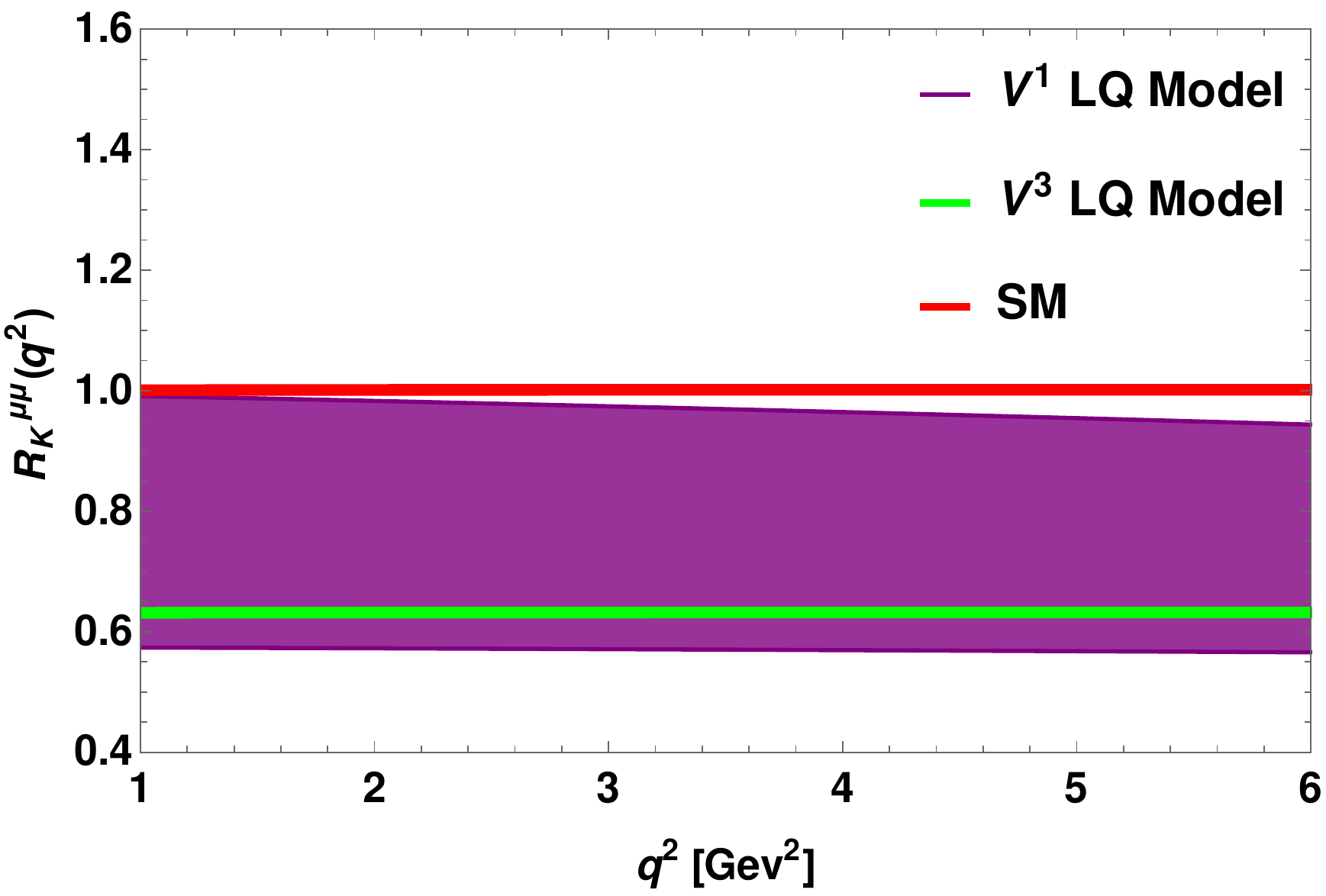}
\quad
\includegraphics[scale=0.4]{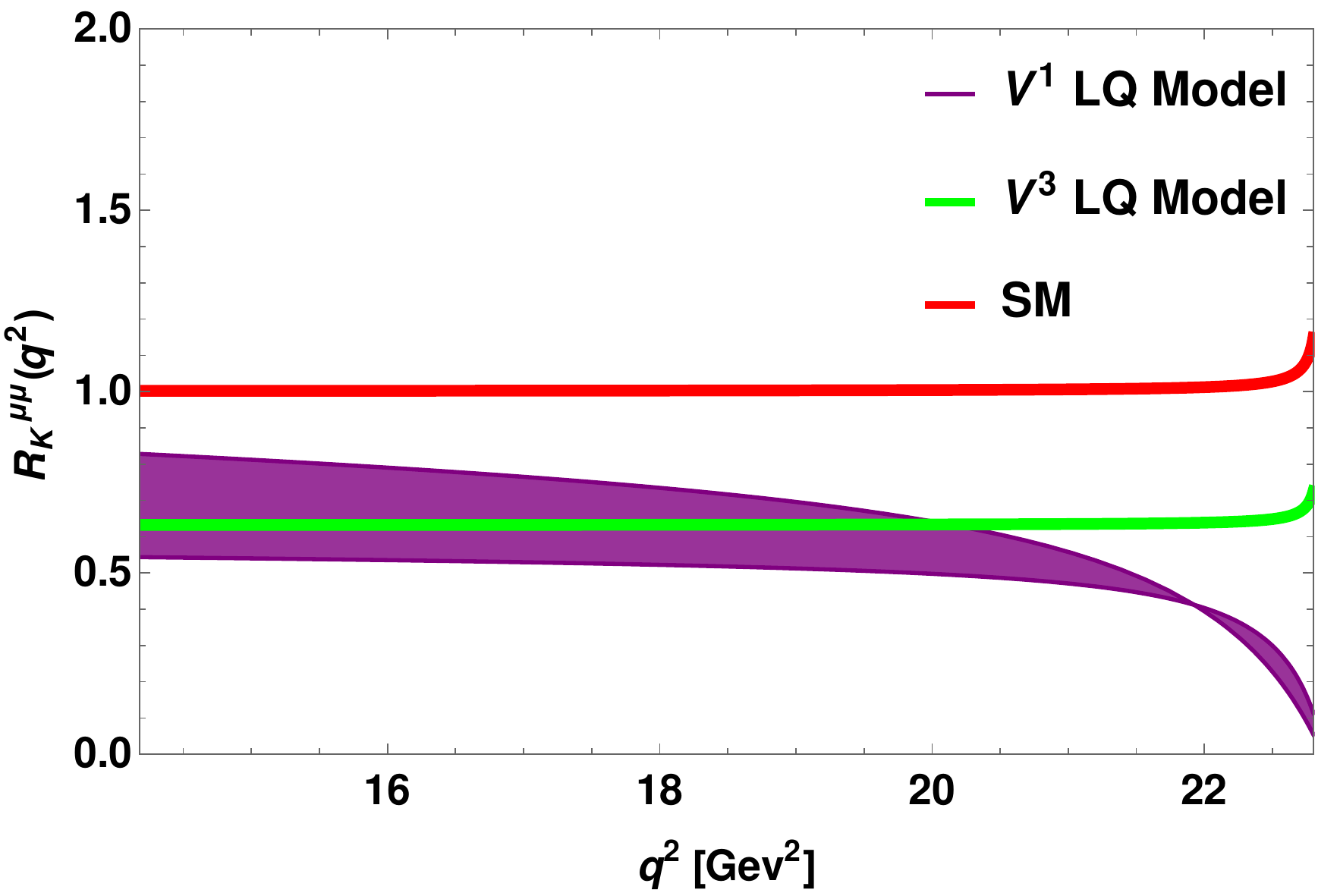}
\caption{The  variation of  lepton non-universality   parameter $R_{K}^{\mu \mu}$  for low $q^2 \in [1, 6]$ (left panel) and high $q^2$ (right panel) regimes. }
\end{figure}

\begin{figure}[h]
\centering
\includegraphics[scale=0.4]{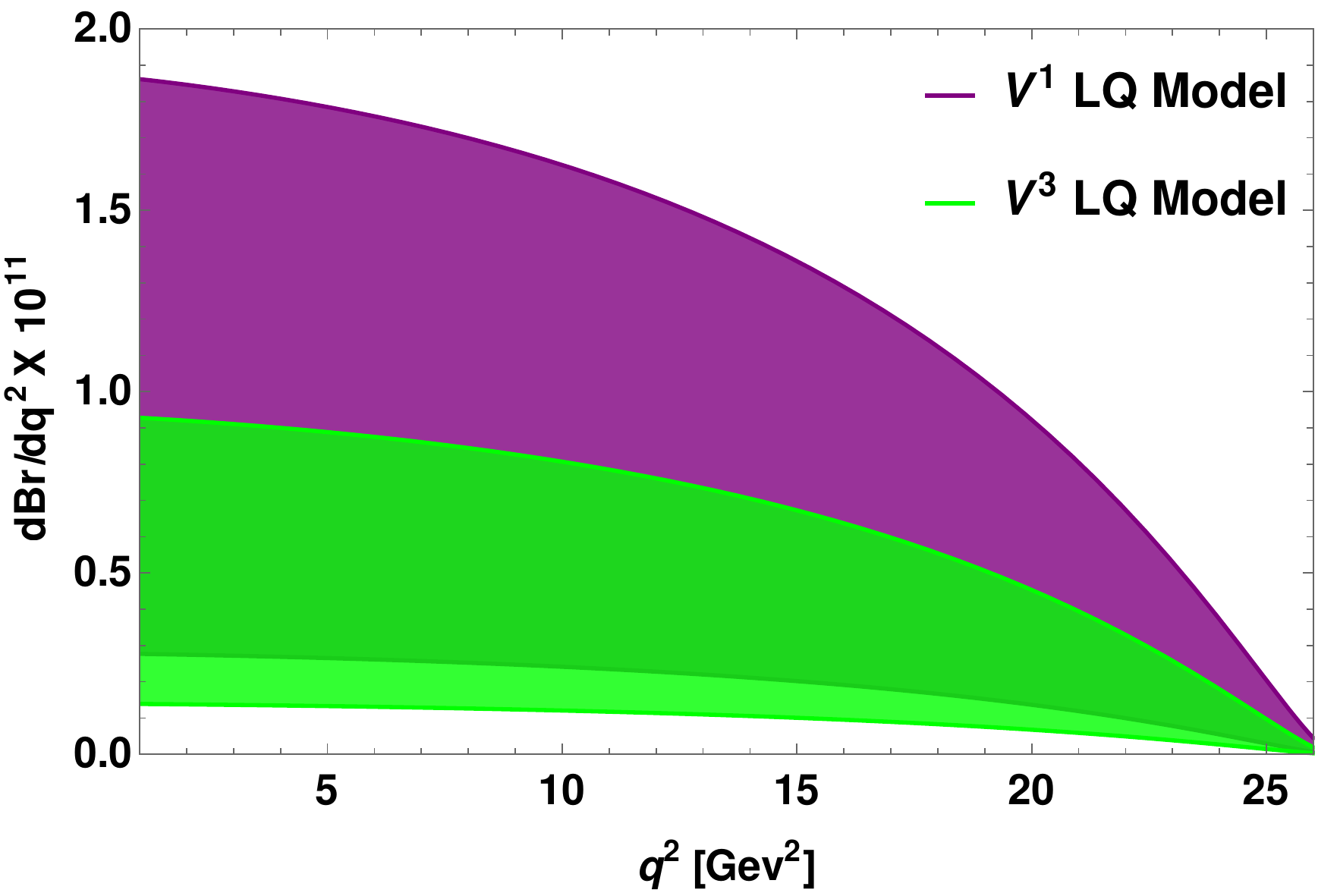}
\quad
\includegraphics[scale=0.4]{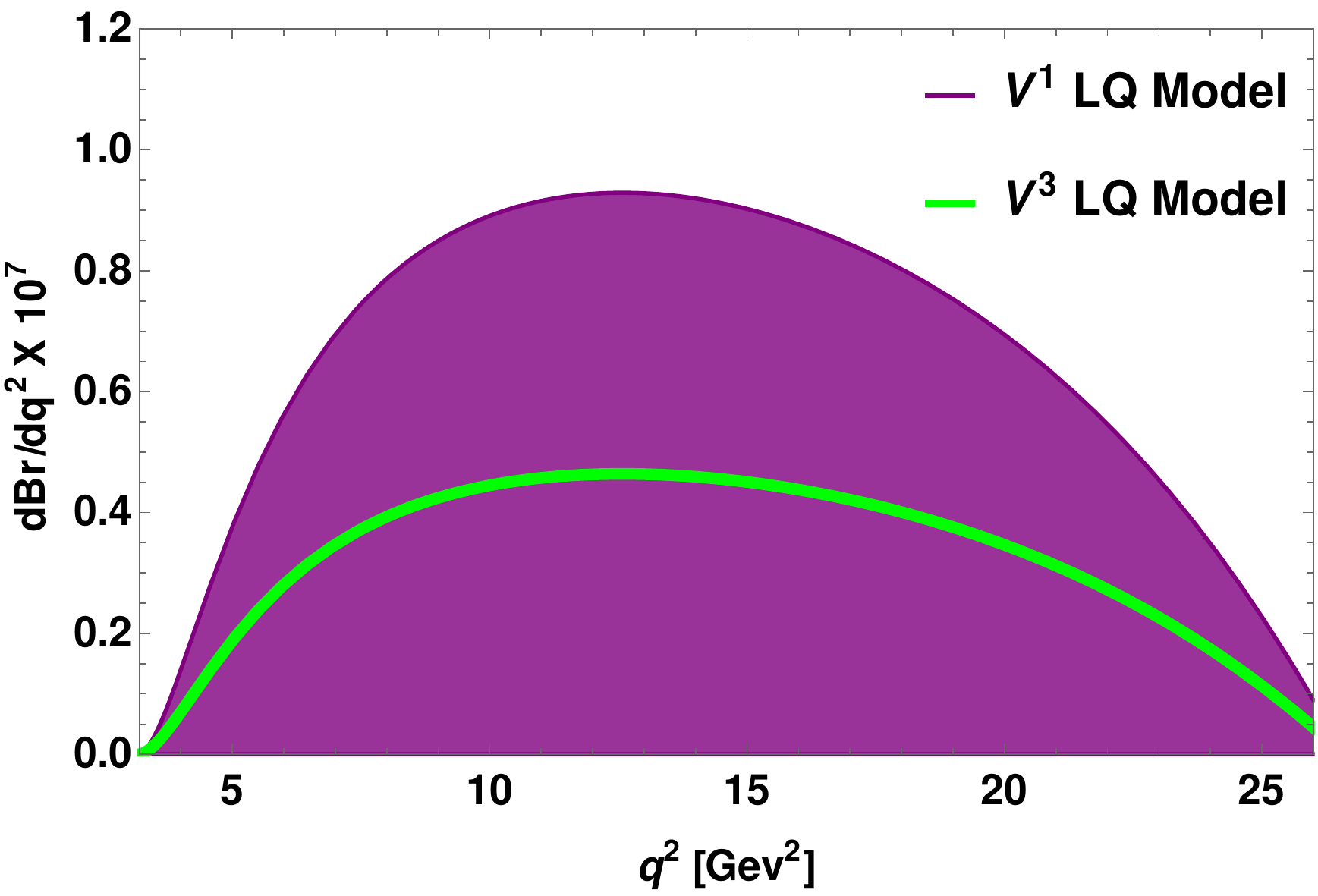}
\quad
\includegraphics[scale=0.4]{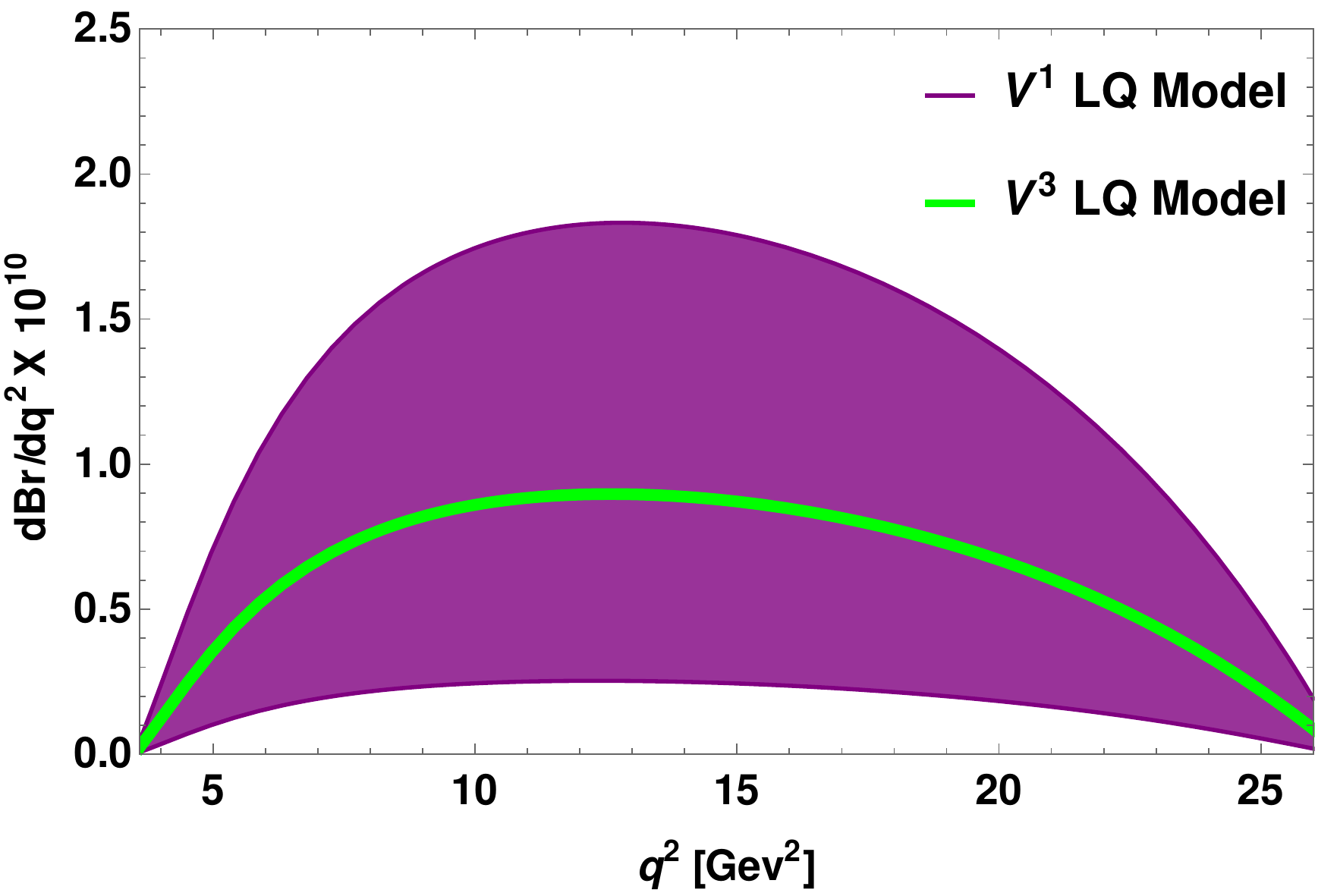}
\caption{The $q^2$ variation of branching ratios of $B^+ \to \pi^+ \mu^- e^+$ (top left panel), $B^+ \to \pi^+ \tau^- e^+$ (top right panel) and $B^+ \to \pi^+ \tau^- \mu^+$ (bottom panel) processes in the $V^{1, 3}$  vector leptoquark model. }
\end{figure}
For numerical analysis we have taken the particle masses and life times of $B_q$ mesons from \cite{pdg}. The form factors $(f_{0, +, T})$  for kaon and pion are taken from \cite{k-form-factor} and \cite{pi-form-factor} respectively. In order to compute the  required LQ  couplings, we use the values of the 
couplings extracted from $B_{s,d} \to l^+ l^-$, as given in Table I and  II. Although the bounds obtained from $K_L \to l^+ l^-$ processes (\ref{KL-bound}) are little stronger   than the bounds obtained from $B_{s,d} \to l^+ l^-$, only the Real part of the couplings can be constrained there. Therefore, in our analysis, we  consider the constraints from Table I and II as basis values and  assume that the LQ  couplings between different generation of quark and lepton follow the simple scaling law, i.e., ${(g_{L(R)})}_{ij} = (m_i/m_j)^{1/2} {(g_{L(R)})}_{ii}$ with $j \textgreater i$. This ansatz has taken from the Ref. \cite{ansatz}, which can explain the decay width of radiative LFV $\mu \to e \gamma$ decay. 
Now using the constrained LQ parameter space, we calculate the branching ratios, forward-backward asymmetries and lepton non-universality in $B \to K (\pi) l_i^- l_j^+$ processes. In Fig. 9, we show the variation of branching ratios of $B^+ \to K^+  \mu^- e^+$  (top left panel), $B^+ \to K^+  \tau^- e^+$ (top right panel) and $B^+ \to K^+  \tau^- \mu^+$ (bottom panel) processes with respect to $q^2$ in both $V^{1,3}$  leptoquark model.  Here the purple  bands represent the  predictions in the $V^1$ vector LQ model and green solid lines are for $V^3$ leptoquark. The predicted branching ratios of $B^+ \to K^+  l_i^- l_j^+$ processes in both the LQ model are presented in Table IV.  The plot for forward-backward asymmetries of $B^+ \to K^+  \mu^- e^+$  (top left panel), $B^+ \to K^+  \tau^- e^+$ (top right panel) and $B^+ \to K^+  \tau^- \mu^+$ (bottom panel) processes in the LQ model are given in Fig. 10.  Fig. 11 shows the $q^2$ variation of $R_{K e}^{\mu e}$ (top right panel),  $R_{K e}^{\tau e}$ (bottom left panel) and $R_{K e}^{\tau \mu}$ (bottom right  panel) parameters   in the high $q^2$  region.  The variation of $R_{K \mu}^{\mu e}$ (top right panel),  $R_{K \mu}^{\tau e}$ (bottom left panel) and $R_{K \mu}^{\tau \mu}$ (bottom right panel) observables are presented in Fig. 12.   The variation of  $R_{K e}^{\mu e}$ and $R_{K \mu}^{\mu e}$ parameters in low $q^2 \in [1,6]$ region  are also shown in the top left panel of Fig. 11 and Fig. 12 respectively. The integrated values of forward-backward asymmetries and lepton non-universality parameters such as  $R_K^{l_i l_j}$ $R_{K e}^{l_i l_j}$, $R_{K \mu}^{l_i l_j}$  are given in Table V. In Fig. 13 we show the plot for lepton non-universality  parameters $R_K^{\mu \mu}$ in low $q^2$ (left panel) and high $q^2$ (right panel)
and the predicted values are presented in Table V.  Here the solid red lines denote the SM contributions and integrated values in the SM are given by
\bea
R_K^{\mu \mu}|_{q^2 \in [1, 6]} = 1.001, ~~~~~R_K^{\mu \mu}|_{q^2 \geq 14.18} = 1.003, ~~~~
R_{K e}^{\tau \tau} = 1.144, ~~~~~R_{K \mu}^{\tau \tau} = 1.14.
\eea

Analogously  we show the variation of the branching ratios of LFV  $B^+  \to \pi^+   \mu^- e^+$  (top left panel), $B^+  \to \pi^+   \tau^- e^+$ (top right panel) and $B^+  \to \pi^+   \tau^- \mu^+$ (bottom panel) decay processes with respect to $q^2$ in Fig. 14 and the predicted branching ratios are given in Table IV.  Fig. 15 shows the variation of  forward-backward asymmetries in $B^+  \to \pi^+   \mu^- e^+$  (top left panel), $B^+  \to \pi^+  \tau^- e^+$ (top right panel) and $B^+  \to \pi^+   \tau^- \mu^+$ (bottom panel) processes.  The lepton non-universality parameters $R_{\pi \mu}^{\mu e}$ (top right panel), $R_{\pi \mu}^{\tau e}$ (bottom left panel) and $R_{\pi \mu}^{\tau \mu}$ (bottom right panel)  are presented in Fig. 16. Also,  we present the  behaviour of  $R_{\pi \mu}^{\mu e}$  parameter (top left panel) in the region $1\leq q^2 \leq 6~{\rm GeV}^2$. In Table VI, we present the predicted  values of  forward-backward asymmetries and   lepton non-universality parameters. The non-universality predictions of $B \to \pi l^+ l^-$ processes in the  SM  are
\bea
R_\pi^{\mu \mu}|_{q^2 \in [1, 6]} = 1.001, ~~~~~R_\pi^{\mu \mu}|_{q^2 \geq 14.18} = 1.003, ~~~~
R_{\pi e}^{\tau \tau} = 1.149, ~~~~~R_{\pi \mu}^{\tau \tau} = 1.146,
\eea
and the values for the parameter $ R_\pi^{\mu \mu}$ in the LQ model are listed in Table VI.

 The predicted values of $R_+^{l_i l_j}$ in $V^{1, 3}$ LQ model respectively are
\bea
&&R_+^{\mu e}|_{V^1 LQ} =0.525 - 53.34, ~~~~~~~~~~~~R_+^{\mu e}|_{V^3 LQ} =0.536, \\
&&R_+^{\tau \mu}|_{V^1 LQ} = 0.536 - 64.1,~~~~~~~~~~~~~~R_+^{\tau \mu}|_{V^3 LQ} =0.578,\\
&&R_+^{\tau e}|_{V^1 LQ} = 0.443 - 0.56,~~~~~~~~~~~~~~R_+^{\tau e}|_{V^3 LQ} =0.56.
\eea

\begin{table}[h]
\caption{The predicted  branching ratios of   $B^+ \to K^+(\pi^+) l_i^- l_j^+$ processes in the $V^{1, 3}$  vector leptoquark model. }
\begin{center}
\begin{tabular}{|c | c | c| c |}
\hline
 Decay process  & Values in $V^1$ LQ model &  Values in $V^3$ LQ model &   Expt. upper limit \cite{pdg}\\
 \hline
 \hline
 $B^+ \to K^+  \mu^- e^+$  & $(0.009-6.16)\times 10^{-10}$  & $\textless ~2.98\times 10^{-10}$ & $\textless ~ 9.1 \times 10^{-8}$ \\
 $B^+ \to K^+  \tau^- e^+$ & $(0.118-1.882)\times  10^{-10}$ & $\textless ~7.12 \times 10^{-11}$    & $\textless ~ 4.3 \times 10^{-5}$  \\
$B^+ \to K^+  \tau^- \mu^+$ & $(0.0064-5.58)\times  10^{-9}$  &  $\textless ~2.52 \times 10^{-9}$ & $\textless ~ 4.5 \times 10^{-5}$  \\
\hline
 $B^+ \to \pi^+  \mu^- e^+$  & $(0.48-3.23)\times 10^{-10}$&  $\textless ~1.6 \times 10^{-10}$ & $\textless ~ 6.4 \times 10^{-3}$\\
 $B^+ \to \pi^+  \tau^- e^+$ & $5.23 \times 10^{-12} - 1.51 \times 10^{-6} $ & $\textless ~7.55 \times 10^{-7}$   & $\textless ~ 7.4 \times 10^{-5} $	 \\
$B^+ \to \pi^+ \tau^- \mu^+$ & $(0.41 - 2.99)\times 10^{-9}$ & $\textless ~1.46\times 10^{-9}$ & $\textless ~ 6.2 \times 10^{-5}$ \\

\hline
\end{tabular}
\end{center}
\end{table}
\begin{figure}[h]
\centering
\includegraphics[scale=0.4]{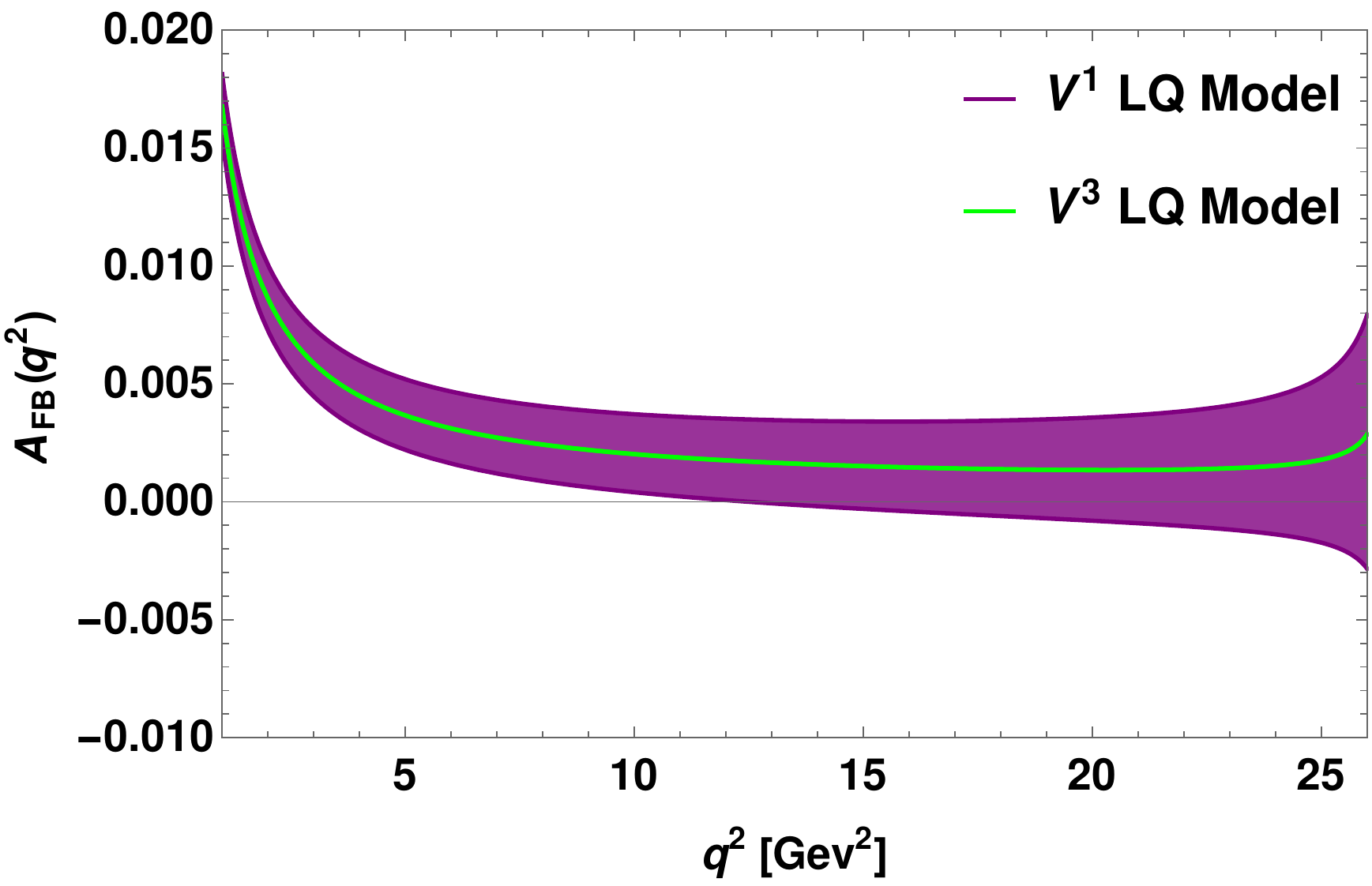}
\quad
\includegraphics[scale=0.4]{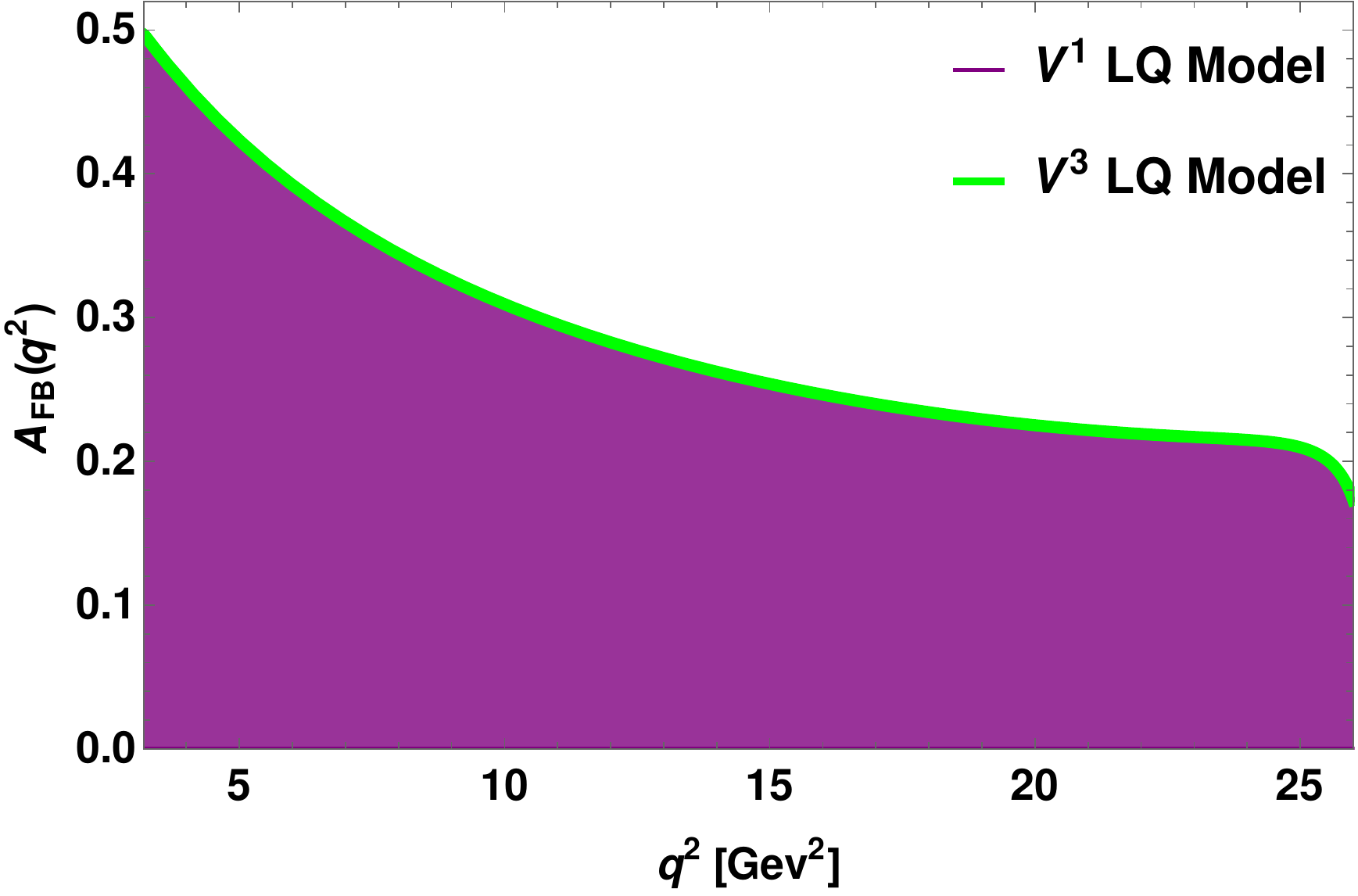}
\quad
\includegraphics[scale=0.4]{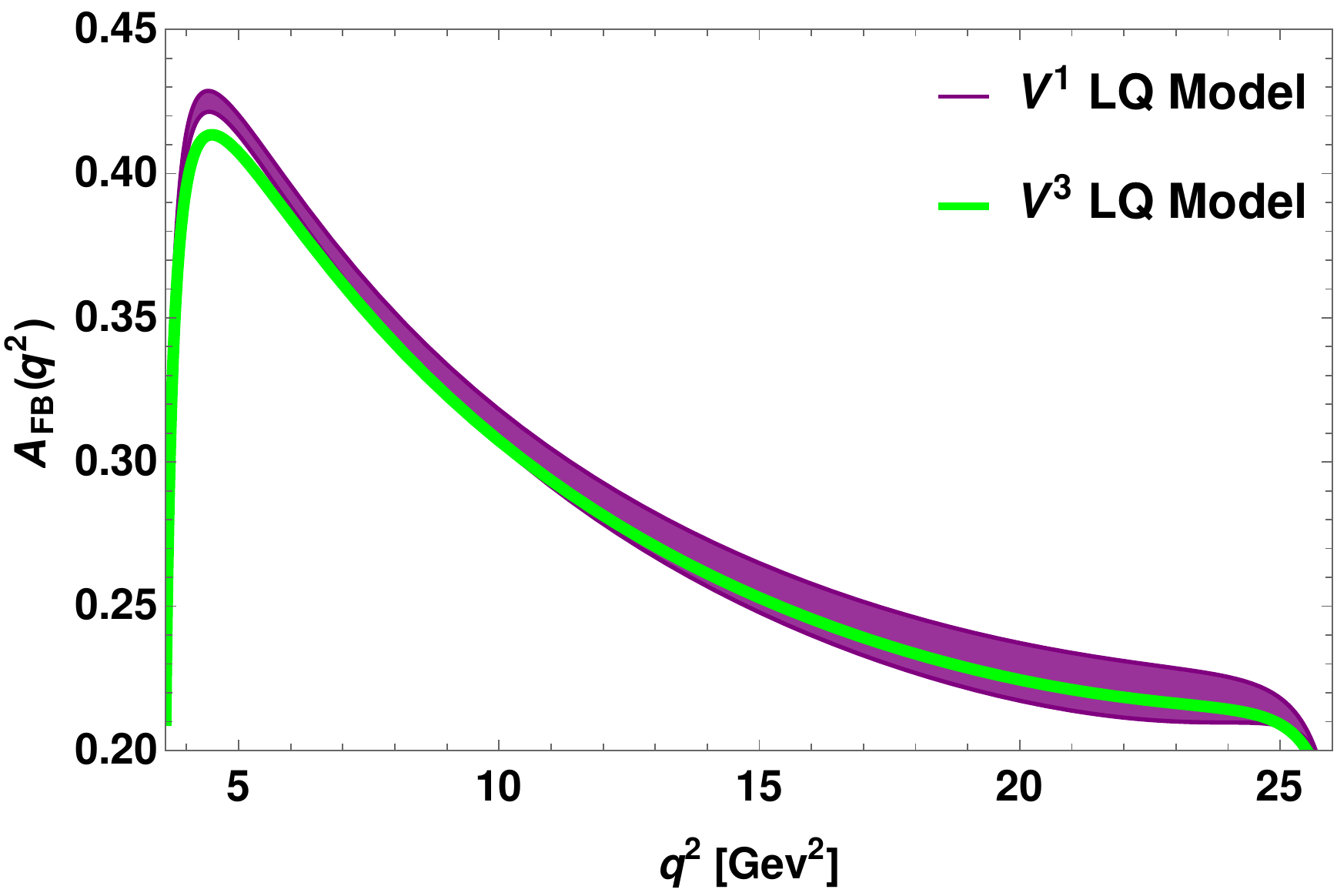}
\caption{The variations of forward-backward asymmetries of   $B^+ \to \pi^+ \mu^- e^+$ (left panel), $B^+ \to \pi^+ \tau^- e^+$ (middle panel) and $B^+ \to \pi^+ \tau^- \mu^+$ (right panel) processes in  leptoquark model. }
\end{figure}
\begin{table}[h]
\caption{The predicted  values of forward backward asymmetries and  lepton non-universality   parameters in $B^+ \to K^+ l_i^- l_j^+$ process   in the $V^{1, 3}$ vector leptoquark model.   Also the  predicted values of $R_{K}^{\mu \mu}$,   $R_{K e}^{l_i l_j}$, $R_{K \mu}^{l_i l_j}$ parameters for muonic (electronic) processes in  low $q^2 \in [1, 6]$ region. }
\begin{center}
\begin{tabular}{|c | c | c| }
\hline
 Observables  & Values in $V^1$ LQ model &  Values in $V^3$ LQ model   \\
 \hline
 \hline
 
 $\langle A_{FB}^{K \mu e}\rangle$ & $7.65\times 10^{-3}$ & $2.82 \times 10^{-3}$  \\
 
$\langle A_{FB}^{K \tau e}\rangle$	& $0.285$ &$0.285$ \\

$\langle A_{FB}^{K \tau \mu}\rangle$  & $(0.039-0.105) \times 10^{-3} $ & $ 4.8 \times 10^{-3}$  \\

\hline

$\langle R_{K e}^{\mu e}\rangle|_{q^2 \in [1,6]}$ & $(0.0038 - 3.5)\times 10^{-4}$ & $2.03 \times 10^{-4}$ \\

$\langle R_{K e}^{\mu e}\rangle$ & $(0.033-3.36)\times 10^{-4}$ & $2.04 \times 10^{-4}$ \\

$\langle R_{K e}^{\tau e}\rangle$ & $0.526 \times 10^{-4} - 2.52 $	& $1.63$  \\

$\langle  R_{K e}^{\tau \mu }\rangle$ & $(0.285- 5.45)\times 10^{-3}$	& $ 3.04 \times 10^{-3}$  \\

 \hline

$\langle R_{K \mu}^{\mu e}\rangle|_{q^2 \in [1,6]}$ & $(0.0039 - 6.1) \times 10^{-4}  $ & $3.2 \times 10^{-4}$  \\

$\langle R_{K \mu}^{\mu e}\rangle$ & $(0.0454-6.44)\times 10^{-4}$ & $3.3 \times 10^{-4}$   \\

$\langle R_{K \mu}^{\tau e}\rangle$ & $0.72\times 10^{-4} - 4.83$ & $2.58$   \\

$\langle  R_{K \mu}^{\tau \mu }\rangle$ & $(0.039-0.1)\times 10^{-3}$ & $4.8 \times 10^{-3}$     \\

\hline

$\langle R_{K}^{\mu \mu} \rangle|_{q^2 \in [1,6]}$ & $0.57-0.9688$ & $0.63$   \\

$\langle R_{K}^{\mu \mu} \rangle$ & $0.521-0.73$ & $0.64$  \\

\hline
\end{tabular}
\end{center}
\end{table}
\begin{figure}[h]
\centering
\includegraphics[scale=0.4]{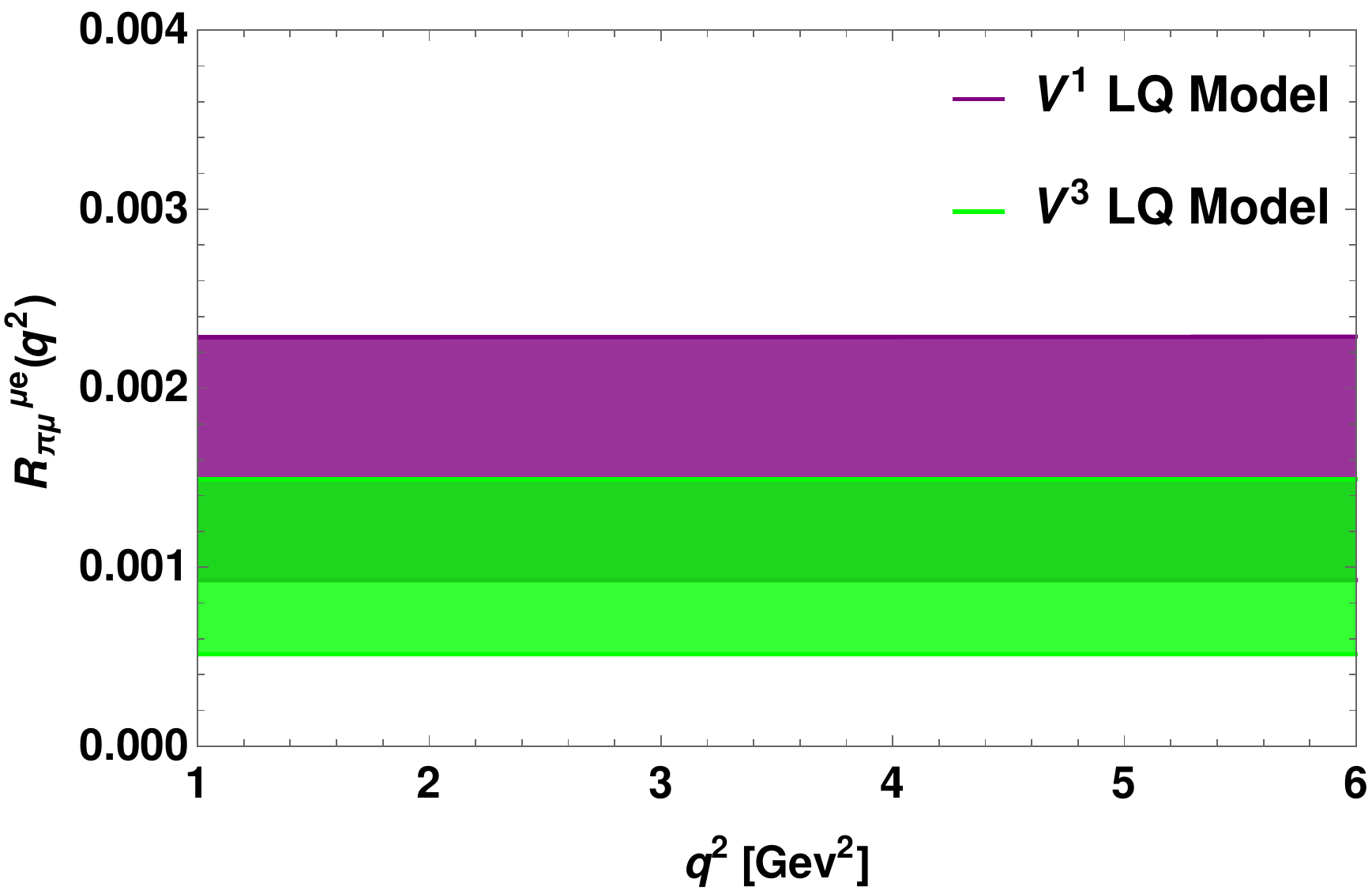}
\quad
\includegraphics[scale=0.4]{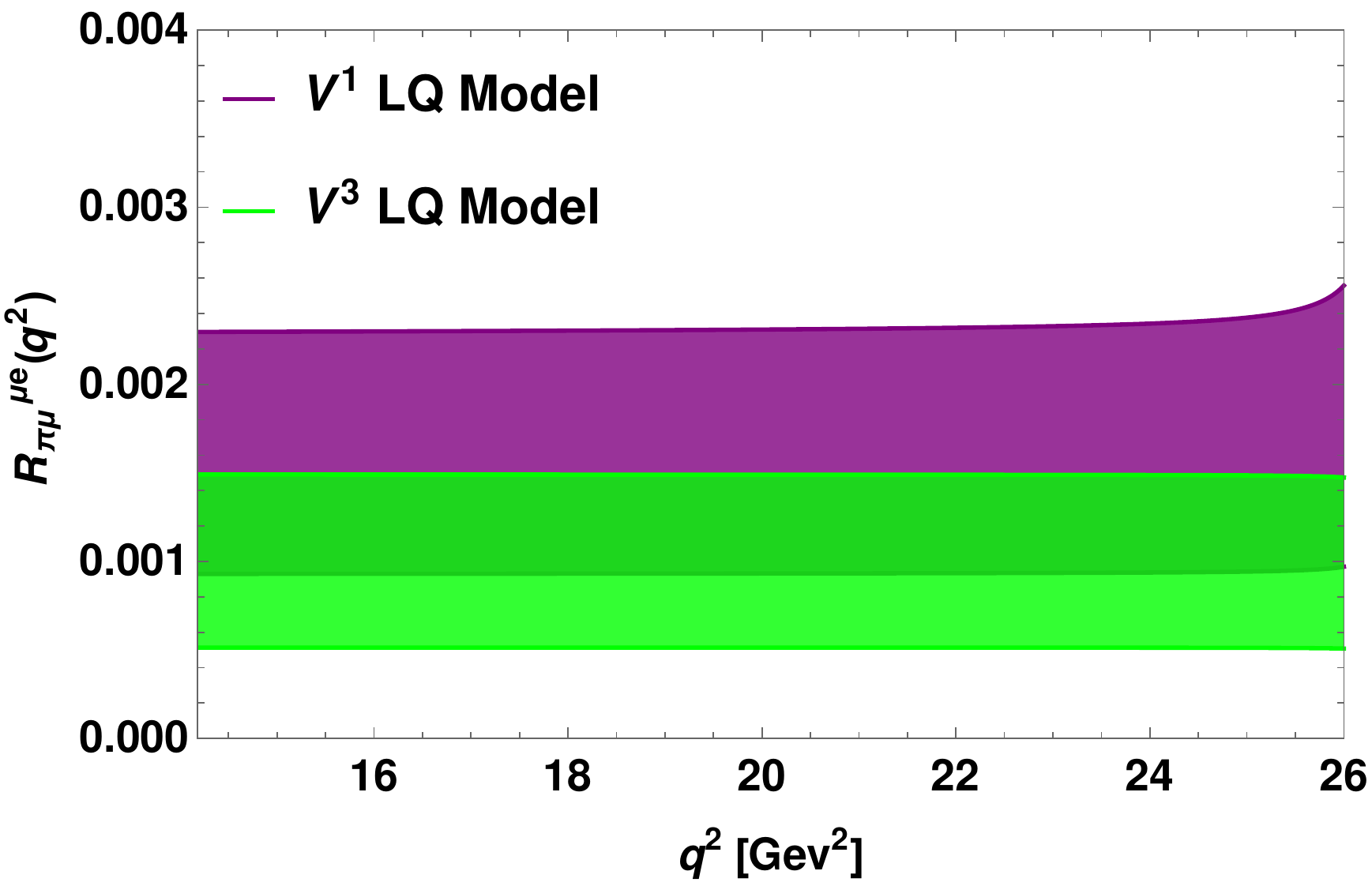}
\quad
\includegraphics[scale=0.4]{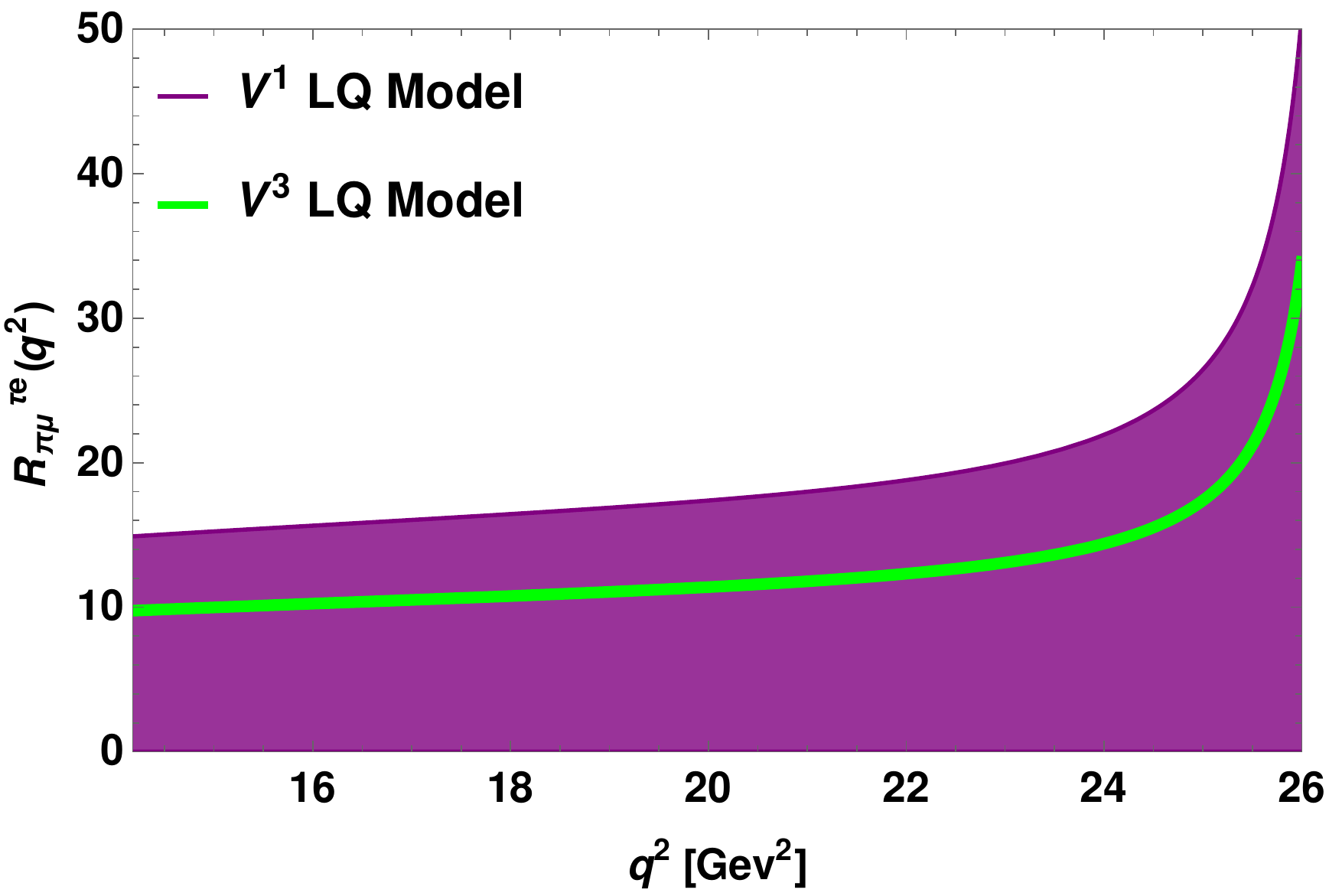}
\quad
\includegraphics[scale=0.4]{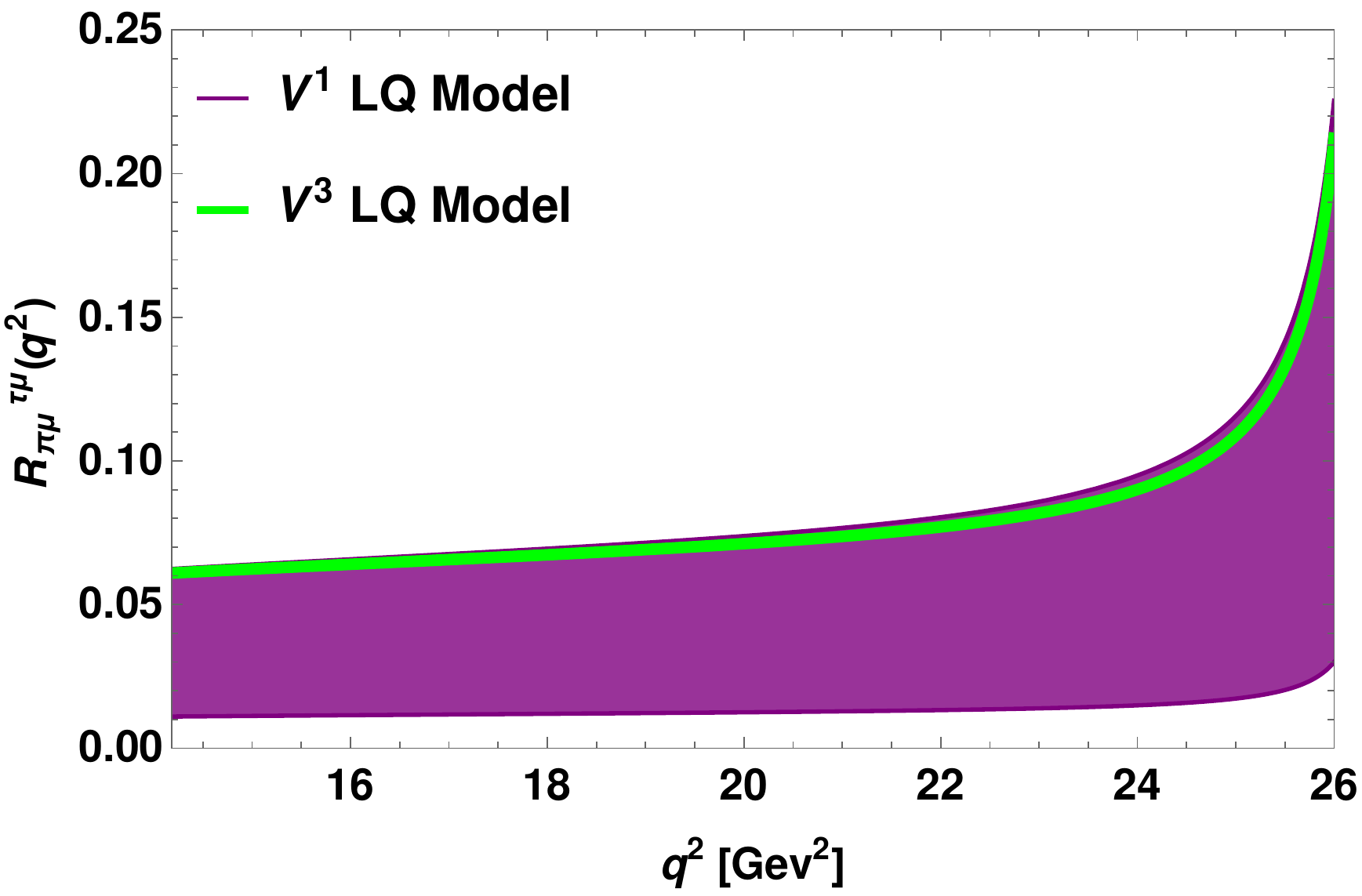}
\caption{The $q^2$ variations of $R_{\pi \mu}^{\mu e}$ (top right panel), $R_{\pi \mu}^{\tau e}$ (bottom left panel) and $R_{\pi \mu}^{\tau \mu}$ (bottom right panel) parameters in high $q^2$ region in the leptoquark model. Here $R_{\pi \mu}^{\mu e}$ (top left panel) shows the lepton non-universality for low $q^2 \in [1, 6]$ region.}
\end{figure}
\begin{table}[h]
\caption{The predicted  values of forward backward asymmetries and  lepton non-universality   parameters in $B \to \pi l_i^- l_j^+$ process   in the $V^{1, 3}$ vector leptoquark model.   Also the  predicted values of $R_{\pi}^{\mu \mu}$,   $R_{\pi e}^{l_i l_j}$, $R_{\pi \mu}^{l_i l_j}$ parameters for muonic (electronic) processes in  low $q^2 \in [1, 6]$ region. }
\begin{center}
\begin{tabular}{|c | c | c| }
\hline
 Observables   & Values in $V^1$ LQ model &  Values in $V^3$ LQ model   \\
 \hline
 \hline

 $\langle A_{FB}^{\pi \mu e}\rangle$  & $(0.432 - 4.42)\times 10^{-3}$& $ 2.4 \times 10^{-3}$  \\
 
 $\langle A_{FB}^{\pi \tau e}\rangle$ & $0.2632 $  & $0.263$ 	 \\
 
$\langle A_{FB}^{\pi \tau \mu}\rangle$ & $0.263 - 0.271$ & $0.26$  \\

\hline
 
$\langle R_{\pi}^{\mu \mu} \rangle|_{q^2 \in [1,6]}$ & $1.084\times 10^{-7} - 1.95$ & $1.655 \times 10^{-7}$  \\

$\langle R_{\pi}^{\mu \mu} \rangle$ & $1.09\times 10^{-7} - 0.6$ & $1.66 \times 10^{-7}$   \\

\hline

$\langle R_{\pi e}^{\mu e}\rangle|_{q^2 \in [1,6]}$ & $2.48 \times 10^{-10} -  1.8 \times 10^{-3}$ & $2.46 \times 10^{-10}$  \\

$\langle R_{\pi e}^{\mu e}\rangle$ &  $ 2.5 \times 10^{-10} - 5.5 \times 10^{-4} $ & $2.47 \times 10^{-10}$   \\

$\langle R_{\pi e}^{\tau e}\rangle$ & $(0.0187 - 1.53)\times 10^{-4} $ & $1.86 \times 10^{-6}$   \\

$\langle  R_{\pi e}^{\tau \mu }\rangle$ & $3.75 \times 10^{-9} - 7.33 \times 10^{-3}$ & $3.6 \times 10^{-9}$  \\

\hline

$\langle R_{\pi \mu}^{\mu e}\rangle|_{q^2 \in [1,6]}$ & $(1.8 - 2.29) \times 10^{-3}$ & $1.492\times 10^{-3}$ \\

$\langle R_{\pi \mu}^{\mu e}\rangle$ & $(0.932-2.3)\times 10^{-3}$ & $1.49 \times 10^{-3}$ \\

$\langle R_{\pi \mu}^{\tau e}\rangle$ & $2.57 \times 10^{-4} - 17.14$	& $11.23$  \\

$\langle  R_{\pi \mu}^{\tau \mu }\rangle$ & $0.0123 - 0.073$	& $0.07$   \\

\hline
\end{tabular}
\end{center}
\end{table}

\section{$(g-2)_\mu$}
The recent experimental measurement \cite{muon-exp} of the anomalous magnetic moment of  muon, i.e., $(g-2)_\mu$ has about $3\sigma$ discrepancies from the SM prediction and has set off a  flurry of excitement amongst theorists.   The experimental result of anomalous magnetic moment of muon is given by \cite{muon-SM}
\bea
a_\mu^{\rm exp} = 116592080(63) \times 10^{-11},
\eea
which when compared to the SM value
\bea
a_\mu^{\rm SM} = 116591785(61) \times 10^{-11},
\eea
has the discrepancy
\bea \label{a-mu}
\Delta a_\mu = a_\mu^{\rm exp} - a_\mu^{\rm SM} = (295 \pm 88) \times 10^{-11}. 
\eea
The absolute magnitude of the discrepancy is small and can be accommodate by adding the new physics contributions. The vector LQ contribution to $a_\mu$ is given by
\bea
\Delta a_\mu &=& -2N_c m_\mu \Bigg [  (g_L)_{b\mu} (g_R)_{b\mu}^* \Big( -\frac{1}{3} \left( f_3(x_b) + f_4(x_b) \right) +\frac{2}{3} \left( \bar{f_3}(x_b) + \bar{f_4}(x_b) \right) \Big ) \nn \\ && ~~~~~~~~~~~~~~ +  \Big( |(g_L)_{b\mu}|^2 + |(g_R)_{b\mu}|^2 \Big ) \Big ( -\frac{1}{3} f_1(x_b) +\frac{2}{3}\bar{f_1}(x_b) \Big ) \Bigg ] \nn \\ &&  -2N_c m_\mu \Big( |(g_L)_{b\mu}|^2 + |(g_R)_{b\mu}|^2 \Big ) \Big ( -\frac{1}{6} f_1(x_b) +\frac{1}{3}\bar{f_1}(x_b) \Big ), 
\eea
where the loop functions 
 are given in sec. V C.  Now using the constrained  leptoquark couplings from $\tau^- \to \mu^- \gamma$ process with the  scaling law as discussed in section VI,  $\Delta a_\mu$ in the leptoquark model  is found as
 \bea
 2.38 \times 10^{-9} \leq \Delta a_\mu \leq 2.95 \times 10^{-9},
 \eea 
which is within its $1\sigma$  range. 
\section{conclusion}
In this work, we have studied the rare lepton flavour violating semileptonic $B$ meson decays in the vector leptoquark model.  These decays occur at loop level with a tiny neutrino mass in one of the loop, thus extremely rare in the SM. Whereas these processes can occur at tree level in the vector leptoquark model. There are three  vector leptoqaurks which are   relevant to study the processes mediated via $b \to (s, d) $ transitions. Of these we consider $(3, 3, 2/3)$ and $(3, 1, 2/3)$  vector leptoquarks in our analysis and constrained the leptoquark couplings from  $B_{s, d} \to l^+ l^-$, $K_L \to l^+ l^-$ and $\tau^- \to l^- \gamma$  processes, where $l$ can be any charged leptons. Using such constrained parameters,  we estimated the branching ratios  and forward-backward  asymmetries of $B \to K l_i^- l_j^+$ and $B \to \pi l_i^- l_j^+$ processes in the vector  leptoquark model.  We also computed some parameters like $R_{K(\pi) e}^{l_i l_j}$, $R_{K(\pi) \mu}^{l_i l_j}$ and $R_{+}^{l_i l_j}$  (the ratios of  various combination of rare  decays) in order to inspect the presence of lepton non-universality. We also study the effect of vector leptoquark on the muon $g-2$ anomaly.  We found that our predicted values are sizeable and within the reach of  currently running/upcoming experimental limits,
the observation of which in the LHCb  experiment would provide univocal signal of new physics.

{\bf Acknowledgments}

RM  and SS would like to thank Science and Engineering Research Board (SERB), Government of India for financial support through grant No. SB/S2/HEP-017/2013.


\begin{thebibliography}{60}


\bibitem{p5p}
S. Descotes-Genon, J. Matias, M. Ramon and J. Virto,  JHEP {\bf 1301}, 048 (2013) 

\bibitem{lhcb1}
R. Aaij et al., [LHCb Collaboration], Phys. Rev. Lett. \textbf{111}, 191801 (2013) [arXiv:1308.1707].


\bibitem{lhcb2}
R. Aaij et al., [LHCb Collaboration], JHEP \textbf{1406}, 133 (2014) [arXiv:1403.8044].


\bibitem{lhcb3}
R. Aaij et al., [LHCb Collaboration], Phys. Rev. Lett. \textbf{113}, 151601 (2014) [arXiv:1406.6482].

\bibitem{bobeth2}
C. Bobeth, G. Hiller and G. Piranishvili, JHEP \textbf{0712}, 040 (2007) [arXiv:0709.4174].

\bibitem{lhcb4}
R. Aaij et al., [LHCb Collaboration], JHEP \textbf{1307}, 084 (2013) [arXiv:1305.2168].

\bibitem{glashow} S. L. Glashow, D. Guadagnoli, K. Lane, Phys. Rev. Lett. {\bf 114}, 091801 (2015)  [arXiv:1411.0565].

\bibitem{h-to-taumu}
V. Khachatryan et al. [CMS Collaboration], Phys. Lett. B \textbf{749}, 337 (2015) [arXiv:1502.07400].

\bibitem{lfv}
 Chao-Jung Lee and
J. Tandean, JHEP {\bf 08}, 123 (2015), [arXiv: 1505.04692]; W. Altmannshofer and I. Yavin,
Phys Rev D. {\bf 92}, 075022 (2015) [arXiv:1508.07009]; A. Crivellin, G. D'Ambrosio, J. Heeck, Phys. Rev. Lett. \textbf{114}, 151801 (2015) [arXiv:1501.00993]; L. Calibbi, A. Crivellin, T. Ota,  	Phys. Rev. Lett. \textbf{115}, 181801 (2015) [arXiv:1506.02661]; R. Alonso, B. Grinstein and J. M. Camalich, [arXiv:1505.05164]; A. Crivellin, L. Hofer, J. Matias, U. Nierste, S. Pokorski, S. Rosiek, Phys. Rev. D {\bf 92}, 054013 (2015) 
[arXiv:1504.07928].
 
\bibitem{mohanta4}
 S. Sahoo and R. Mohanta, Phys. Rev. D \textbf{93}, 114001 (2016)  [arXiv:1512.04657].
 
 \bibitem{mohanta2}
 S. Sahoo and R. Mohanta, Phys. Rev. D \textbf{91}, 094019 (2015) [arXiv:1501.05193].
  
 
 \bibitem{kosnik-new}
 D. Becirevic, N. Kosnik, O. Sumensari, and R. Zukanovich Funchal,  [arXiv:1608.07583].

\bibitem{pdg}
 K.A. Olive et al. (Particle Data Group), Chin. Phys. C  \textbf{38}, 090001 (2014).

 \bibitem{georgi}
  H. Georgi and S. L. Glashow, Phys. Rev. Lett.    \textbf{32}, 438 (1974); J. C. Pati and A. Salam, Phys. Rev. D \textbf{10}, 275 (1974).
  
\bibitem{georgi2}
   H. Georgi, AIP Conf. Proc. \textbf{23} 575 (1975); H. Fritzsch and P. Minkowski, Annals Phys. \textbf{93}, 193 (1975); P. Langacker, Phys. Rep. \textbf{72}, 185 (1981).
   
 \bibitem{kaplan}
  D. B. Kaplan, Nucl. Phys. B \textbf{365}, 259 (1991).
  
 \bibitem{schrempp}
B. Schrempp and F. Shrempp, Phys. Lett. B \textbf{153}, 101 (1985); B. Gripaios, JHEP \textbf{1002}, 045 (2010) [arXiv:0910.1789].

 \bibitem{mohanta1}
 R. Mohanta, Phys. Rev. D \textbf{89}, 014020 (2014) [arXiv:1310.0713].

\bibitem{leptoquark}
   S. Davidson, D. C. Bailey and B. A. Campbell, Z. Phys. C \textbf{61}, 613 (1994) [arXiv:hep-ph/9309310]; I. Dorsner, S. Fajfer, J. F. Kamenik, N. Kosnik, Phys. Lett. B \textbf{682}, 67 (2009) [arXiv:0906.5585]; S. Fajfer, N. Kosnik, Phys. Rev. D \textbf{79}, 017502 (2009) [arXiv:0810.4858];
R. Benbrik, M. Chabab, G. Faisel, [arXiv:1009.3886]; 
A. V. Povarov, A. D. Smirnov, [arXiv:1010.5707]; J. P Saha, B. Misra and A. Kundu, Phys. Rev. D \textbf{81}, 095011 (2010) [arXiv:1003.1384]; 
I. Dorsner, J. Drobnak, S. Fajfer, J. F. Kamenik, N. Kosnik, JHEP \textbf{11}, 002 (2011) [arXiv:1107.5393]; F. S. Queiroz, K. Sinha, A. Strumia, Phys. Rev. D \textbf{91}, 035006 (2015) [arXiv:1409.6301]; B. Allanach,A. Alves, F. S. Queiroz, K. Sinha,
A. Strumia,  Phys. Rev. D \textbf{92}, 055023 (2015) [arXiv:1501.03494]; R. Alonso, B. Grinstein and J. M. Camalich, [arXiv:1505.05164];
  Ivo de M. Varzielas, G. Hiller, JHEP, \textbf{1506}, 072 (2005), [arXiv:1503.01084];  
  M. Bauer and M. Neubert, Phys. Rev. Lett. \textbf{116}, 141802 (2016) [arXiv:1511.01900];
S. Fajfer and N. Kosnik,  Phys. Lett. B, \textbf{755}, 270 (2016)  [arXiv:1511.06024];  I. Dorsner, S. Fajfer, A. Greljo, J. F. Kamenik, and  N. Kosnik,  Phys. Rep.,  \textbf{641}, 1 (2016) [arXiv:1603.04993]; S.-w. Wang and  Ya-dong Yang, [arXiv:1608.03662]; D. Aristizabal Sierra, M. Hirsch, S. G. Kovalenko, Phys. Rev. D \textbf{77}, 055011 (2008), [arXiv:0710.5699]; 
K.S. Babu, J. Julio, Nucl. Phys. B \textbf{841}, 130 (2010), [arXiv:1006.1092]; S. Davidson, S. Descotes-Genon,  JHEP \textbf{1011}, 073 (2010), [arXiv:1009.1998]; S. Fajfer, J. F. Kamenik, I. Nisandzic, J. Zupan, Phys. Rev. Lett. \textbf{109}, 161801, 
(2012) [arXiv:1206.1872]; K. Cheung,  D. A. Camargo, [arXiv:1509.04263]; 
S. Baek, K. Nishiwaki, Phys. Rev. D \textbf{93}, 015002 (2016) [arXiv:1509.07410];
 J. M. Arnold, B. Fornal and M. B. Wise, Phys. Rev. D \textbf{88}, 035009 (2013), [arXiv:1304.6119];
 D.  A.  Faroughy, A.  Greljo and  J.  F.  Kamenik, [arXiv:1609.07138]; D.   Becirevic, S.  Fajfer, N.  Kosnik and  O.  Sumensari, [arXiv:1608.08501]; 
 C.-Hung Chen,  T. Nomura, and H.  Okada, [arXiv:1607.04857]; G. Kumar, Phys. Rev. D \textbf{94}, 014022 (2016) [arXiv:1603.00346];  R. Barbieri, G. Isidori, A. Pattori and F. Senia, Euro. Phys. Jour. C {\bf 76}, 67 (2016) [arXiv:1512.01560].
    
\bibitem{mohanta3}
S. Sahoo and R. Mohanta, Phys. Rev. D. \textbf{93}, 034018  [arXiv:1507.02070];  New J. Phys. \textbf{18}, 013032 (2016) [arXiv:1509.06248]; New J. Phys. \textbf{18}, 093051 (2016) [arXiv:1607.04449]; [arXiv:1612.02543].

\bibitem{RD-star-LQ}
     M. Freytsis, Z. Ligeti and J. T. Ruderman, Phys. Rev. D
\textbf{92}, 054018 (2015) [arXiv:1506.08896]; I. Dorsner, S. Fajfer, J. F. Kamenik and N. Kosnik, Phys. Lett. B \textbf{682}, 67 (2009) [arXiv:0906.5585];   Xin-Q. Li, Ya-D. Yang, X. Zhang,  [arXiv:1605.09308]; B. Dumont, K. Nishiwaki, R. Watanabe,  Phys. Rev. D \textbf{94}, 034001 (2016) [arXiv:1603.05248]; S.Sahoo, R. Mohanta, A. K. Giri, [arXiv:1609.04367]; G. Hiller, D. Loose, K. Sch\"{o}nwald, [arXiv:1609.08895]; B. Bhattacharya, A. Datta, J. Guevin, D. London, R. Watanabe, [arXiv:1609.09078]

\bibitem{10-LQ}
W. Buchmuller, R. Ruckl and D. Wyler, Phys. Lett. B \textbf{191}, 442 (1987); Erratum-ibid. B \textbf{448},
320 (1997).

\bibitem{kosnik}
  N. Kosnik, Phys. Rev. D \textbf{86}, 055004 (2012), [arXiv:1206.2970].

\bibitem{tau-mu-gamma}
I. Dorsner, S. Fajfer, A. Greljo, J.  F. Kamenik, N.  Kosnik, I. Nisandzic, doi:10.1007/JHEP06(2015)108, [arXiv:1502.07784]; K. Cheung, W.-Y. Keung, P.-Y. Tseng, Phys. Rev. D \textbf{93}, 015010 (2016) [arXiv:1508.01897]; S. Baek, K. Nishiwaki, Phys. Rev. D \textbf{93}, 015002 (2016) [arXiv:1509.07410].

\bibitem{Hamiltonian}
A. J. Buras and M. Munz, Phys. Rev. D \textbf{52}, 186 (1995); M. Misiak, Nucl. Phys. B\textbf{ 393}, 23
(1993); {\it ibid}. \textbf{439}, 461 (E) (1995).


\bibitem{Rplus}
R. Aaij et al. [LHCb Collaboration] JHEP \textbf{12}, 135 (2012) [arXiv:1210.2645].

\bibitem{BR-leptonic}
A. J. Buras, R. Fleischer, J. Girrbach, R. Knegjens,  	JHEP \textbf{1307}, 77 (2013), [arXiv:1303.3820].

\bibitem{BR-leptonic-2}
 K. De Bruyn,  R. Fleischer, R. Knegjens,  P. Koppenburg,  M. Merk,  A. Pellegrino and N. Tuning, Phys. Rev. Lett. \textbf{109}, 041801 (2012).

 \bibitem{bobeth1}
C. Bobeth, M. Gorbahn, T. Hermann, M. Misiak, E. Stamou, M. Steinhauser, Phys. Rev.Lett.
\textbf{112}, 101801 (2014) [arXiv:1311.0903].

\bibitem{average}
LHCb, CMS Collaboration, V. Khachatryan et al., Nature \textbf{522}, 68 (2015) [arXiv:1411.4413].

\bibitem{CDF}
T. Aaltonen et al. (CDF Collaboration), Phys. Rev. Lett. \textbf{102}, 201801 (2009).

\bibitem{tau-expt}
LHCb Collaboration, LHCb-CONF-2016-011, https://cds.cern.ch/record/2220757.

\bibitem{K-Isdori}
G. Isdori and R. Unterdorfer, JHEP \textbf{0401}, 009 (2004) [arXiv:hep-ph/0311084].

\bibitem{K-Buchalla}
G. Buchalla and A. J. Buras, Nucl. Phys. B \textbf{412}, 106 (1994), [arXiv:hep-ph/9308272].

\bibitem{K-Misiak}
M. Misiak and J. Urban, Phys. Lett. B \textbf{451}, 161 (1999), [arXiv:hep-ph/9901278]; G. Buchalla
and A. J. Buras, Nucl. Phys. B \textbf{548}, 309 [arXiv:hep-ph/9901288].



\bibitem{Lavoura}
L. Lavoura, Eur. Phys. J. C \textbf{29}, 191 (2003) [arXiv:hep-ph/0302221].

\bibitem{k-form-factor}
 J. A. Bailey \textit{et  al}. (Fermilab Lattice and MILC Collaborations), (2015) [arXiv:1509.06235].

\bibitem{pi-form-factor}
A. Khodjamirian, T. Mannel, N. Offen, and Y.-M. Wang, Phys. Rev. D \textbf{83}, 094031 (2011) 

\bibitem{ansatz}
B. Gripaios, M. Nardecchia, S. A. Renner, JHEP \textbf{1505}, 006 (2015) [arXiv:1412.1791]; S. Davidson, G. Isidori, and S. Uhlig, Phys. Lett. B \textbf{663}, 73 (2008)  [arXiv:0711.3376];
M. Redi, JHEP \textbf{1309}, 060 (2013)  [arXiv:1306.1525].

\bibitem{muon-exp}
G. W. Bennett {\it et  al}.
[Muon g-2 Collaboration], Phys. Rev. Lett. \textbf{92}, 161802 (2004) [arXiv:hep-
ex/0401008].

\bibitem{muon-SM}
J. P. Miller, E. de Rafael and B. L. Roberts, Rept. Prog. Phys. \textbf{70}, 795
(2007) [arXiv:hep-ph/0703049].



\end{thebibliography}
\end{document}